\documentclass[a4paper,12pt]{article}

\usepackage{amsmath}
\usepackage{amssymb}
\usepackage{amsthm}
\usepackage{authblk}
\usepackage{bm}
\usepackage{bbm}
\usepackage{color}
\usepackage[margin=2cm]{geometry}
\usepackage{graphicx}
\usepackage{hyperref}
\usepackage[authoryear]{natbib}
\usepackage{textcomp}
\usepackage{gensymb}
\usepackage{multicol}

\setcitestyle{round}
\hypersetup{
  colorlinks=true,
  linkcolor=blue,
  filecolor=blue,
  urlcolor=blue,
  citecolor=blue
}

\newcommand{\Gau}{\mathrm{Gau}}

\newcommand{\Phivec}{\bm{\Phi}}
\newcommand{\Psivec}{\bm{\Psi}}
\newcommand{\Sigmavec}{\boldsymbol{\Sigma}}

\newcommand{\Avec}{\mathbf{A}}

\newcommand{\Dvec}{\mathbf{D}}
\newcommand{\Hvec}{\mathbf{H}}
\newcommand{\Ivec}{\mathbf{I}}
\newcommand{\Rvec}{\mathbf{R}}
\newcommand{\Vvec}{\mathbf{V}}

\newcommand{\Zvec}{\mathbf{Z}}

\newcommand{\zerovec}{\bm{0}}
\newcommand{\alphavec}{\bm{\alpha}}

\newcommand{\epsilonvec}{\bm{\epsilon}}
\newcommand{\ellvec}{\bm{\ell}}
\newcommand{\etavec}{\bm{\eta}}
\newcommand{\gammavec}{\bm{\gamma}}
\newcommand{\kappavec}{\bm{\kappa}}
\newcommand{\muvec}{\bm{\mu}}
\newcommand{\phivec}{\bm{\phi}}
\newcommand{\psivec}{\bm{\psi}}
\newcommand{\rhovec}{\bm{\rho}}
\newcommand{\tauvec}{\bm{\tau}}
\newcommand{\thetavec}{\bm{\theta}}

\newcommand{\xivec}{\bm{\xi}}

\newcommand{\avec}{\mathbf{a}}
\newcommand{\bvec}{\mathbf{b}}

\newcommand{\svec}{\mathbf{s}}
\newcommand{\wvec}{\mathbf{w}}
\newcommand{\xvec}{\mathbf{x}}

\newcommand{\var}{\mathrm{var}}

\newcommand{\intd}{\textrm{d}}

\newcommand{\nospellcheck}[1]{#1}

\title{
  Inferring changes to the global carbon cycle with WOMBAT \lowercase{v}2.0, a hierarchical flux-inversion framework
}

\date{}

\author[,1]{Michael Bertolacci\footnote{michael\_bertolacci@uow.edu.au}}
\author[1]{Andrew Zammit-Mangion}
\author[2]{Andrew Schuh}
\author[3]{Beata Bukosa}
\author[4]{Jenny Fisher}
\author[1]{Yi Cao}
\author[5,6]{Aleya Kaushik}
\author[1,7]{Noel Cressie}

\affil[1]{\normalsize School of Mathematics and Applied Statistics, University of Wollongong}
\affil[2]{\normalsize Cooperative Institute for Research in the Atmosphere (CIRA), Colorado State University}
\affil[3]{\normalsize National Institute of Water and Atmospheric Research (NIWA)}
\affil[4]{\normalsize Centre for Atmospheric Chemistry, School of Earth, Atmospheric and Life Sciences, University of Wollongong}
\affil[5]{\normalsize Global Monitoring Laboratory, National Oceanic and Atmospheric Administration}
\affil[6]{\normalsize Cooperative Institute for Research in Environmental Sciences, University of Colorado}
\affil[7]{\normalsize Jet Propulsion Laboratory, California Institute of Technology}

\begin{document}

\maketitle

\begin{abstract}
  The natural cycles of the surface-to-atmosphere fluxes of carbon dioxide (CO$_2$) and other important greenhouse gases are changing in response to human influences. These changes need to be quantified to understand climate change and its impacts, but this is difficult to do because natural fluxes occur over large spatial and temporal scales and cannot be directly observed. Flux inversion is a technique that estimates the spatio-temporal distribution of a gas' fluxes using observations of the gas' mole fraction and a chemical transport model. To infer trends in fluxes and identify phase shifts and amplitude changes in flux seasonal cycles, we construct a flux-inversion system that uses a novel spatially varying time-series decomposition of the fluxes. We incorporate this decomposition into the Wollongong Methodology for Bayesian Assimilation of Trace-gases (WOMBAT, Zammit-Mangion et al., \emph{\nospellcheck{Geosci. Model Dev.}}, 15, 2022), a hierarchical flux-inversion framework that yields posterior distributions for all unknowns in the underlying model. We also extend WOMBAT to accommodate physical constraints on the fluxes, and to take direct \emph{in situ} and flask measurements of trace-gas mole fractions as observations. We apply the new method, which we call WOMBAT~v2.0, to a mix of satellite observations of CO$_2$ mole fraction from the Orbiting Carbon Observatory-2 (OCO-2) satellite and direct measurements of CO$_2$ mole fraction from a variety of sources. We estimate the changes in the natural cycles of CO$_2$ fluxes that occurred from January 2015 to December 2020, and compare our posterior estimates to those from an alternative method based on a bottom-up understanding of the physical processes involved. We find substantial trends in the fluxes, including that tropical ecosystems trended from being a net source to a net sink of CO$_2$ over the study period. We also find that the amplitude of the global seasonal cycle of ecosystem CO$_2$ fluxes increased over the study period by 0.11 PgC/month (an increase of 8\%), and that the seasonal cycle of ecosystem CO$_2$ fluxes in the northern temperate and northern boreal regions shifted earlier in the year by 0.4--0.7 and 0.4--0.9 days, respectively (2.5th to 97.5th posterior percentiles), consistent with expectations for the carbon cycle under a warming climate.
\end{abstract}

\section{Introduction}
\label{sec:introduction}

The increasing atmospheric concentrations of carbon dioxide (CO$_2$), methane, and nitrous oxide are the primary drivers of climate change \citep{ipcctechnical2021}. The increases are largely due to anthropogenic emissions, but the surface-to-atmosphere fluxes of these gases also have natural cycles. These cycles are influenced by human activities such as land usage \citep{vitouseketal1997,tagessonetal2020}, by increasing surface temperatures \citep{barichivichetal2013}, and by the direct response of ecosystems to increased atmospheric concentrations \citep{friedlingstein2015}, and it is important to gauge the scale of the influences. For CO$_2$ in particular, the land and ocean have responded to increased concentrations by absorbing more than half of the anthropogenic emissions \citep{shevliakovaetal2013}, and the seasonal cycles of important source and sink processes such as respiration and photosynthesis are changing \citep{gonsamoetal2018,parketal2019}. Understanding how much the natural cycles of fluxes of climate-influencing trace gases have changed, and where those changes have occurred, is therefore critical in the study of climate change.

The net surface-to-atmosphere flux of a trace gas is the sum of its fluxes due to different processes at Earth's surface. For CO$_2$, these processes include the burning of fossil fuels, assimilation of CO$_2$ through photosynthesis (often called gross primary productivity, or GPP), respiration of CO$_2$ by plants and soils, biomass burning, human use of biofuels, and CO$_2$ exchange between the ocean and the atmosphere (known as air--sea fluxes). The net flux of a trace gas can thus be modelled as
\begin{equation}
  X(\svec, t)
  = \sum_{c \in \mathcal{C}} X_c(\svec, t)
  ;
  \quad
  \svec \in \mathbb{S}^2,\ 
  t \in \mathcal{T},
  \label{eqn:total_decomposition}
\end{equation}
where $X(\svec, t)$ is the net flux of the trace gas at location $\svec$ and time $t$, $\mathbb{S}^2$ is the surface of Earth, $\mathcal{T} \equiv [t_0, t_1]$ is the time period of interest, and $\{ X_c(\cdot, \cdot) : c \in \mathcal{C} \}$ is a collection of component flux fields of size $C = |\mathcal{C}|$. It is impossible to measure flux fields directly over large spatial and temporal scales, and hence estimation of $X(\cdot, \cdot)$ and $\{ X_c(\cdot, \cdot) : c \in \mathcal{C} \}$ is indirect. There are two broad categories of methods for doing so: bottom-up and flux inversion (which is top-down).

Bottom-up methods typically target a single component field, $X_c(\cdot, \cdot)$, such as GPP for CO$_2$, rather than the net flux \citep[e.g., see][for natural ocean and land fluxes, respectively]{landschutzeretal2016,haynesetal2019a}. These methods use a mechanistic understanding of the physical processes relevant to the component, such as the relationship between temperature and plant decay, and they are informed by ancillary information such as estimates of surface temperatures. Bottom-up methods can estimate the spatio-temporal distribution of fluxes at relatively fine scales, but their accuracy varies depending on the target field \citep{crowelletal2019}. Estimates of CO$_2$ fluxes from fossil-fuel usage, for example, are considered to be much more accurate \citep{odaetal2011,vanderwerf2017} than those of natural CO$_2$ fluxes \citep{basuetal2013}.

Flux inversion, the approach taken in this paper, uses atmospheric mole-fraction data to infer fluxes at Earth's surface. A gas' mole-fraction field, which we denote by $Y(\svec, h, t)$ at location $\svec \in \mathbb{S}^2$, geopotential height $h \geq 0$, and time $t \in \mathcal{T}$, is related to its flux field, $X(\svec, t)$, through the process of atmospheric transport. Flux inversion involves inverting the forward transport relationship and hence is an ill-posed problem that usually requires a significant injection of prior information \citep{enting2002}. Most inversion methods take a Bayesian approach where prior information comes from bottom-up estimates of fluxes; typically these are used to specify the prior mean of the flux field \citep[e.g.,][]{chevallieretal2005,basuetal2013,zammitmangionetal2022}. Estimates from flux inversions are usually most accurate at coarse scales, while inference at finer scales is less accurate.

The majority of operational CO$_2$ flux-inversion systems \citep[e.g.,][]{crowelletal2019} are parameterised over a partition of space and time, without parameters that explicitly govern the trend and seasonality of fluxes. Exceptions include the works of \citet{rayneretal2005} and related studies \citep[e.g.,][]{scholzeetal2016,heetal2022} that directly adjust the parameters of a bottom-up land biosphere model using observations of atmospheric CO$_2$ mole fractions (which in turn directly affect the seasonal behaviour of the biosphere model). With this approach, the parameterisation is rather parsimonious and lends itself to inferences that are physically interpretable. Another method that does not use a direct space--time partition is the system labelled ``\nospellcheck{CSU}'' in \citet{crowelletal2019}, which adjusts the bottom-up estimates through multiplication by harmonics at different frequencies; this also allows for borrowing of strength between observations. Although based on harmonics, the parameterisation does not explicitly separate the trend and seasonal parts of the fluxes, which are features of primary scientific interest and a key feature of our approach.

In this article, we present a new flux-inversion method with several innovations that facilitate the study of the trends and changing seasonal cycles of the component flux processes in \eqref{eqn:total_decomposition}. Our method builds on the Bayesian hierarchical flux-inversion framework of \citet{zammitmangionetal2022}, called the \nospellcheck{WOllongong} Methodology for Bayesian Assimilation of Trace-gases, version 1.0 (WOMBAT~v1.0), and hence we call it WOMBAT~version 2.0 (WOMBAT~v2.0).

A major contribution of WOMBAT~v2.0 is the incorporation of a novel spatially varying time-series decomposition of the component fluxes built on a spatially-varying regression coefficient model \citep{brunsdonetal1996,wikleetal1998,gelfandetal2003}. Our decomposition allows for direct inference of the flux trends and of phase shifts and amplitude changes in the flux seasonal cycle, through which we are able to answer questions of scientific importance. The decomposition is preserved under aggregation, so that inferences on the changes can be made at different spatial scales. The underlying component processes can also be projected into the future. Inference on the trend and seasonal cycle of the flux at a location is parsimonious and efficient, as it is made through a small number of parameters. This is in contrast to a conventional flux-inversion setup where the component flux is parameterised over a space--time partition \citep[e.g., the flux in a grid cell over eight days, as in][]{chevallieretal2005}, so that seasonal cycles must be inferred anew every year. Our parameterisation also helps with gaps in observational coverage over time through borrowing of strength between years: if observations for some time period are missing, observations from the same time period in other years help fill the gap.

A second major contribution of WOMBAT~v2.0 builds on an extension to WOMBAT~v1.0 that we implemented in \citet{stelletal2022} to allow for fully-Bayesian inference on flux fields that must satisfy physical constraints. For example, in our application to CO$_2$ flux, one component field is always negative (GPP) and another is always positive (respiration). We show how these constraints can be approximated using a system of just over half a million linear constraints, which allows the Hamiltonian Monte Carlo (HMC) method of \citet{pakmanpaninski2014} to sample the unknown parameters in the model with the constraints preserved. Other improvements to WOMBAT~v1.0 include the ability to estimate CO$_2$ fluxes over a longer time period (79 months rather than WOMBAT~v1.0's 31 months), and the flexibility to augment the satellite observations with \emph{in situ} and flask measurements. Other important features of WOMBAT~v1.0 are retained, including the capacity to combat model misspecification by specifying a parameter model and inferring these unknowns within the Bayesian hierarchical framework.

The paper is organised as follows. In Section~\ref{sec:model} we present the novel Bayesian hierarchical model for WOMBAT~v2.0, along with details on how we make inferences with the model. In Section~\ref{sec:co2_estimation} we discuss how we configure WOMBAT~v2.0 to estimate land biosphere fluxes and ocean (air--sea) fluxes of CO$_2$ using a mix of satellite observations of CO$_2$ mole fraction from the Orbiting Carbon Observatory-2 (OCO-2) satellite and direct measurements of CO$_2$ mole fraction from a variety of sources, with a focus on estimating changes to the natural cycle of the land fluxes and quantifying their uncertainties. The results, which are given in Section~\ref{sec:results}, both reinforce our understanding of and shed new light on Earth's carbon cycle. In agreement with the bottom-up estimates, we find that the overall global seasonal cycle of CO$_2$ is increasing in amplitude over the six years in our study. We find evidence of linear trends in the fluxes in all regions of the globe. Significantly, we find that over 2015--2020 the land ecosystems of the northern extratropics (defined as 23$\degree$N--90$\degree$N) are a net sink of CO$_2$, but a positive trend indicates that the sink is decreasing. Equally significantly, over the same time period the ecosystems of the tropics (23$\degree$S--23$\degree$N) trend from a source of CO$_2$ to a sink. This latter finding is important as tropical ecosystems have been described as the ``lungs'' of our planet \citep{wood1990}, and this increased sink is likely a response to increased CO$_2$ in the atmosphere. The decomposition used by WOMBAT~v2.0 is able to isolate the impact of the 2015/16 \nospellcheck{El Ni\~{n}o} event, which is found to have caused a reduction in ecosystem productivity leading to a net positive flux (i.e., source) of CO$_2$ over northern tropical South America. We also find changes to regional CO$_2$ flux seasonal cycles, including that the seasonal cycles of ecosystem CO$_2$ fluxes in the northern temperate and northern boreal regions are shifting earlier in the year.

Section~\ref{sec:conclusion} concludes the paper with a discussion of our framework, the results, and future research directions. Appendices~\ref{sec:remaining_hierarchical_model} through to \ref{sec:additional_figures} contains extra details on the framework and how it was configured to estimate fluxes of CO$_2$, as well as additional figures and tables to support the results. Software that implements the method is available online at \url{https://github.com/mbertolacci/wombat-v2-workflow}.

\section{Model}
\label{sec:model}

Our top-down framework for estimating the surface fluxes of a trace gas has its foundations in the WOMBAT~v1.0 framework \citep{zammitmangionetal2022}. As in WOMBAT~v1.0, we construct a  hierarchical Bayesian statistical model consisting of a process model for the fluxes, a process model for the mole-fraction field linked to the flux field, a mole-fraction data model, and a parameter model for the unknown parameters in the process and data models. Sections~\ref{sec:decomposition}--\ref{sec:alpha_prior} present the new process model that is needed for WOMBAT~v2.0 to answer important questions about a changing carbon cycle. Section~\ref{sec:full_hierarchical_model} outlines the parts of WOMBAT~v1.0 that are retained in WOMBAT~v2.0. Section~\ref{sec:inference} discusses inference in the framework.

\subsection{Decomposition of the flux process model}
\label{sec:decomposition}

Recall from \eqref{eqn:total_decomposition} that the net flux of a trace gas, $X(\svec, t)$, is modelled as the sum of the component flux fields $\{ X_c(\cdot, \cdot) : c \in \mathcal{C} \}$. The component fluxes depend on the physical processes that are relevant to the trace gas being studied; for an example of a set of components, see the CO$_2$ application in Section~\ref{sec:co2_estimation}. It is common for some of these components, such as the fossil-fuel contribution for CO$_2$, to be assumed known and fixed to a bottom-up estimate \citep{crowelletal2019}.

In WOMBAT~v1.0, and in the vast majority of flux inversion systems in use today, the component fluxes in $\{ X_c(\cdot,\cdot) : c \in \mathcal{C} \}$ that are not fixed are modelled as rescaled bottom-up estimates, where the scaling factors are allowed to vary slowly in both space and time. As discussed in Section~\ref{sec:introduction}, this model requires adjustments to the seasonal cycle of fluxes to be made anew each year, and does not allow explicit inference on the characteristics of the natural cycles of the fluxes. To remedy this, the new process model in WOMBAT~v2.0 posits that $X_c(\cdot, \cdot)$ has the following spatially varying time-series decomposition:
\begin{multline}
  X_c(\svec, t)
  = \beta_{c, 0}(\svec) + \beta_{c, 1}(\svec) t
    + \sum_{k = 1}^{K_c} (\beta_{c, 2, k}(\svec) + \beta_{c, 3, k}(\svec) t) \cos(2\pi k t / 365.25) \\
    + \sum_{k = 1}^{K_c} (\beta_{c, 4, k}(\svec) + \beta_{c, 5, k}(\svec) t) \sin(2\pi k t / 365.25)
      + \epsilon_c(\svec, t),
\label{eqn:component_decomposition}
\end{multline}
for $\svec \in \mathbb{S}^2$ and $t \in \mathcal{T}$, where $\beta_{c, \cdot}(\svec)$, $\beta_{c, \cdot, \cdot}(\svec)$, and $\epsilon_c(\svec, t)$ are treated as unknown. The joint distribution of the unknown parameters in \eqref{eqn:component_decomposition} is constructed through a combination of bottom-up estimates and a basis-function representation; see Section~\ref{sec:basis_function_decomposition}. 

The decomposition in \eqref{eqn:component_decomposition} comprises an intercept, a linear temporal trend, $2K_c$ harmonics of the solar-cycle with a period of 365.25 days, and a residual term, $\epsilon_c(\cdot, \cdot)$. The intercept and the linear trend capture the long-term changing level of the fluxes, while the harmonics capture the seasonal (intra-annual) component, which is also allowed to change between years. The residual term, $\epsilon_c(\cdot, \cdot)$, accommodates any deviations from the flux trend and seasonality at location $\svec$ and time $t$. These deviations are largely unpredictable in the long term, are often substantial, and are themselves of scientific interest. For example, in Section~\ref{sec:results} an \nospellcheck{El Ni\~{n}o} event that occurred during our study period leads to substantial residual CO$_2$ fluxes in some regions of Earth. Some component fluxes also have a diurnal cycle with a period of 24 hours, which cannot be represented by the annual harmonics. The diurnal cycle will therefore appear in the residual term; this high-frequency cycle is not the focus of this study, so we make no further effort to isolate it.

The decomposition in \eqref{eqn:component_decomposition} fits into the class of spatially-varying coefficient models \citep{brunsdonetal1996,wikleetal1998,gelfandetal2003}, in which the coefficients of a series of covariates are assigned a spatial structure. In this context, previous studies have also used harmonics with temporally and spatially-varying coefficients to accommodate changing seasonality. A recent example is given by \citet{northetal2021}, who examine the changing seasonality of temperature in North America over 40 years. In their case, the coefficients of the harmonics evolve in time according to a dynamical model, while in \eqref{eqn:component_decomposition} their change in time depends on a random trend term.

\subsubsection{Time-varying phase and amplitude}
\label{eqn:time_varying_phase}

A useful property of the decomposition in \eqref{eqn:component_decomposition} for the study of natural cycles of the fluxes is that the harmonic terms accommodate \emph{time-varying periodicity}. This is because the seasonal component of the decomposition can be written as
\begin{multline}
  \sum_{k = 1}^{K_c} (\beta_{c, 2, k}(\svec) + \beta_{c, 3, k}(\svec) t) \cos(2\pi k t / 365.25) \\
    + \sum_{k = 1}^{K_c} (\beta_{c, 4, k}(\svec) + \beta_{c, 5, k}(\svec) t) \sin(2\pi k t / 365.25)
  = \sum_{k = 1}^{K_c} A_{c, k}(\svec, t) \cos(2\pi k t / 365.25 + P_{c, k}(\svec, t)),
  \label{eqn:harmonic_phase_amplitude}
\end{multline}
where
\begin{align*}
  A_{c, k}(\svec, t)
  & \equiv \sqrt{(\beta_{c, 2, k}(\svec) + \beta_{c, 3, k}(\svec) t)^2 + (\beta_{c, 4, k}(\svec) + \beta_{c, 5, k}(\svec) t)^2}\ \text{ and} \\
  P_{c, k}(\svec, t)
  & \equiv \tan^{-1}\left(-\frac{
      \beta_{c, 4, k}(\svec) + \beta_{c, 5, k}(\svec) t
    }{
      \beta_{c, 2, k}(\svec) + \beta_{c, 3, k}(\svec) t
    }\right),
\end{align*}
for $\svec \in \mathbb{S}^2$ and $t \in \mathcal{T}$. Equation \eqref{eqn:harmonic_phase_amplitude} expresses the seasonal component as a sum of $K_c$ harmonics, each with its own time-varying amplitude, $A_{c, k}(\cdot, \cdot)$, and phase, $P_{c, k}(\cdot, \cdot)$. Therefore, by making inference on the $\beta$ coefficients in \eqref{eqn:component_decomposition}, one is implicitly making inference on the changing amplitude and phase of the seasonal fluxes.

In the typical space--time parameterisation of fluxes used in most flux-inversion systems (such as WOMBAT~v1.0), fluxes are often assigned prior means derived from bottom-up estimates of the fluxes and marginal prior standard deviations proportional to the magnitude of the prior means. This can complicate the ability of the system to shift the phase of the fluxes relative to the prior mean, since fluxes at time points where the prior mean is small (e.g., at positive-to-negative transitions) are assigned a low prior uncertainty, and hence not updated in the inversion. By contrast, \eqref{eqn:harmonic_phase_amplitude} shows that phase shifts are easily accommodated in our framework, and in Section~\ref{sec:results} we show the utility of this modelling choice for making inference on changes to the natural cycles of CO$_2$ fluxes.

\subsubsection{Decomposition of linear aggregates}
\label{sec:linear_aggregates}

Another useful property of the decomposition in \eqref{eqn:component_decomposition} is that it has the same form under linear aggregation. For example, given two components $c$ and $c'$ for which $K_c = K_{c'}$, the sum of their fluxes $X_{c + c'}(\svec, t) \equiv X_c(\svec, t) + X_{c'}(\svec, t)$, is equal to
\begin{multline}
  X_{c + c'}(\svec, t)
  = \beta_{c + c', 0}(\svec) + \beta_{c + c', 1}(\svec) t
     + \sum_{k = 1}^{K_c} (\beta_{c + c', 2, k}(\svec) + \beta_{c + c', 3, k}(\svec) t) \cos(2\pi k t / 365.25) \\
     + \sum_{k = 1}^{K_c} (\beta_{c + c', 4, k}(\svec) + \beta_{c + c', 5, k}(\svec) t) \sin(2\pi k t / 365.25)
     + \epsilon_{c + c'}(\svec, t),
  \label{eqn:sum_aggregate}
\end{multline}
where $\beta_{c + c', 0}(\svec) \equiv \beta_{c, 0}(\svec) + \beta_{c', 0}(\svec)$, $\beta_{c + c', 1}(\svec) \equiv \beta_{c, 1}(\svec) + \beta_{c', 1}(\svec)$, and so on. That is, the coefficients of the decomposition of the sum can be calculated by summing the corresponding coefficients for the two components $c$ and $c'$. Similarly, suppose $\mathcal{S} \subseteq \mathbb{S}^2$, and consider the aggregate flux at time $t$ in the region $\mathcal{S}$, given by
\begin{equation}
  \tilde{X}_c(\mathcal{S}, t) \equiv \int_\mathcal{S} a(\svec) X_c(\svec, t)\,\intd\svec,
  \label{eqn:spatial_aggregate}
\end{equation}
where $a(\svec)$ corresponds to a weighting by area. For example, if $\mathcal{S} = \mathbb{S}^2$, $\tilde{X}_c(\mathcal{S}, t)$ is the global total flux of component $c$ at time $t$. Through linearity, $\tilde{X}_c(\mathcal{S}, t)$ can be decomposed by applying the aggregation operation to each individual $\beta$ and $\epsilon$ coefficient in the original point-level decomposition \eqref{eqn:component_decomposition}. For a discretisation of $\mathbb{S}^2$, such as a latitude--longitude grid, \eqref{eqn:spatial_aggregate} can be rewritten as an area-weighted sum.

A consequence of \eqref{eqn:sum_aggregate} and \eqref{eqn:spatial_aggregate} is that the seasonal component \eqref{eqn:harmonic_phase_amplitude} is well defined under aggregation, so the time-varying phase and amplitude is well defined for an aggregated flux. This is used in Section~\ref{sec:seasonal_cycle_changes} to examine changes to the annual cycle of fluxes at various spatial scales.

\subsection{Basis-function representation of the flux decomposition}
\label{sec:basis_function_decomposition}

The terms of \eqref{eqn:component_decomposition}, $\{\beta_{c, \cdot}(\svec)\}$, $\{\beta_{c, \cdot, \cdot}(\svec)\}$, and $\epsilon_c(\svec, t)$, are highly variable spatial or spatio-temporal processes that generally span the whole globe $\mathbb{S}^2$ and multiple years. For computations to remain tractable, we impose on these terms a lower-dimensional structure using a basis-function representation of the individual flux components. We do this by partitioning the globe into $R$ disjoint regions, which we denote by $\{ D_r \subset \mathbb{S}^2 : r = 1, \ldots, R \}$, and by partitioning the study's time period into $Q$ disjoint sequential time periods, which we denote by $\{ E_q \subset \mathcal{T} : q = 1, \ldots, Q \}$. Define the $R$-dimensional vector of spatial indicators,
\begin{equation*}
  \wvec_\mathrm{S}(\svec)
  \equiv (
    \mathbbm{1}(\svec \in D_1),
    \mathbbm{1}(\svec \in D_2),
    \ldots,
    \mathbbm{1}(\svec \in D_R)
  )',\quad \svec \in \mathbb{S}^2,
\end{equation*}
and the $QR$-dimensional spatio-temporal vector of indicators,
\begin{multline*}
  \wvec_\mathrm{ST}(\svec, t)
  \equiv (
    \mathbbm{1}(\svec \in D_1 \cap t \in E_1),
    \mathbbm{1}(\svec \in D_1 \cap t \in E_2),
    \ldots,
    \mathbbm{1}(\svec \in D_R \cap t \in E_Q)
  )', \\
  \quad \svec \in \mathbb{S}^2,~~t \in \mathcal{T},
\end{multline*}
on the spatial and spatio-temporal partitions, respectively, where $\mathbbm{1}(\cdot)$ is the indicator function. For $c \in \mathcal{C}$, we model the unknown spatial and spatio-temporal processes that comprise $X_c(\cdot, \cdot)$ in \eqref{eqn:component_decomposition} as follows:
\begin{equation}
  \begin{split}
    \beta_{c, j}(\svec)
    & = (1 + \wvec_\mathrm{S}(\svec)' \alphavec_{c, j}) \beta_{c, j}^0(\svec),
        \quad j = 0, 1, \\
    \beta_{c, j, k}(\svec)
    & = (1 + \wvec_\mathrm{S}(\svec)' \alphavec_{c, j, k}) \beta_{c, j, k}^0(\svec),
        \quad j = 2, \ldots, 5, \ k = 1, \ldots, K_c, \\
    \epsilon_c(\svec, t)
    & = (1 +  \wvec_\mathrm{ST}(\svec, t)' \alphavec_{c, 6}) \epsilon_c^0(\svec, t),
  \end{split}
  \label{eqn:beta_model}
\end{equation}
for $\svec \in \mathbb{S}^2$ and $t \in \mathcal{T}$, where $\alphavec_{c, j} \equiv (\alpha_{c, j, 1}, \ldots, \alpha_{c, j, R})'$ for $j = 0, 1$; $\alphavec_{c, j, k} \equiv (\alpha_{c, j, k, 1},$ $\ldots, \alpha_{c, j, k, R})'$ for $j = 2, \ldots, 5$ and $k = 1, \ldots, K_c$; and $\alphavec_{c, 6} \equiv (\alpha_{c, 6, 1, 1}, \alpha_{c, 6, 1, 2}, \ldots, \alpha_{c, 6, R, Q})'$. The $\alphavec$ vectors are random and unknown, while the fields $\beta_{c, \cdot}^0(\cdot)$, $\beta_{c, \cdot, \cdot}^0(\cdot)$, and $\epsilon_c^0(\cdot, \cdot)$ are (known) bottom-up estimates of the coefficients that define the flux trend, seasonality, and residual. In Section~\ref{sec:co2_flux_model}, these are obtained through a deterministic decomposition of a bottom-up estimate of the component flux. At location $\svec$, $\beta_{c, 0}(\svec)$ is equal to $(1 + \alpha_{c, 0, r}) \beta_{c, 0}(\svec)$ if $\svec \in D_r$, and is equal to zero otherwise, and a similar relationship holds for all other processes. The unknown and random $\alphavec$ vectors therefore serve to spatially adjust the bottom-up estimate of the trend and seasonality, and to spatio-temporally adjust the bottom-up estimate of the residual.

The degree to which spatial or spatio-temporal adjustment is done is determined by the resolution of the partitioning of $\mathbb{S}^2$ and $\mathcal{T}$. Although adjustments happen over partitions, within each spatial partition the trend and seasonality still vary spatially according to their bottom-up estimates, and within each spatio-temporal partition the residual varies spatio-temporally according to its bottom-up estimate. Thus, fine-scale effects that are present in the bottom-up estimates (such as differing carbon assimilation and respiration among land types within a spatial partition) are represented in our flux process model.

Equations \eqref{eqn:component_decomposition} and \eqref{eqn:beta_model} can be combined to write the component fluxes as a basis function model. For $c \in \mathcal{C}$,
\begin{equation}
  X_c(\svec, t)
  = \phivec_c(\svec, t)' (\bm{1} + \alphavec_c)
  = X_c^0(\svec, t) + \phivec_c(\svec, t)' \alphavec_c,
  \quad \svec \in \mathbb{S}^2, \enskip t \in \mathcal{T},
  \label{eqn:flux_basis_c}
\end{equation}
where $\phivec_c(\cdot, \cdot)$ is a vector of basis functions of dimension $(2R + 4K_c R + QR)$; $X_c^0(\cdot, \cdot) \equiv \phivec_c(\svec, t)' \bm{1}$ is equal to the sum of the bottom-up estimates of the flux trend, seasonality, and residual (i.e., the bottom-up estimate of the net flux); and $\alphavec_c \equiv (\alphavec_{c, 0}', \alphavec_{c, 1}',$ $\alphavec_{c, 2, 1}', \alphavec_{c, 2, 2}',$ $\ldots, \alphavec_{c, 5, K_c}', \alphavec_{c, 6}')'$. Each element of $\phivec_c(\cdot, \cdot)$ corresponds to one term of \eqref{eqn:component_decomposition} evaluated over a spatial or spatio-temporal partition, and is equal to zero outside the partition. For example, the first set of $R$ elements of $\phivec_c(\svec, t)$ are equal to $\beta_{c, 0}^0(\svec)$ for $\svec \in D_r$ and zero otherwise, for $r = 1, \ldots, R$, and the second set of $R$ elements are equal to $\beta_{c, 1}^0(\svec) t$ for $\svec \in D_r$ and zero otherwise, for $r = 1, \ldots, R$. The last $QR$ elements are equal to $\epsilon_c^0(\svec, t)$ for $\svec \in D_r$ and $t \in E_q$, and equal to zero otherwise, for $r = 1, \ldots, R$, $q = 1, \ldots, Q$.

Combining \eqref{eqn:total_decomposition} and \eqref{eqn:flux_basis_c} gives a basis-function representation of the total flux,
\begin{equation}
  X(\svec, t) = X^0(\svec, t) + \phivec(\svec, t)' \alphavec,
  \label{eqn:flux_basis}
\end{equation}
where $X^0(\cdot, \cdot) \equiv \sum_{c \in \mathcal{C}} X_c^0(\cdot, \cdot)$ is the total combined flux from the bottom-up estimates, $\phivec(\cdot, \cdot) \equiv (\phivec_{c_1}(\cdot, \cdot)', \ldots, \phivec_{c_C}(\cdot, \cdot)')'$, $\alphavec \equiv (\alphavec_{c_1}', \ldots, \alphavec_{c_C}')'$, and $c_1, \ldots, c_C$ are the $C$ elements in $\mathcal{C}$.

\subsection{Constrained distribution on the basis-function coefficients}
\label{sec:alpha_prior}

In WOMBAT~v1.0, the distribution of $\alphavec$ in \eqref{eqn:flux_basis} was set to be Gaussian with mean zero, and the only dependence assumed was temporal dependence. These assumptions are practical and have the useful property that the prior mean of the flux field is equal to the bottom-up estimate. However, they are unrealistic when some component fields in \eqref{eqn:total_decomposition} are correlated and subject to sign constraints, such as GPP and respiration in the context of CO$_2$ (see Section~\ref{sec:co2_flux_model}). In WOMBAT~v2.0, we consider a more elaborate distribution over $\alphavec$, which (i) allows for prior correlations between component fluxes, and (ii) constrains the inferred fluxes to satisfy pre-determined physical constraints.

Our general model for the basis-function coefficients is
\begin{equation}
  \alphavec
  \sim \mathrm{ConstrGau}(\zerovec, \Sigmavec_\alpha, F_\alpha),
  \label{eqn:alpha_prior}
\end{equation}
where $\mathrm{ConstrGau}(\muvec, \Sigmavec, F)$ is the constrained multivariate Gaussian distribution, defined as a multivariate Gaussian distribution with mean vector $\muvec$ and covariance matrix $\Sigmavec$ constrained to the set $F$. The constraint set $F_\alpha$ in \eqref{eqn:alpha_prior} is specified through $M > 0$ linear constraints of the form $\avec + \Phivec \alphavec \geq \zerovec$, applied element-wise, where $\Phivec$ is an $M \times B$ matrix, $B$ is the number of elements in $\alphavec$, and $\avec$ is a known $M$-dimensional vector. We show in Section~\ref{sec:constraints} how the non-positivity and non-negativity of the GPP and respiration component flux fields, respectively, can be represented in approximate fashion using a large but finite number of linear constraints.

The constraint set in \eqref{eqn:alpha_prior} means that $\alphavec = \zerovec$ will typically not be the mean of the distribution over $\alphavec$, but in most applications $\alphavec = \zerovec$ will be its mode. This is because $\alphavec = \zerovec$ in \eqref{eqn:flux_basis} corresponds to $X_c(\cdot, \cdot) = X_c^0(\cdot, \cdot)$ for $c \in \mathcal{C}$, and the bottom-up estimate $X_c^0(\cdot, \cdot)$ will almost always satisfy the physical constraints embodied in $F_\alpha$. This is distinct from WOMBAT~v1.0 and most other flux inversions systems, where the flux prior mean is equal to the bottom-up estimate.

The parameterisation of the covariance matrix $\Sigmavec_\alpha$ is described in Appendix~\ref{sec:alpha_covariance_matrix}. The parameters involved are $\tau_c^\beta$ and $\tau_c^\epsilon$, $c \in \mathcal{C}$, the marginal precisions of the component $c$ trend/seasonality and the residual scaling factors, respectively; $\rho_{c, c'}^\beta$ and $\rho_{c, c'}^\epsilon$, $c, c' \in \mathcal{C}$, the correlation between the scaling factors of component $c$ and $c'$ for the trend/seasonality and for the residual, respectively; and $\kappa_c^\epsilon$, $c \in \mathcal{C}$, the temporal correlation between residual scaling factors for component $c$.

\subsection{Mole-fraction process and data models}
\label{sec:full_hierarchical_model}

Sections~\ref{sec:decomposition}--\ref{sec:alpha_prior} outlined the innovations in WOMBAT~v2.0 that accommodate a scientifically interpretable and flexible flux process model tailored to the study of the natural cycles of fluxes. The remaining two levels of the hierarchy are the mole-fraction process model and the mole-fraction data model, both of which remain largely unchanged from WOMBAT~v1.0 \citep{zammitmangionetal2022}. The mole-fraction process model links the flux field, $X(\svec, t)$, to the mole-fraction field, $Y(\svec, h, t)$, where $h > 0$ corresponds to geopotential height, and is described in Appendix~\ref{sec:mole_fraction_process_model}. The mole-fraction data model encodes the error properties of the mole-fraction observations used as data. Observations are split into groups $g = 1, \ldots, G$, within which observations are assumed to have similar error properties. Observations are prescribed known ``error budgets'' that determine the total marginal variance of their errors, and the error budgets for group $g$ are scaled by unknown factors $(\gamma_g^Z)^{-1}$. The total error associated with each observation is split into temporally correlated and uncorrelated components, and the proportion of the error budget assigned to the correlated component for group $g$ is given by the unknown parameter $\rho_g^Z$. The $e$-folding length of the temporal correlation for group $g$ is given by the unknown parameter $\ell_g^Z$. Observations in each group are also assumed to be biased, and we use a linear model for the bias terms  where the unknown coefficients are denoted as $\bm{\eta}_g, g = 1, \dots, G$. Full mathematical details of the mole-fraction data model, including the priors on the unknown parameters, are given in Appendix~\ref{sec:mole_fraction_data_model}.

\subsection{Inference}
\label{sec:inference}

The unknown quantities in our model are $\alphavec$, $\{ \tau_c^\beta : c \in \mathcal{C} \}$, $\{ \rho_{c, c'}^\beta : c, c' \in \mathcal{C} \}$, $\{ \tau_c^\epsilon : c \in \mathcal{C} \}$, $\{ \kappa_c^\epsilon : c \in \mathcal{C} \}$, $\{ \rho_{c, c'}^\epsilon : c, c' \in \mathcal{C} \}$, $\etavec$, $\gammavec^Z \equiv (\gamma_1^Z, \ldots, \gamma_G^Z)'$, $\rhovec^Z \equiv (\rho_1^Z, \ldots, \rho_G^Z)'$, and $\ellvec^Z \equiv (\ell_1^Z, \ldots, \ell_G^Z)'$. The parameters $\rhovec^Z$ and $\ellvec^Z$ are the most computationally demanding to estimate, and \citet{zammitmangionetal2022} use graphics processing units (GPUs) to alleviate some of the computational burden. However, the dimensionality of the problem considered in Section~\ref{sec:co2_estimation} precludes their estimation in a full Bayesian setting, even with the use of GPUs. We therefore estimate these parameters via a two-stage process.

The first stage is a preliminary inversion in which, for $c, c' \in \mathcal{C}$ and $g = 1, \ldots, G$, all the parameters are fixed to reasonable starting values, in our case to $\tau_c^\beta = 1$, $\rho_{c, c'}^\beta = 0$, $\tau_c^\epsilon = 1$, $\kappa_c^\epsilon = 0$, $\rho_{c, c'}^\epsilon = 0$, $\gamma_g^Z = 1$, and $\rho_g^Z = 0$ (the value of $\ell_g^Z$ is therefore irrelevant in this case because errors are assumed to be uncorrelated). Denote by $\tilde{\alphavec}$ and $\tilde{\etavec}$ the posterior means of $\alphavec$ and $\etavec$ in this first-stage inversion, respectively. Conditional on the first-stage estimates $\tilde{\alphavec}$ and $\tilde{\etavec}$, we estimate $\rhovec^Z$ and $\ellvec^Z$ by maximising the full conditional density function, $p(\rhovec^Z, \ellvec^Z, \gammavec^Z \mid \alphavec = \tilde{\alphavec}, \etavec = \tilde{\etavec})$. This maximisation yields the estimates $\hat{\rhovec}^Z \equiv (\hat{\rho}_1^Z, \ldots, \hat{\rho}_G^Z)'$ and $\hat{\ellvec^Z} \equiv (\hat{\ell}_1^Z, \ldots, \hat{\ell}_G^Z)'$ of $\rhovec^Z$ and $\ellvec^Z$, respectively, which are then treated as known in the second stage. The maximisation also yields an estimate of $\gammavec^Z$ but, since this parameter does not lead to any computational issues, $\gammavec^Z$ is re-estimated in the second stage.

The second stage is a full inversion where $\rhovec^Z$ and $\ellvec^Z$ are fixed to their estimates $\hat{\rhovec}^Z$ and $\hat{\ellvec^Z}$, respectively. The MCMC sampling scheme for estimating the remaining parameters is described in Appendix~\ref{sec:mcmc}. This sampling scheme is mostly standard, except for its handling of the basis-function coefficients, $\alphavec$, which in this study are constrained to the set $F_\alpha$ (see Section~\ref{sec:alpha_prior}). To handle this, we use the exact HMC scheme for constrained multivariate Gaussian distributions proposed by \citet{pakmanpaninski2014}. In the application to CO$_2$ in Section~\ref{sec:co2_estimation}, this technique is able to draw samples of $\alphavec$ in a reasonable time frame, even when using 623,485 linear constraints across 6,325 variables. Once the posterior samples of $\alphavec$ are computed, posterior samples of the $\beta$ coefficients in \eqref{eqn:component_decomposition}, and of the flux component fields, $\{ X_c(\svec, t) \}$, can be computed by substituting the samples of $\alphavec$ into \eqref{eqn:beta_model} and \eqref{eqn:flux_basis_c}, respectively.

\section{\texorpdfstring{Top-down estimation of CO$_2$ fluxes}{Top-down estimation of CO2 fluxes}}
\label{sec:co2_estimation}

We use the hierarchical framework of Section~\ref{sec:model} to make inference on the natural cycles of surface CO$_2$ fluxes over the period from December 2015 to January 2020 from OCO-2 and \emph{in-situ}/flask data. In this section we show how we tailor the framework for this purpose by describing the bottom-up estimates we use, the data we use, and the prior distributions we apply to the unknown parameters.

We base our configuration on the protocol of the current round of the OCO-2 model intercomparison project \citep[MIP; see][for past rounds]{crowelletal2019,peiroetal2022}, an organised effort to compare flux-inversion systems under a common protocol. This round of the MIP is called the v10 MIP since it centres around the use of version 10 OCO-2 data. To make results comparable between teams, the MIP protocol prescribes the mole-fraction data to use in the inversions, a common bottom-up fossil-fuel flux field that is assumed fixed and known, and a common time period over which to report inferred fluxes. The data for the OCO-2 v10 MIP span September 2014 to March 2021 and, while fluxes are inferred over this full time period, flux estimates are reported for January 2015 to December 2020, inclusive. The extra months of data on either side of the reporting period are considered necessary to reliably estimate the fluxes at the start and end of the reporting period.

\subsection{Flux model and bottom-up estimates}
\label{sec:co2_flux_model}

The net CO$_2$ surface flux at a given time and location is the sum of the component fluxes due to a variety of physical processes, the most important of which are: the burning of fossil fuels; the burning of wood, charcoal, and agricultural waste for energy (collectively termed biofuel fluxes); biomass burning; GPP and respiration of CO$_2$ from the terrestrial biosphere; and air--sea exchanges due to differences in the ocean and the atmospheric partial pressure of CO$_2$. We represent this decomposition of the net flux through \eqref{eqn:total_decomposition}, which here is
\begin{multline}
  X(\svec, t)
  = X_\mathrm{fossil}(\svec, t)
    + X_\mathrm{biofuel}(\svec, t)
    + X_\mathrm{bioburn}(\svec, t)
    + X_\mathrm{gpp}(\svec, t)
    + X_\mathrm{resp}(\svec, t) \\
    + X_\mathrm{ocean}(\svec, t)
    + X_\mathrm{other}(\svec, t),
  \label{eqn:total_flux_model_co2}
\end{multline}
for $\svec \in \mathbb{S}^2, t \in \mathcal{T}$, where the fields $\{ X_c(\cdot, \cdot) \}$ match those named above. The fields satisfy the following constraints: $X_\mathrm{fossil}(\svec, t)$, $X_\mathrm{biofuel}(\svec, t)$, $X_\mathrm{bioburn}(\svec, t)$, and $X_\mathrm{resp}(\svec, t)$ are always zero or positive; $X_\mathrm{gpp}(\svec, t)$ is always zero or negative; and $X_\mathrm{ocean}(\svec, t)$ and $X_\mathrm{other}(\svec, t)$ can be positive or negative. At ocean locations, the biofuel, biomass-burning, GPP, and respiration fluxes are zero, while ocean fluxes are zero at land locations (note that, in practice, we work with a discretisation of the flux field in which a grid cell can contain both land and ocean, and in these cells any component flux can be non-zero; for details on the discretisation, see Appendix~\ref{sec:transport}). Fossil-fuel fluxes and ``other'' fluxes can be non-zero anywhere. The sum of GPP and respiration is called the net ecosystem exchange (NEE), which we define in our framework as $X_\mathrm{nee}(\svec, t) \equiv X_\mathrm{gpp}(\svec, t) + X_\mathrm{resp}(\svec, t)$. The NEE represents the total flux from the terrestrial biosphere, excluding those fluxes that come from biomass burning.

We assume that the fossil-fuel, biofuel, and biomass-burning CO$_2$ fluxes are known, and set them equal to their respective bottom-up estimates. This is a common assumption in flux inversion, made because bottom-up estimates of these fluxes have relatively low uncertainty compared to the natural fluxes \citep{basuetal2013}. We also assume that the ``other'' fluxes are negligible and set them equal to zero everywhere. The fluxes from the remaining components (GPP, respiration, and air--sea fluxes), are to be estimated. Our working model for the flux at location $\svec$ and time $t$ is therefore
\begin{multline}
  X(\svec, t)
  = X_\mathrm{fossil}^0(\svec, t)
    + X_\mathrm{biofuel}^0(\svec, t)
    + X_\mathrm{bioburn}^0(\svec, t)
    + X_\mathrm{gpp}(\svec, t) \\
    + X_\mathrm{resp}(\svec, t)
    + X_\mathrm{ocean}(\svec, t).
  \label{eqn:flux_process_model}
\end{multline}
where the terms with a zero superscript are fixed at their bottom-up estimates. The three unknown terms, $X_\mathrm{gpp}(\cdot, \cdot)$, $X_\mathrm{resp}(\cdot, \cdot)$, and $X_\mathrm{ocean}(\cdot, \cdot)$, are modelled using the decomposition and basis-function representation given in \eqref{eqn:component_decomposition} and \eqref{eqn:flux_basis}.

Both the fixed and the unknown flux components are informed by bottom-up estimates, and Figure~\ref{fig:global_bottom_up} shows the monthly-average global total fluxes for the bottom-up estimates we use for each component. The fossil-fuel component is prescribed by the OCO-2 v10 MIP protocol, and its estimate comes from the Open-source Data Inventory for Anthropogenic CO$_2$ monthly fossil-fuel emissions \citep[ODIAC;][]{odaetal2011, odaetal2018} with weekly scaling factors from Temporal Improvements for \nospellcheck{Modeling} Emissions by Scaling (TIMES) \citep{nassaretal2013}. The remaining components are not prescribed by the v10 MIP protocol. For biofuel fluxes we use the estimated fluxes of \citet{yevichlogan2003}, whose estimates are based on data from 1985. Unlike all the other estimates used, the biofuel fluxes vary in space but not in time. Biomass burning emissions come from the Global Fire Emissions Database, version 4.1s (GFED4.1s), described by \citet{vanderwerf2017}. The unknown flux components are modelled according to \eqref{eqn:component_decomposition} and \eqref{eqn:beta_model}, which also require bottom-up estimates. Those for GPP and respiration fluxes are provided by the SiB4 model, which is described below. The ocean bottom-up estimates are given by \citet{landschutzeretal2016}, and they are also described below. 

SiB4 is a mechanistic land--surface model that simulates several features of the terrestrial biosphere, including the terrestrial carbon cycle \citep{haynesetal2019a,haynesetal2019b}. Carbon GPP fluxes, respiration fluxes, and biomass are modelled simultaneously. Inputs to SiB4 include the biomass-burning emissions from GFED4.1s (used to model changes to biomass), meteorological drivers such as solar radiation and temperature, and satellite data that inform the model of the type of plants in a grid cell, while the timing of plant life-cycle events (known as plant phenology) is modelled dynamically. The simulations we use were made offline, and cover the period from January 1, 2000 to December 31, 2020, and they are done with a 10-minute time resolution on a latitude--longitude grid of resolution $0.5\degree \times 0.5\degree$. To form a bottom-up estimate suitable for our purposes, we aggregated the simulated fluxes to a $1\degree \times 1\degree$ latitude--longitude grid and an hourly time resolution.

The bottom-up estimates of ocean fluxes by \citet{landschutzeretal2016} are derived from observations of the partial pressure difference of CO$_2$ at the air--sea boundary; the CO$_2$ flux is proportional to this difference. Other inputs include estimates of sea-ice cover and wind speed. The bottom-up estimates of ocean fluxes cover the period from January 1982 to December 2019, and they are available on a latitude--longitude grid of $1\degree \times 1\degree$ at a monthly time resolution.

\subsection{Decomposition of bottom-up flux estimates}
\label{sec:bottom_up_decomposition}

SiB4 and \citet{landschutzeretal2016} together give bottom-up estimates of the net flux in each grid cell and time period for the GPP, respiration, and air-sea flux components. To get bottom-up estimates of the flux trend, seasonality, and residual suitable for use in the basis-function representation in \eqref{eqn:flux_basis}, we decompose the estimates of the net flux according to the spatially varying time-series decomposition in \eqref{eqn:component_decomposition}. That is, the bottom-up estimates of the flux for components $c \in \{ \mathrm{gpp}, \mathrm{resp}, \mathrm{ocean} \}$ are decomposed as
\begin{multline}
  X_c^0(\svec, t)
  = \beta_{c, 0}^0(\svec) + \beta_{c, 1}^0(\svec) t
    + \sum_{k = 1}^{K_c} (\beta_{c, 2, k}^0(\svec) + \beta_{c, 3, k}^0(\svec) t) \cos(2\pi k t / 365.25) \\
    + \sum_{k = 1}^{K_c} (\beta_{c, 4, k}^0(\svec) + \beta_{c, 5, k}^0(\svec) t) \sin(2\pi k t / 365.25)
    + \epsilon_c^0(\svec, t),
  \label{eqn:bottom_up_decomposition}
\end{multline}
for $\svec \in \mathbb{S}^2$ and $t \in \mathcal{T}$. The decomposition is performed separately for each grid cell in the GPP, respiration, and ocean bottom-up estimates by applying least-squares regression to the available bottom-up estimates of the fluxes \citep[21 years for SiB4 and 38 years for][]{landschutzeretal2016}. This procedure gives the spatially resolved coefficients $\beta_{c, \cdot}^0(\cdot)$ and $\beta_{c, \cdot, \cdot}^0(\cdot)$ at the grid resolution, which are then used to characterise each term (including the residual term) appearing in \eqref{eqn:bottom_up_decomposition}.

\begin{figure}[t]
  \begin{center}
    \includegraphics{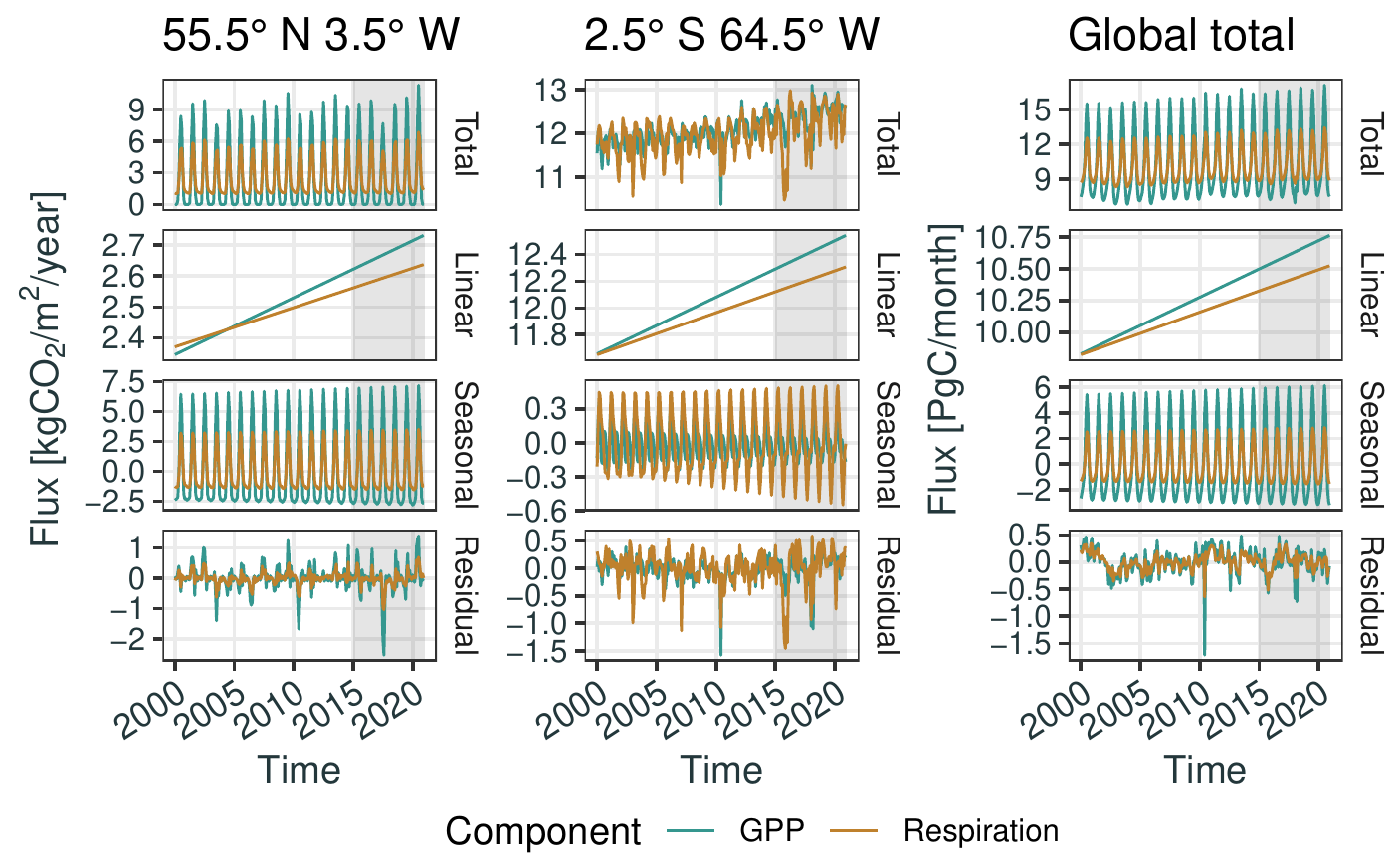}
  \end{center}

  \caption{
    Monthly-aggregate bottom-up flux estimates from SiB4 for $X_\mathrm{gpp}^0(\cdot, \cdot)$ (green) and $X_\mathrm{resp}^0(\cdot, \cdot)$ (brown). From left to right, fluxes are shown, respectively, for a grid cell in the northern extratropics (coordinates $55.5\degree$ N, 3$.5 \degree$ W), a grid cell in the southern tropics (coordinates $2.5\degree$ S, $64.5\degree$ W), and the global total aggregate flux. The first row gives the total fluxes, the second row the linear component of the fluxes, the third row the seasonal component, and the last row gives the residual fluxes, $\epsilon_c^0(\cdot, \cdot)$. GPP fluxes are multiplied by $-1$ for comparison with respiration fluxes. For the grid cell fluxes, the scale is in $\mathrm{kgCO}_2/\mathrm{m}^2/\mathrm{year}$, while the global total flux is given in $\mathrm{PgC}/\mathrm{month}$. The shaded region indicates the flux reporting period, January 2015 to December 2020.
  }
  \label{fig:sib4_fluxes}
\end{figure}

We applied the decomposition in \eqref{eqn:bottom_up_decomposition} to the SiB4 GPP and respiration fluxes with the number of harmonics set to $K_\mathrm{gpp} = K_\mathrm{resp} = 3$. Our time period (September 2014 to March 2021) ends after the last month of the SiB4 simulation (December 2020), so we extend the estimates by setting the residual terms for January to March 2021 equal to those for January to March 2020. For illustration, some of the resulting decomposed fluxes are shown in Figure~\ref{fig:sib4_fluxes}, where the GPP fluxes are shown in green and the respiration fluxes are shown in brown. The first two columns show the fluxes, aggregated to a monthly resolution, at a northern extratropical grid cell (coordinates $55.5\degree$ N, 3$.5 \degree$ W) and at a southern tropical grid cell (coordinates $3.5\degree$ S, $135.5\degree$ E), respectively. The first row depicts the total fluxes, the second row the linear component of the fluxes, the third row the seasonal component, and the last row gives the residual fluxes. Over the 21 years shown, trends in the fluxes and changes to the seasonality can be seen in both grid cells, especially for the fluxes in the tropical grid cell, where a linear trend is visible for both GPP and respiration, and where the amplitude of the seasonal cycle of the respiration fluxes changes substantially in time. The residual fluxes have no apparent seasonal structure, so the choice of three harmonics ($K_{\mathrm{bio}} = 3$) appears to be sufficient.

The bottom-up flux decomposition in \eqref{eqn:bottom_up_decomposition} has the same form as that in \eqref{eqn:component_decomposition}, so the useful properties of time-varying phase and amplitude (Section~\ref{eqn:time_varying_phase}) and preservation under linear aggregation (Section~\ref{sec:linear_aggregates}) also apply to the bottom-up fluxes. Using the latter property, we apply \eqref{eqn:sum_aggregate} to define the SiB4 bottom-up estimate of the NEE, which we denote by $X_\mathrm{nee}^0(\svec, t) \equiv X_\mathrm{gpp}^0(\svec, t) + X_\mathrm{resp}^0(\svec, t)$.

The decomposition in \eqref{eqn:bottom_up_decomposition} was also applied to the bottom-up air-sea fluxes given by \citet{landschutzeretal2016}, $X_\mathrm{ocean}^0(\cdot, \cdot)$, with $K_\mathrm{ocean} = 2$ harmonics. Like the SiB4 estimates, the end of the \citeauthor{landschutzeretal2016} estimates is before the end of our study period (December 2019 and March 2021, respectively). We extend the estimates by assuming that the residual terms for January to December 2020 are equal to those for January to December 2019, and those for January to March 2021 equal to those for January to March 2019. Figure~\ref{fig:landschutzer_fluxes} shows the estimated fluxes and the resulting decomposition for two ocean grid cells and for the global total aggregate ocean flux.

\subsection{\texorpdfstring{Basis functions and process model for $\bm{\alpha}$}{Basis functions and process model for alpha}}
\label{sec:co2_basis_functions}

For the basis-function representation described in Section~\ref{sec:basis_function_decomposition}, we discretise space and time across $R = 23$ regions and $Q = 79$ monthly time periods (September 2014 to March 2021, inclusive). The regions are the 22 TransCom3 regions \citep{gurneyetal2002}, plus an extra region comprising the land area of New Zealand, which will be the target of a separate analysis reported elsewhere. The spatial extents of the regions are shown in Figure~\ref{fig:region_map}, and their full names are given in Table~\ref{tab:region_table}. Of the 22 disjoint TransCom3 regions, 11 are composed principally of land, and 11 of ocean; the boundaries of the regions include some coastal areas that contain both land and ocean.

For the process model on the basis-function coefficients $\alphavec$, we assume zero correlation between the coefficients for the biosphere basis functions and the coefficients for the ocean basis functions, that is, that $\rho_{\mathrm{gpp},\mathrm{ocean}}^\beta = \rho_{\mathrm{resp},\mathrm{ocean}}^\beta = \rho_{\mathrm{gpp},\mathrm{ocean}}^\epsilon = \rho_{\mathrm{resp},\mathrm{ocean}}^\epsilon = 0$. The assumption that land and ocean fluxes are uncorrelated is commonly made in flux inversions \citep[e.g.,][]{basuetal2013}.

One challenge when estimating component fluxes of CO$_2$ is that the mole-fraction field. $Y(\svec, h, t)$, is a function of the spatio-temporal distribution of the net flux of CO$_2$. For the biosphere, this means that it can be hard to attribute fluxes to either the GPP or the respiration component, as attribution must rely on sources of information such as the spatial and temporal distribution of the bottom-up estimates and the sign constraints of the component fluxes. This problem is particularly acute for the intercept terms ($\beta_{c, 0}(\cdot)$) and the trend terms ($\beta_{c, 1}(\cdot)$) in \eqref{sec:decomposition}, because their bottom-up estimates for GPP and respiration are very similar, and their flux basis functions lack complex temporal structure. To address this identifiability issue, we fix these terms for the respiration fluxes to be known and equal to the bottom-up estimates by setting their basis-function coefficients equal to zero. Any change relative to the bottom-up estimate in the linear component of respiration will therefore likely appear in the GPP linear component. Consequently, when making inference, we limit ourselves to examining the linear component of NEE (the sum of GPP and respiration), which can be more reliably estimated.

We also fix the trend and seasonality of the land fluxes to their bottom-up estimates in the 11 predominantly ocean regions and the New Zealand region, which have small land areas, as well as the trend and seasonality of the air--sea fluxes in all regions. We also introduce further information in the form of a multivariate pseudo observation to more reliably constrain the ocean fluxes. A discussion of the rationale for these choices, and the details of their implementation, are given in Appendix~\ref{sec:additional_prior_information}.

The 5,737 basis-function coefficients that we make inference on are therefore:
\begin{itemize}
  \item
    $\alpha_{c, 0, r}$ and $\alpha_{c, 1, r}$ for $c = \mathrm{gpp}$ and $r = 1, \ldots, 11$;

  \item
    $\alpha_{c, j, k, r}$ for $c \in \{ \mathrm{gpp}, \mathrm{resp} \}$, $j = 2, \ldots, 5$, $k = 1, 2, 3$, and $r = 1, \ldots, 11$;

  \item
    $\alpha_{c, 6, r, q}$ for $c \in \{ \mathrm{gpp}, \mathrm{resp}, \mathrm{ocean} \}$, $r = 1, \ldots, 23$, and $q = 1, \ldots, 79$.
\end{itemize}
These coefficients correspond, respectively, to: the intercept and trend for GPP in the land regions; the harmonic terms for GPP and respiration in the land regions; and the residual terms for all components, regions, and months. While some parts of the decompositions of the respiration and ocean components are fixed, the residual terms are allowed to vary for all components. This means that the net flux for $c \in \{ \mathrm{gpp}, \mathrm{bioresid}, \mathrm{ocean} \}$ is not fixed at any time nor at any location.

\subsection{Constraints on terrestrial biosphere fluxes}
\label{sec:constraints}

The constrained multivariate Gaussian prior on the basis-function coefficients, $\alphavec$, given in \eqref{eqn:alpha_prior} in Section~\ref{sec:alpha_prior}, allows for the specification of a constraint set, $F_\alpha$. We use this set to represent two separate types of physical constraints, which both apply to the GPP and respiration fluxes.

The first type of constraint we impose is a sign constraint. The GPP and respiration flux fields satisfy the natural constraints $X_\mathrm{gpp}(\cdot, \cdot) \leq 0$ (always a sink) and $X_\mathrm{resp}(\cdot, \cdot) \geq 0$ (always a source), and it is reasonable to expect that the sum of the trend and seasonality of these fields satisfies the same constraints. We enforce these constraints in approximate fashion by aggregating $X_\mathrm{gpp}(\cdot, \cdot)$ and $X_\mathrm{resp}(\cdot, \cdot)$ onto a space-time grid with a spatial resolution of $2\degree$ latitude by $2.5\degree$ longitude (with half-size polar cells of $1\degree \times 2.5 \degree$; this matches the spatial resolution of the transport model described in Appendix~\ref{sec:transport}) and a monthly temporal resolution that spans the study period $\mathcal{T}$. Let the resulting spatio-temporally aggregated values be given by the vectors $\xvec_\mathrm{gpp}$ and $\xvec_\mathrm{resp}$ for GPP and respiration, respectively. We aggregate the sum of the trend and seasonality of the two fields in the same manner to yield $\xvec_\mathrm{gpp}^\beta$ and $\xvec_\mathrm{resp}^\beta$, respectively. For $c \in \{ \mathrm{gpp}, \mathrm{resp} \}$, denote by $\xvec_c^0$ and $\xvec_c^{\beta,0}$ the same aggregation applied to the bottom-up estimates. Through \eqref{eqn:flux_basis_c}, we can write
\begin{equation}
  \begin{split}
    \xvec_c
    & = \xvec_c^0 + \Phivec_c \alphavec_c, \\
    \xvec_c^\beta
    & = \xvec_c^{\beta,0} + \Phivec_c^\beta \alphavec_c,
  \end{split}
  \label{eqn:aggregated_flux_basis}
\end{equation}
where $\Phivec_c$ is a matrix whose rows are computed by applying the aggregation operation to the basis functions $\phivec_c(\cdot, \cdot)$, and $\Phivec_c^\beta$ is the same as $\Phivec_c$ but with the columns corresponding to the basis-function coefficients for the residual terms set to zero. The sign constraints for the aggregated quantities are
\begin{equation}
  \begin{split}
    \xvec_\mathrm{gpp}^0 + \Phivec_{x,\mathrm{gpp}} \alphavec_\mathrm{gpp}
    & \leq 0
      \quad \text{(\nospellcheck{GPP total} $\leq 0$)}, \\
    \xvec_\mathrm{gpp}^{0,\beta} + \Phivec_{x,\mathrm{gpp}}^\beta \alphavec_\mathrm{gpp}
    & \leq 0
      \quad \text{(\nospellcheck{GPP trend and seasonality}\ $\leq 0$)}, \\
    \xvec_\mathrm{resp}^0 + \Phivec_{x,\mathrm{resp}} \alphavec_\mathrm{resp}
    & \geq 0
      \quad \text{(\nospellcheck{Resp.\ total} $\geq 0$)}, \\
    \xvec_\mathrm{resp}^{0,\beta} + \Phivec_{x,\mathrm{resp}}^\beta \alphavec_\mathrm{resp}
    & \geq 0
      \quad \text{(\nospellcheck{Resp.\ trend and seasonality}\ $\geq 0$)}, \\
  \end{split}
  \label{eqn:constraint_inequalities}
\end{equation}
where the inequality is applied element-wise. These constraints are necessary but not sufficient to satisfy the physical constraints, because they allow for the possibility that the field occasionally violates the sign at spatial and temporal scales finer than those of the aggregates. In this study, however, we only report results that are themselves aggregates of the aggregation, and the constraints in \eqref{eqn:constraint_inequalities} are sufficient for those aggregates. In practice, a slight relaxation of \eqref{eqn:constraint_inequalities} is required, where fluxes are allowed to violate the constraint by up to $10^{-10}$ $\mathrm{kgCO}_2/\mathrm{m}^2/\mathrm{s}$. Without the relaxation, some values of the aggregation are numerically very close to the boundary. This causes the HMC algorithm that samples from the conditional distribution of $\alphavec$ to spend too much time reflecting off the boundary of the constrained space, which makes the time needed to sample a single value of $\alphavec$ infeasible \citep[see Appendix~\ref{sec:mcmc}, and][for more details on this algorithm]{pakmanpaninski2014}. The allowable violations are small in practical terms: if the sign constraints were violated by the maximum allowable amount for every element of the discretisation, the total flux for the component field over the full study period would violate the constraint by around 0.1\% of its bottom-up estimate.

The second type of constraint we impose pertains to the diurnal cycle of the GPP and respiration fluxes which, as discussed in Section~\ref{sec:decomposition}, appears in the residual term $\epsilon_c(\cdot, \cdot)$. The phase of the diurnal cycle in these components is principally a function of the local solar time and is well understood. However, without any constraints, the parameterisation of the basis-function coefficients in \eqref{eqn:beta_model} allows the diurnal cycle to be flipped within a given month and region, implying an unrealistic 12-hour phase shift in the cycle. We enforce ``no-flip'' constraints to avoid this problem by requiring that $\alphavec_{c, 6} \geq -1$ for $c \in \{\mathrm{gpp}, \mathrm{resp}\}$.

The different types of constraints are all linear and can therefore be jointly represented in matrix-vector form as $\avec + \Phivec \alphavec \geq \zerovec$. In total, there are 1,136,810 constraints across 6,325 variables. The matrix $\Phivec$ is therefore high-dimensional, although it is also rather sparse with only 0.2\% of its entries non-zero. To reduce the computational burden of working with this system of constraints, we used the linear programming method of \citet{caronetal1989} to identify constraints that are implied by the others and that can therefore be deemed redundant, and hence removed. Through this method, the number of constraints was reduced by approximately a factor of two, to 623,485.

\subsection{Mole-fraction observations}
\label{sec:data}

The CO$_2$ mole-fraction observations used for our inversion include retrievals of column-averaged CO$_2$ mole-fraction by NASA's OCO-2 satellite \citep{elderingetal2017}, and point-referenced \emph{in situ} and flask measurements of CO$_2$ mole-fraction from a variety of sources \citep{schuldtetal2021b,schuldtetal2021a,tohjimaetal2005,naraetal2017}. These data sets are prescribed in the OCO-2 v10 MIP protocol, and full descriptions of their characteristics are given in Appendix\ref{sec:data_details}. 

\begin{figure}
  \begin{center}
    \includegraphics{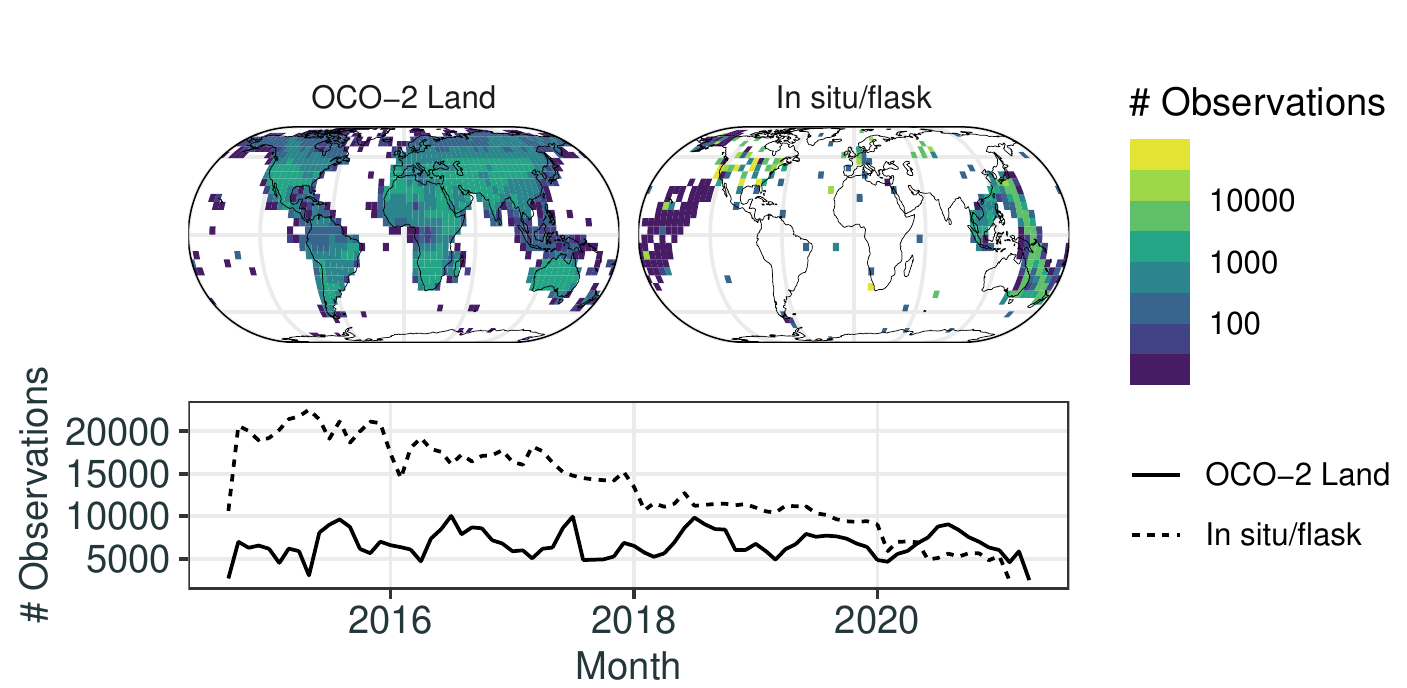}
  \end{center}
  \caption{
    The total number of observations in each $5\degree \times 5\degree$ grid cell for each observation type (OCO-2 or \emph{in situ}/flask) in our study period (top panel), and the total number of observations per month for each type in our study period (bottom panel). In the top panel, unshaded cells have zero observations, the colour scale has been log-transformed, and the number of observations at different vertical levels is summed together.
  }
  \label{fig:observation_map}
\end{figure}

In the context of flux inversion, satellite retrievals of column-average CO$_2$ have different strengths and weaknesses to \emph{in situ} and flask measurements of CO$_2$. \emph{In situ} and flask measurements are considered to be unbiased and to have negligible measurement error. On the other hand, these measurements only provide a snapshot of the mole-fraction field at a specific altitude, latitude, and longitude, which may not be representative of the average of the coarse 3-D grid cell used in the transport model. By contrast, column-average CO$_2$ retrievals contain information on CO$_2$ averaged across all vertical levels (though its sensitivity varies with altitude and the satellite retrieval usually contains the equivalent information of only one or two observations), which can make them less sensitive to errors in modelling transport \citep{deutscheretal2010}. Satellite retrievals have wide spatial and temporal coverage, though they are less abundant in cloudy regions such as parts of the tropics, and in high latitudes during winter. By contrast, \emph{in situ} and flask measurements are available in great abundance in some regions such as North America, Western Europe, and parts of the Pacific Ocean, and are absent in others; see Figure~\ref{fig:observation_map}, which shows the spatial and temporal density of the observations of each type. The temporal coverage of the OCO-2 data is even throughout the study period, while the number of \emph{in situ} and flask measurements has declined over time. This decrease is largely due to the time it takes for observations to be validated and incorporated into the \emph{in situ} and flask data sets.

The error characteristics for the different data sources are encoded in the mole-fraction data model (see Section~\ref{sec:full_hierarchical_model} above and Appendix~\ref{sec:mole_fraction_data_model}). We describe how the data model is tailored to the CO$_2$ mole-fraction observations in Appendix\ref{sec:error_budgets}.

\subsection{\texorpdfstring{Remaining details of the CO$_2$ application}{Remaining details of the CO2 application}}

The remaining details of how WOMBAT~v2.0 is configured for the CO$_2$ application are given in appendices: Appendix~\ref{sec:transport} describes the chemical transport model \citep[GEOS-Chem;][]{beyetal2001} and the mole-fraction initial condition that we use, Appendix~\ref{sec:co2_parameter_model} describes the priors on the parameters of the process model and the data model, and Appendix~\ref{sec:computation} gives details of how the computations were performed.

\section{Results}
\label{sec:results}

We now present the results of applying the WOMBAT~v2.0 framework described in Section~\ref{sec:model} to the OCO-2 and \emph{in situ}/flask data described in Section~\ref{sec:co2_estimation}. In Section~\ref{sec:flux_linear_trends}, we consider the net flux over the study period, as well as linear trends in the fluxes. The broad-scale characteristics of the seasonal part of the fluxes are discussed in Section~\ref{sec:flux_seasonal_cycle}. Estimates of the residual, $\{ \epsilon_c(\cdot, \cdot) : c \in \mathcal{C} \}$, are discussed in Section~\ref{sec:flux_residual}. In Section~\ref{sec:seasonal_cycle_changes} we discuss evidence for changes to the flux seasonal cycles. Finally, in Section~\ref{sec:parameter_estimates} we present posterior estimates of the model parameters and discuss their physical relevance. When considering the flux estimates in this section, it is important to note that the posterior estimates for the NEE are likely to be the most reliable, as the mole-fraction process, and hence the observed mole fraction, is driven by the net flux.

\subsection{Net flux and linear trends}
\label{sec:flux_linear_trends}

The linear trend in \eqref{eqn:component_decomposition} captures the long-term changes in the fluxes. As described in Section~\ref{sec:co2_basis_functions}, we limit ourselves to examining the linear part of NEE, because the individual linear parts of the GPP and respiration components are largely unidentifiable.

\begin{figure}[t]
  \begin{center}
    \includegraphics{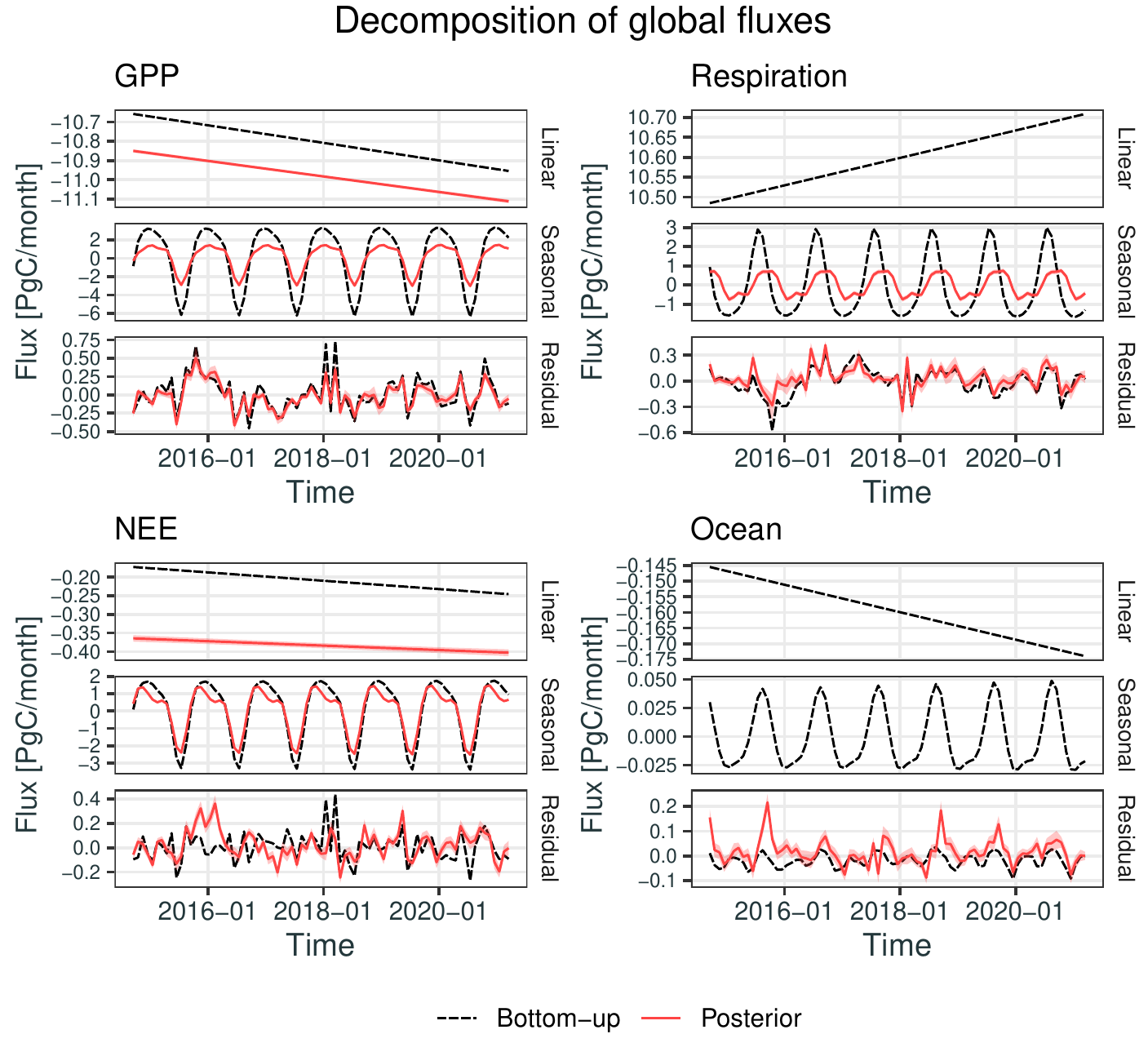}
  \end{center}
  \caption{
    Bottom-up and posterior-mean monthly global totals of the linear, seasonal, and residual parts of the flux components for GPP, respiration, ocean fluxes, and NEE (the sum of GPP and respiration). The shaded areas depict the values between the 2.5th and 97.5th posterior percentiles (i.e., a 95\% prediction interval). Panels that do not show posterior estimates of fluxes correspond to terms that are fixed to bottom-up estimates.
  }
  \label{fig:global_components}
\end{figure}

\emph{Global land sink is increasing:} Figure~\ref{fig:global_components} shows estimates of the monthly global total fluxes for each part of the decomposition in \eqref{eqn:component_decomposition} for the GPP, respiration, and ocean flux components, as well as for the NEE. Note that the posterior uncertainty of the linear component is difficult to see in Figure~\ref{fig:global_components} because it is small. The posterior estimate of the linear component of the global NEE flux indicates a deeper sink of CO$_2$ (i.e., it is more negative) than the bottom-up estimate. The trend of the posterior net global sink is $-0.008$ to $-0.005$ PgC/year (2.5th and 97.5th posterior percentiles), while the bottom-up estimate (SiB4) of the trend is $-0.011$ PgC/year. The sink is therefore estimated to be increasing in magnitude over time in both the posterior and the bottom-up estimate; this is thought to be mostly driven by the ``CO$_2$ fertilisation'' effect, where an increase in atmospheric CO$_2$ directly causes increased uptake of CO$_2$ by the terrestrial biosphere \citep{norbyetal2005}. However, the trend is more negative in the bottom-up estimate than in our posterior estimate. This suggests that either the CO$_2$ fertilisation effect modelled by SiB4 is too strong on average, or that other phenomena not modelled by SiB4 may be influencing the trend, such as changes to land usage.

\emph{Northern extratropics (boreal and temperate) are a decreasing sink:} Figure~\ref{fig:flux_components_n_extratropics} shows the estimated fluxes for the northern boreal band (50$\degree$N--90$\degree$N) and the northern temperate band (23$\degree$N--50$\degree$N) in the same format as Figure~\ref{fig:global_components}. The posterior estimates of the NEE for these regions show that the northern extratropics are a net sink of CO$_2$ (i.e., fluxes are negative) but, unlike the global fluxes, the trends for these regions are positive at 0.0003 to 0.0009 PgC/year and 0.0018 to 0.0031 PgC/year for the boreal and temperate band, respectively (2.5th and 97.5th posterior percentiles). This indicates the magnitude of the sink is decreasing, and is in contrast to the bottom-up estimates where the trend is negative. This suggests that the capacity for ecosystems in this region to act as a net sink of CO$_2$ may be reducing over time.

\begin{figure}
  \begin{center}
    \includegraphics{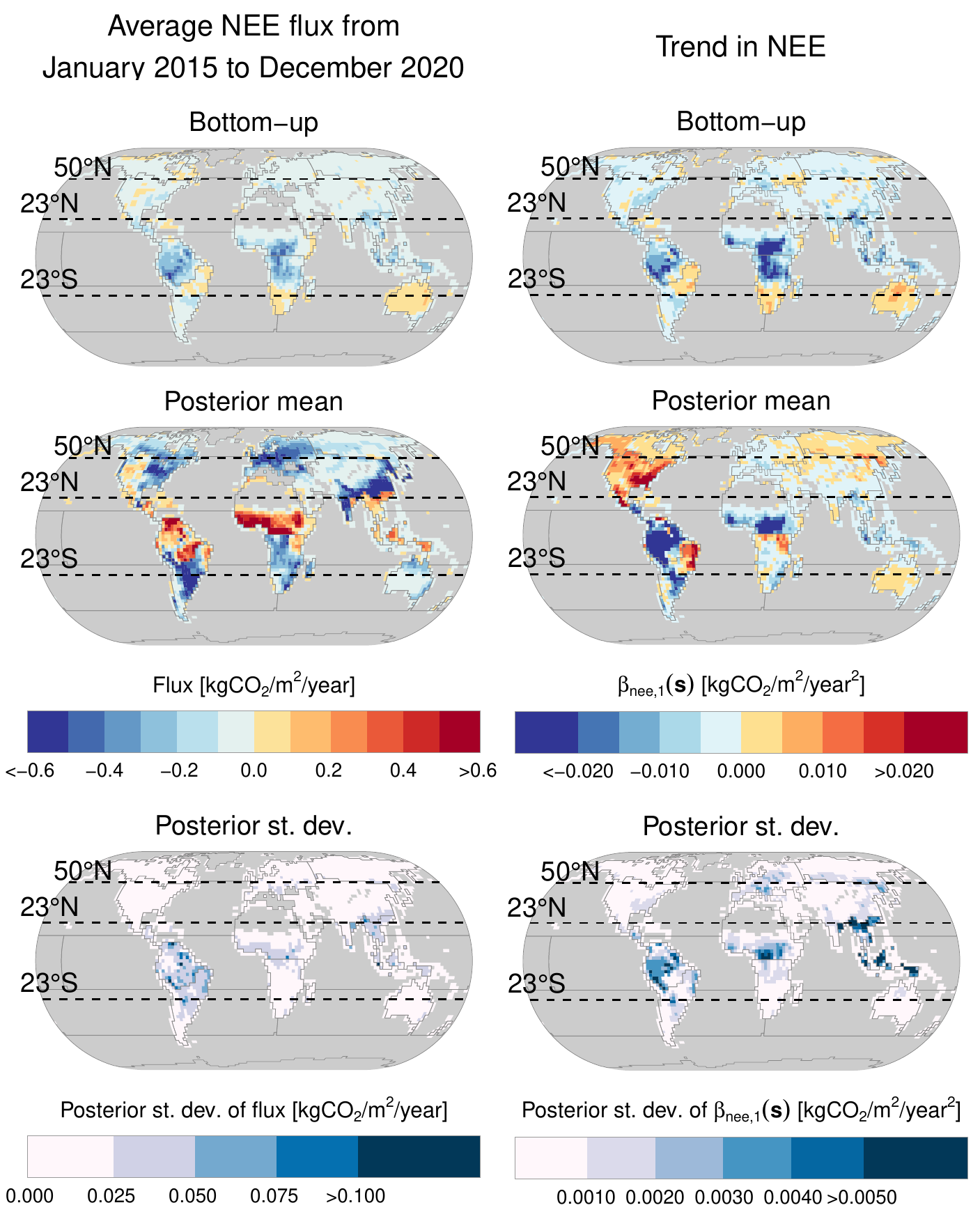}
  \end{center}
  \caption{
    Estimated average NEE flux over January 2015--December 2020 (left) and trend in NEE fluxes $\beta_{\mathrm{nee}, 1}(\cdot)$ (right). The grid cells are $2\degree \times 2.5\degree$, the resolution of the transport model (see Appendix~\ref{sec:transport}). The top row shows the bottom-up estimates, the middle panel shows the posterior mean estimates, and the bottom panel shows the posterior standard deviation. Cells shown in grey have zero NEE flux, and the colour scales are truncated to their maximum values. Grey lines mark the boundaries of the 23 regions used for the basis functions (see Section~\ref{sec:co2_basis_functions}).
  }
  \label{fig:average_and_trend_map}
\end{figure}

To investigate sub-zonal spatial variability in the net flux and the trends leading to this decrease over time, Figure~\ref{fig:average_and_trend_map} shows grid-scale bottom-up and posterior mean estimates of the average NEE flux over January 2015--December 2020 (which indicates whether a region is a net source or sink), as well as estimated trends in the NEE fluxes. The net sink of CO$_2$ observed for the northern extratropics is present at almost all locations, except for parts of the western United States. The majority of the net positive trend in NEE observed for the region appears to arise from temperate and boreal North America.

\emph{Tropics (northern and southern) transition from source to sink:} The bottom-up and posterior estimates of the fluxes for the northern tropics (0$\degree$--23$\degree$N) and southern tropics (23$\degree$S--0$\degree$) are shown separately in Figure~\ref{fig:flux_components_tropical}, and estimates for the tropics as a whole (23$\degree$S--23$\degree$N) are shown in Figure~\ref{fig:flux_components_tropical_s_extratropics}. The bottom-up estimates of the NEE attribute both the northern and southern regions as a sink of CO$_2$, while the posterior estimates of the NEE indicate a net source in the northern tropics and a net sink in the southern tropics. The tropics as a whole transition in 2018 from being a net source to a net sink of CO$_2$, with the 2.5th and 97.5th posterior percentiles for the trend at -0.010 and -0.007 PgC/year, respectively. The negative trend is present in both the northern and southern tropics, indicating that ecosystems in both regions trended towards increased carbon uptake over the study period. This increasing absorption of CO$_2$ by tropical ecosystems is important, as tropical rainforests are the ``lungs'' of our planet \citep{wood1990}, and our results indicate that they may play an increasing role over time in offsetting anthropogenic emissions of CO$_2$.

The grid-scale estimates in Figure~\ref{fig:average_and_trend_map} show strong net source regions in tropical northern Africa and parts of tropical South America. This is in disagreement with the bottom-up estimate (SiB4), but for North Africa it is in agreement with observational studies such as that  of \citet{palmeretal2019}, who use very similar data to us to conclude that this region is the cause of the majority of tropical emissions. In both tropical northern Africa and northern tropical South America, the flux trend is negative, indicating that these sources are getting smaller. One caveat to our findings in these regions is that both tropical northern Africa and northern tropical South America are major biomass burning sources; biomass burning fluxes are considered fixed and known in our inversion, so inaccuracies in the bottom-up biomass burning estimates could impact the posterior estimates of the NEE. This is particularly relevant for the Amazon, where \citet{gattietal2021} find that climate change and deforestation are impacting the frequency and intensity of burns.

\emph{Southern extratropics are an increasing sink:} Estimates of the fluxes for the southern extratropics are shown in Figure~\ref{fig:flux_components_tropical_s_extratropics}. These indicate a net sink in the region that is deepening over time. The grid-scale estimates in Figure~\ref{fig:average_and_trend_map} locate this sink primarily in southern South America and southern Africa, and the largest increases in the sink occur in southern South America. 

\subsection{Flux seasonal cycle}
\label{sec:flux_seasonal_cycle}

In this section we examine the broad-scale characteristics of our estimates of the seasonal cycles of the components of CO$_2$ flux through the seasonal component in \eqref{eqn:component_decomposition}; consideration of \emph{changes} to these cycles is given later, in Section~\ref{sec:seasonal_cycle_changes}.

\emph{Posterior global seasonal cycles have smaller amplitudes than bottom-up}: The posterior and bottom-up estimates of the seasonal cycles of the component CO$_2$ global total fluxes are shown in Figure~\ref{fig:global_components} (note that the posterior uncertainty of the seasonal cycle is difficult to see in the plots because it is small). The posterior estimate of the global seasonal cycle has a smaller amplitude than the bottom-up estimate for NEE, GPP, and respiration. This feature also appears in the seasonal cycles of the northern and southern extratropics, shown in Figure~\ref{fig:flux_components_n_extratropics} and Figure~\ref{fig:flux_components_tropical_s_extratropics}, respectively.

\emph{Posterior tropical NEE seasonal cycle has larger amplitude than bottom-up}: In contrast to the extratropical regions, the posterior estimates of the seasonal cycles for the tropics, shown in Figure~\ref{fig:flux_components_tropical}, have larger amplitudes than the bottom-up estimates. However, the individual GPP and respiration seasonal cycles in the tropics have smaller amplitudes. The difference between the posterior and bottom-up estimates of the amplitude is most striking for the respiration seasonal cycle in the southern tropics, which has relatively little seasonality in the posterior estimates.

\emph{Global NEE and respiration seasonal cycles have two peaks:} The posterior estimates of the global respiration and NEE fluxes (Figure~\ref{fig:global_components}) both have a small second peak in March that does not appear in the bottom-up estimates. The second peak in NEE has been observed in a number of other flux inversions \citep{peiroetal2022}.

\subsection{Residual flux}
\label{sec:flux_residual}

The residual term in \eqref{eqn:component_decomposition} incorporates any features of the fluxes that are not due to either seasonality or to long-term changes. This includes disturbances to the GPP and respiration fluxes due to external influences such as a strong \nospellcheck{El Ni\~{n}o} event.

\emph{Emissions during the 2015/16 \nospellcheck{El Ni\~{n}o} event}: The posterior estimates in Figure~\ref{fig:global_components} indicate that there was a large positive residual NEE flux from December 2015 to February 2016 during the major \nospellcheck{El Ni\~{n}o} event in 2015/16. The positive NEE flux is primarily due to an unusually large positive residual GPP flux (corresponding to reduced photosynthetic activity) which, unlike in the bottom-up estimate, is not balanced out by an equivalently large negative respiration flux residual. The zonal-scale estimates of the residual in Figures~\ref{fig:flux_components_n_extratropics}--Figure~\ref{fig:flux_components_tropical_s_extratropics} indicate that these residual fluxes arose from the tropical bands between $23\degree$S and $23\degree$N (shown in Figure~\ref{fig:flux_components_tropical}). To see how the residual flux in this period was distributed in space, Figure~\ref{fig:el_nino_2015_residual} shows the estimated residual flux in each grid cell averaged over the three months from December 2015 to February 2016. The majority of the NEE flux during this event occurs in tropical South America. Since this arises from a positive GPP flux anomaly (that is, a decrease in primary productivity), this suggests that the effect of the \nospellcheck{El Ni\~{n}o} event on this ecosystem was to decrease plant growth without attenuating other processes, such as plant decay, that drive respiration.

\citet{liuetal2017} also study the 2015/16 \nospellcheck{El Ni\~{n}o}. They used a two-stage process in which a flux-inversion framework is applied to OCO-2 data to estimate the NEE flux, and then the NEE is separated into GPP and respiration components using auxiliary data in the form of retrievals of solar-induced fluorescence \citep[SIF;][]{sunetal2018}, a quantity that is linked to GPP. Our estimates during the 2015/16 \nospellcheck{El Ni\~{n}o} event partially corroborate those of \citeauthor{liuetal2017}, who find that the tropical ecosystem responded strongly to this event. \citeauthor{liuetal2017} also found the strongest net source to be in tropical South America, and they also attribute this to decreased GPP. However, unlike us, \citeauthor{liuetal2017} also found a net source during this event in tropical Africa, which they attribute to increased respiration. This disagreement may arise from differences in inversion methodology, the priors over the fluxes, or due to the different technique \citeauthor{liuetal2017} use to isolate GPP.

\emph{Ocean residual may contain land fluxes:} The residual for the global total ocean fluxes in Figure~\ref{fig:global_components} contains the adjustments to the ocean fluxes from the bottom-up estimate. The residual shows some evidence of seasonality, which is unexpected. This might be occurring because of small problems in the ocean flux seasonality (which is fixed \emph{a priori}), or it might indicate that some seasonality that ought to be attributed to the land has been attributed to the ocean. The posterior estimates of the ocean residual also indicate a positive CO$_2$ flux during the 2015/16 \nospellcheck{El Ni\~{n}o} event. This is unexpected, as the scientific consensus is that sea-to-air fluxes of CO$_2$ decrease during an \nospellcheck{El Ni\~{n}o} event \citep{feelyetal2002,liaoetal2021}. This may be another instance of land fluxes being inferred as ocean fluxes.

\subsection{Phase shifts and amplitude changes in flux seasonal cycles}
\label{sec:seasonal_cycle_changes}

In this section we discuss our estimates of phase shifts and amplitude changes of the annual cycle of natural land carbon fluxes. As shown in \eqref{eqn:harmonic_phase_amplitude}, the harmonic part of the component flux $X_c(\svec, t)$, $c \in \mathcal{C}$, or of a spatial aggregate $\tilde{X}_c(\mathcal{S}, t)$ over a region $\mathcal{S}$, can be written as the sum of $K_c$ sine curves at multiples of the annual frequency. Each harmonic is associated with its own spatio-temporally varying phase and amplitude, $P_{c, k}(\svec, t)$ and $A_{c, k}(\svec, t)$, respectively, for $k = 1, \ldots, K_c$. Denote by $\Delta P_{c, k}(\svec) \equiv P_{c, k}(\svec, t_1) - P_{c, k}(\svec, t_0)$ and $\Delta A_{c, k}(\svec) \equiv A_{c, k}(\svec, t_1) - A_{c, k}(\svec, t_0)$ the phase shift and amplitude change, respectively, over the study period (January 2015 to December 2020) at location $\svec \in \mathbb{S}^2$. For a spatial aggregate over a region $\mathcal{S}$ we analogously denote the phase shift and amplitude change as $\Delta \tilde{P}_{c, k}(\mathcal{S})$ and $\Delta \tilde{A}_{c, k}(\mathcal{S})$, respectively.

We omit discussion of changes to the seasonal cycles in the tropics because the tropical seasonal cycle is relatively weak and observations in the tropics are relatively sparse. We focus on changes to the first annual harmonic ($k = 1$, with a period of 12 months) associated with the GPP, respiration, and NEE fluxes, as this first harmonic largely dominates the seasonality of the land fluxes outside the tropics. For the first annual harmonic, a positive phase shift (left shift) indicates the peak and trough of the harmonic moved earlier in the year, and a negative phase shift (right shift) indicates the peak and trough moved later.

\begin{figure}[t]
  \begin{center}
    \includegraphics{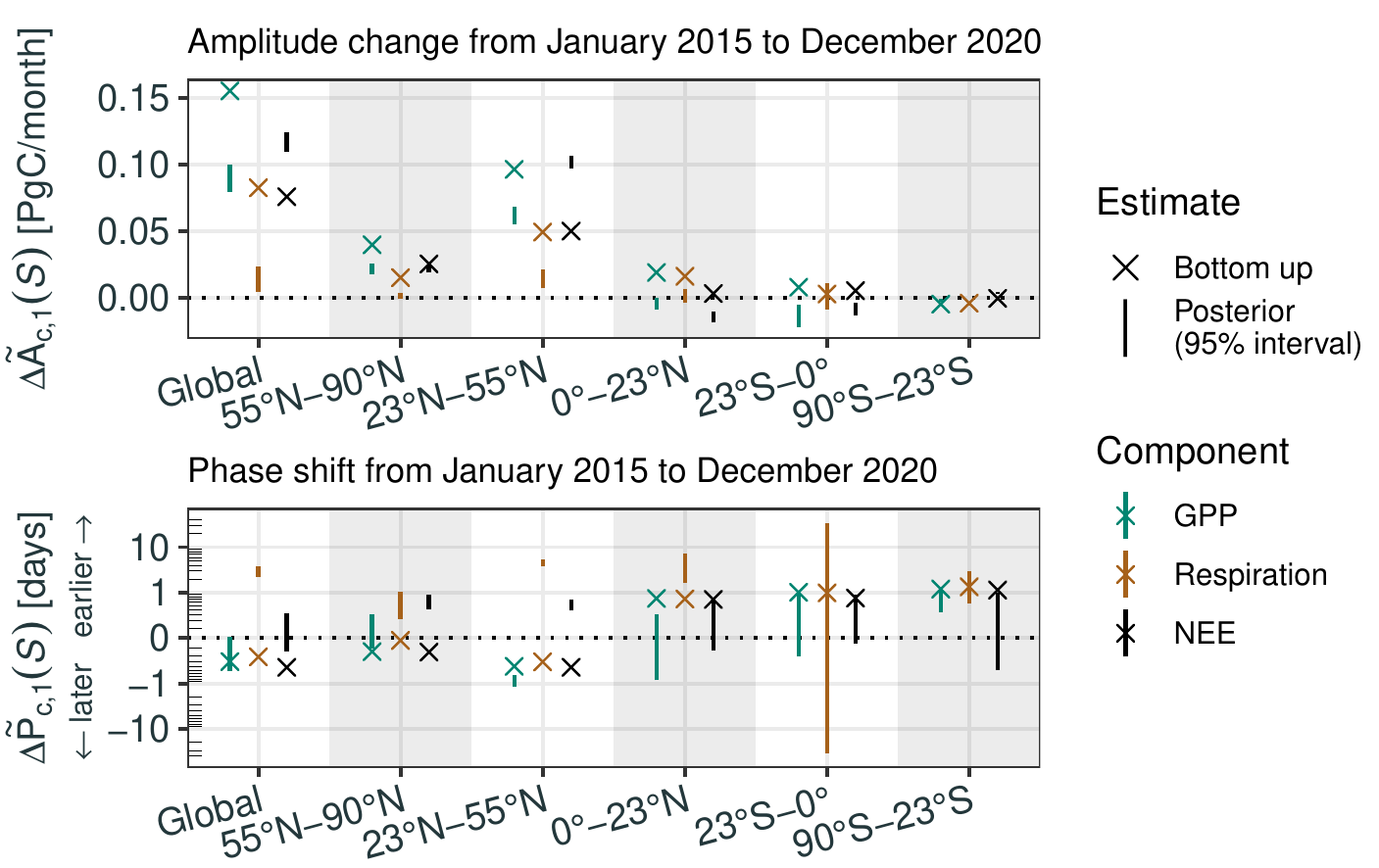}
  \end{center}
  \caption{
    Amplitude changes (top) and phase shifts (bottom) over the period from January 2015 to December 2020 for the first annual harmonic of the GPP, respiration, and NEE fluxes. Changes are shown for the global aggregate fluxes and for the fluxes for five zonal bands: the northern boreal region (55$\degree$N--90$\degree$N), the northern temperate region (23.5$\degree$N--55$\degree$N), the northern tropics (0$\degree$--23$\degree$N), the southern tropics (23$\degree$S--0$\degree$), and the southern extratropics (90$\degree$S--23$\degree$S). The crosses depict the bottom-up estimate, and the vertical line segments span the 2.5th posterior percentile to the 97.5th posterior percentile (i.e., a 95\% prediction interval). The vertical axis for the phase shifts is nonlinear to allow for easier interpretation of the results.
  }
  \label{fig:harmonic_shift_global_zonal}
\end{figure}

\emph{Increasing global seasonal cycle amplitude driven by northern regions:} The upper panel of Figure~\ref{fig:harmonic_shift_global_zonal} shows bottom-up estimates and the 2.5th and 97.5th posterior percentiles for $\Delta \tilde{A}_{c, 1}(\mathcal{S})$ for the GPP, respiration, and NEE fluxes. Estimates are shown for the global total ($\mathcal{S} = \mathbb{S}^2$) and for the five zonal bands discussed previously. The seasonal amplitude of the global NEE flux increased in the posterior by 0.11--0.12 PgC/month (2.5th and 97.5th posterior percentiles), which corresponds to a 7.7\%--8.8\% increase over the six-year study period, or 1.24\%--1.41\% per year. This change is driven by an increase in the amplitude of the NEE fluxes in the northern temperate and boreal regions, which aligns with \citet{gravenetal2013}, who find that the amplitude of the seasonal cycle of atmospheric CO$_2$ has increased over the 50 years since 1960, and who also attribute this change to the northern ecosystems. The GPP and respiration global seasonal amplitudes also increase over the study period, but the change to respiration is much smaller than the change to GPP and NEE. This agrees with \citet{forkeletal2016}, who find that changes to ecosystem productivity drive the changing seasonal amplitude of NEE.

\begin{figure}
  \begin{center}
    \includegraphics{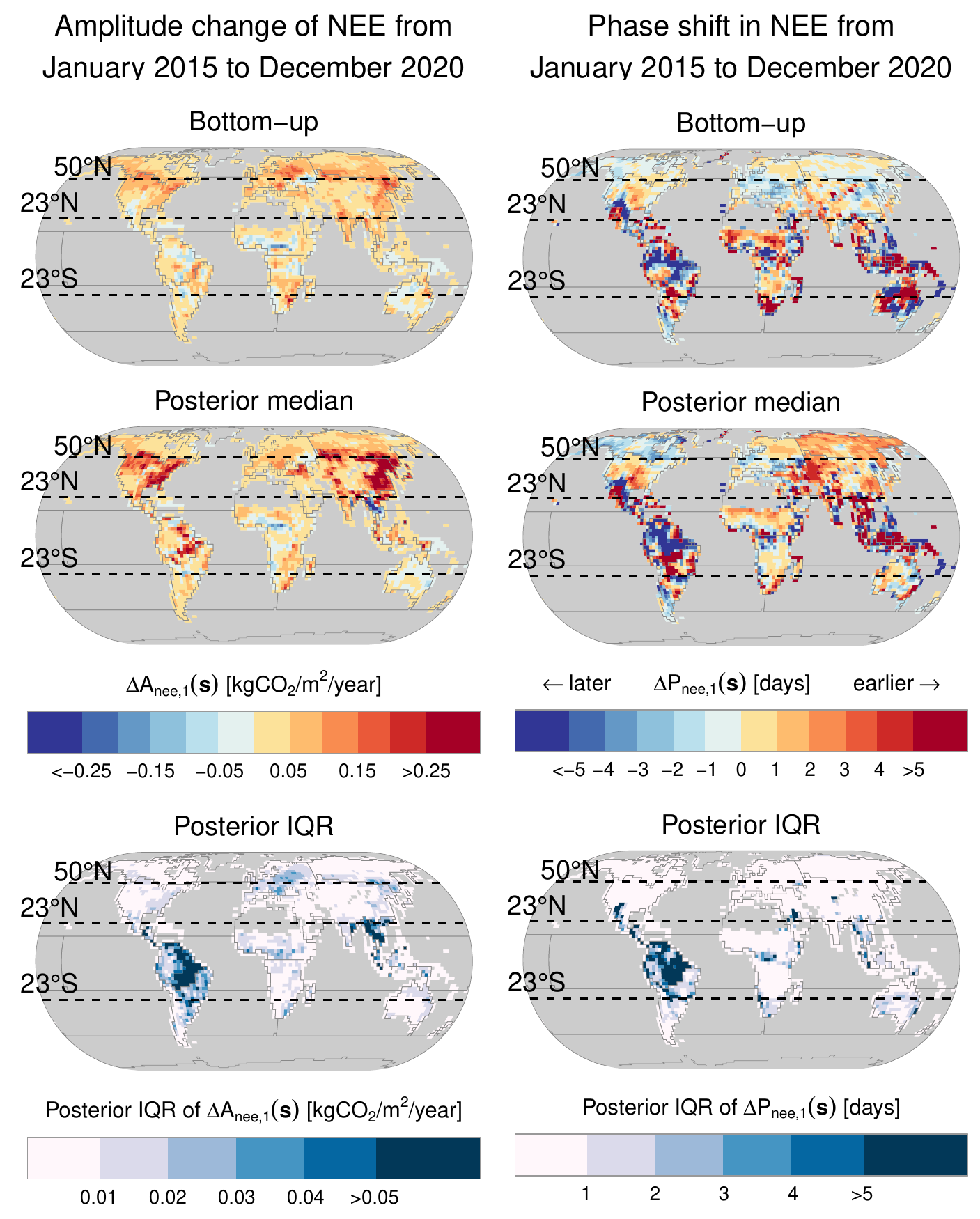}
  \end{center}
  \caption{
    Grid-scale phase shifts (left column) and amplitude changes (right column) from January 2015 to December 2020 for the first annual harmonic of the NEE flux component. The grid cells are $2\degree \times 2.5\degree$, the resolution of the transport model (see Appendix~\ref{sec:transport}). The first row shows the bottom-up estimate of the quantity, the second row the posterior median, and the third row the posterior interquartile range (IQR). Areas with zero flux are coloured grey. Grey lines mark the boundaries of the 23 regions used for the basis functions (see Section~\ref{sec:co2_basis_functions}), and the colour scales are truncated to their maximum values.
  }
  \label{fig:harmonic_shift_map_nee}
\end{figure}

To investigate amplitude changes at the sub-zonal scale, the left column of Figure~\ref{fig:harmonic_shift_map_nee} shows grid-scale bottom-up estimates, posterior medians, and posterior interquartile ranges (IQRs) for $A_{c, 1}(\svec, t)$ for the NEE flux component. Figures~\ref{fig:harmonic_shift_map_gpp} and \ref{fig:harmonic_shift_map_resp} repeat Figure~\ref{fig:harmonic_shift_map_nee} but for the GPP and respiration fluxes, respectively. Across the northern extratropics, the amplitude changes for NEE and GPP in the posterior estimates are uniformly positive, and those for respiration are either positive or small in magnitude. The bottom-up estimates of the amplitude changes largely agree in sign with the posterior, but the magnitude of the posterior changes are mostly larger.

\emph{Global NEE phase is not changing:} The upper panel of Figure~\ref{fig:harmonic_shift_global_zonal} shows bottom-up estimates and 2.5th and 97.5th posterior percentiles for the phase shift, $\Delta \tilde{P}_{c, 1}(\mathcal{S})$, in the same format and for the same regions as for the amplitude changes. The posterior 95\% interval for the phase shift in NEE includes zero, and spans $-$0.1 days to $+$0.3 days; this contrasts with the bottom-up estimate of a -0.4 day shift.

\emph{Northern extratropical (boreal and temperate) NEE cycle shifting earlier:} The posterior estimate for the phase shift in the NEE fluxes is 0.4--0.7 days earlier (2.5th and 97.5th posterior percentiles) for the northern temperate latitudes (23$\degree$N--55$\degree$N) and 0.4--0.9 days earlier for the northern boreal latitudes (55$\degree$N--90$\degree$N). Consistent with our results, \citet{gravenetal2013} found earlier shifts of 0.17 days/year in the seasonal cycle of CO$_2$ concentrations observed at a location in the northern boreal latitudes (Barrow, Alaska). The right column of Figure~\ref{fig:harmonic_shift_map_nee} shows grid-scale bottom-up estimates, posterior medians, and IQRs for the phase shift, $P_{c, 1}(\svec, t)$, for the NEE flux component. The grid-scale posterior estimates of the phase shift are consistently earlier in the northern temperate region, but in the northern boreal region the shift is later in North America and earlier in Eurasia.

\emph{Regionally-varying phase shifts in GPP in the northern extratropics:} Several studies have found evidence that climate change is influencing the timing of growing seasons, linked to GPP, in northern regions \citep[e.g.,][]{zhuetal2012,buitenwerfetal2015,gonsamoetal2018,parketal2019}. The posterior estimate of the phase shift for GPP in the northern boreal region is $-$0.1 to $+$0.3 days (2.5th and 97.5th posterior percentiles; see Figure~\ref{fig:harmonic_shift_global_zonal}); at the regional scale, the shift is towards later seasons in boreal North America, and earlier in boreal Eurasia (Figure~\ref{fig:harmonic_shift_map_gpp}). For the northern temperate region, the prediction interval of the phase shift for GPP is $-$1.2 to $-$0.7 days (an earlier shift). Like the boreal region, however, the grid-scale estimates reveal substantial variation in the sign of the phase shifts within the temperate region. Both \citet{gonsamoetal2018} and \citet{parketal2019} find a net phase shift towards earlier peak photosynthesis (linked to GPP) in the northern extratropics; our results suggest that there is substantial regional variation in changes to the phase of the GPP seasonal cycle.

\emph{Respiration shifting earlier in the northern and southern extratropics}: The posterior estimate for the phase shift of respiration is 0.2--1.0 days earlier in the northern boreal region, 3.7--5.3 days earlier in the northern temperate region, and 0.6--3 days earlier in the southern extratropics (2.5th and 97.5th posterior percentiles; see Figure~\ref{fig:harmonic_shift_global_zonal}). The grid-scale estimates in Figure~\ref{fig:harmonic_shift_map_resp} show that the posterior phase shift is uniformly earlier across the northern boreal region, but the phase shift varies in sign within the northern temperate region and in the southern extratropics.

\subsection{Parameter estimates}
\label{sec:parameter_estimates}

Recall from Section~\ref{sec:inference} that we use a two-stage process to get estimates $\hat{\rhovec}^Z$ and $\hat{\ellvec}^Z$ of the parameters $\rhovec^Z$ and $\ellvec^Z$, respectively, and these are held fixed in the MCMC sampling scheme. For the OCO-2 land observations, this gives $\hat{\rho}_\mathrm{OCO\textrm{-}2\;land}^Z = 0.934$, indicating that the majority of the error is correlated; for the remaining groups, we assume that $\rho_g^Z = 1$ (see Appendix\ref{sec:error_budgets} for our justification). The estimated length scales, $\ell_g^Z$, for each group are 53.9 seconds for OCO-2 land retrievals, 7.4 minutes for aircraft measurements, 4.9 hours for shipboard measurements, 16 hours for surface measurements, and 14.9 hours for tower measurements. It appears that the inferred length scales are proportional to the speeds of the moving instruments. The OCO-2 satellite's ground track moves very quickly (24,480 km hour$^{-1}$) and its observations have the shortest length scale. Aircraft observations are the next fastest (around 600 km hour$^{-1}$) and have the next shortest length scale, followed by shipboard observations (around 20 km hour$^{-1}$). The stationary instruments at surface and tower sites have the longest length scales. Taking into account instrument velocity, the implied spatial correlation length scale is on the order of a hundred kilometres, which is similar to the resolution of our flux/transport model (between 100--200 km, depending on latitude). This similarity suggests that the correlated components of the error budgets are dominated by errors in the transport model.

Table~\ref{tab:hyperparameter_table} shows posterior 2.5th, 50th, and 97.5th percentiles for the parameters that govern the prior on $\alphavec$. The estimated precisions for the GPP basis-function coefficients are larger than those for respiration for both the trend and seasonality ($\tau_c^\beta$) and the residual ($\tau_c^\epsilon$). This indicates that smaller adjustments from the prior are needed for GPP than for respiration. The posterior median of the correlation between the GPP and respiration basis-function coefficients is 0.78 for the trend and seasonality ($\rho_\mathrm{gpp,resp}^\beta$), and 0.20 for the residual ($\rho_\mathrm{gpp,resp}^\epsilon$). The respiration fluxes are positive and the GPP fluxes are negative, so the positive correlation parameters indicate that adjustments to these two flux fields are anti-correlated, which is to be expected. The adjustments to the trend and seasonality of the components may be more correlated than the residuals because they are averaged over longer periods and are driven by similar processes such as the temperature and insolation of Earth's surface. The posterior median temporal correlation between the monthly basis-function coefficients for the residual fluxes for GPP and respiration ($\kappa_\mathrm{bio}^\epsilon$) is 0.52.

\begin{table}[t]
  \begin{center}
    \bgroup
    \def\arraystretch{1.3}
    \begin{tabular}{llll||llll}
  \hline\
  Variable & 2.5\% & 50\% & 97.5\% & Variable & 2.5\% & 50\% & 97.5\% \\ \hline
$\tau_\text{gpp}^\beta$ & 0.774 & 1.001 & 1.253 & $\kappa_\text{bio}^\epsilon$ & 0.465 & 0.516 & 0.563 \\
$\tau_\text{resp}^\beta$ & 0.377 & 0.488 & 0.624 & $(\gamma_\text{OCO-2 land}^Z)^{-1}$ & 1.318 & 1.323 & 1.328 \\
$\tau_\text{gpp}^\epsilon$ & 3.076 & 3.509 & 3.950 & $(\gamma_\text{Aircraft}^Z)^{-1}$ & 1.393 & 1.423 & 1.454 \\
$\tau_\text{resp}^\epsilon$ & 1.665 & 1.912 & 2.187 & $(\gamma_\text{Surface}^Z)^{-1}$ & 1.433 & 1.440 & 1.446 \\
$\rho_\text{gpp,resp}^\beta$ & 0.700 & 0.778 & 0.841 & $(\gamma_\text{Tower}^Z)^{-1}$ & 1.406 & 1.412 & 1.418 \\
$\rho_\text{gpp,resp}^\epsilon$ & 0.114 & 0.201 & 0.285 & $(\gamma_\text{Shipboard}^Z)^{-1}$ & 1.355 & 1.363 & 1.371 \\
\hline
\end{tabular}
    \egroup
  \end{center}
  \caption{
    Posterior percentiles for the parameters that govern the hierarchical prior on $\alphavec$ (see Section~\ref{sec:alpha_prior}), and for the error-budget scaling factors $\{ (\gamma_g^Z)^{-1} \}$ (see Appendix~\ref{sec:mole_fraction_data_model}).
  }
  \label{tab:hyperparameter_table}
\end{table}

Posterior quantiles for the error-budget scaling factors $(\gamma_\text{OCO-2 land}^Z)^{-1}$, $(\gamma_\text{Aircraft}^Z)^{-1}$, $(\gamma_\text{Shipboard}^Z)^{-1}$, $(\gamma_\text{Surface}^Z)^{-1}$, and $(\gamma_\text{Tower}^Z)^{-1}$ are also shown in Table~\ref{tab:hyperparameter_table}. These are all around 1.3--1.5, indicating that our prescribed error budgets are too small.

\section{Conclusions}
\label{sec:conclusion}

The WOMBAT~v2.0 framework allows for the estimation of changes to the natural cycles of the fluxes of a greenhouse gas during a time when the impacts of climate change are becoming increasingly apparent. In WOMBAT~v2.0, fluxes are modelled using the spatially varying time-series decomposition in \eqref{eqn:component_decomposition}, which includes terms for the trend, seasonal cycle, and inter-annual variability (called the residual) of the fluxes. The statistical distributions of the terms in the decomposition are constructed using bottom-up estimates of the same quantities, and posterior estimates of the flux decomposition are derived from observations of trace-gas mole fractions. Component flux fields with physical constraints (e.g., non-negativity) are dealt with in a fully-Bayesian manner using a constrained prior and an HMC sampling scheme. Software that implements the WOMBAT~v2.0 method is available online at \url{https://github.com/mbertolacci/wombat-v2-workflow}.

Sections~\ref{sec:co2_estimation} and \ref{sec:results} discussed how we applied WOMBAT~v2.0 to estimate global CO$_2$ fluxes from remote sensing, \emph{in situ}, and flask mole-fraction data. We found substantial trends in the fluxes, including that the tropics trend from a net source to a net sink of CO$_2$ over the 2015--2020 time period. The 2015/16 \nospellcheck{El Ni\~{n}o} event was found to have been associated with anomalous net CO$_2$ emissions over tropical South America. We found that the amplitude of the global seasonal cycle of CO$_2$ fluxes is increasing, and that there are phase shifts in different directions in different regions.

Adjustments to the bottom-up estimates in WOMBAT~v2.0 are made for the flux trend and seasonality over a spatial partition, and for the residual over a space--time partition. For the CO$_2$ application described in this paper, the surface of Earth was partitioned into 23 regions, and the time horizon in the study was partitioned into months. A similar partitioning is used by \citet{zammitmangionetal2022} for WOMBAT~v1.0, which is relatively coarse compared to some other flux-inversion systems. For example, \citet{basuetal2013} use a latitude--longitude grid of $6\degree \times 4\degree$ and a monthly time period, while \citet{chevallieretal2005} use a latitude--longitude grid of $3.75\degree \times 2.5\degree$ and an 8-day time period; see \citet{peiroetal2022} for several other examples of how flux inversion systems are parameterised in practice. It is reasonable to expect better estimates of fluxes at finer scales when using a model with a finer resolution, but studies that achieve much higher dimensionality tend to use iterative techniques that (as it stands) preclude full uncertainty quantification in a Bayesian hierarchical framework. The main barrier to using a finer-resolution partition in WOMBAT is the impact on computation time from the resulting increase in dimensionality. Exploring ways to deal with this in WOMBAT, such as through the use of iterative techniques to draw samples from the posterior distribution of the flux field \citep{chevallieretal2005}, is a promising avenue for future work. The novel decomposition into trend, seasonality, and residual at the core of WOMBAT~v2.0 can also be applied to parameterise fluxes outside of a Bayesian hierarchical setting; this approach would allow for increased dimensionality, but possibly at the cost of some untenable modelling assumptions and inaccurate uncertainty quantification.

The decomposition of the fluxes in \eqref{eqn:component_decomposition} cleanly identifies the flux annual cycle. However, the diurnal cycle was not modelled explicitly and, as noted in Section~\ref{sec:decomposition}, it will show up in the residual term. This limits the ability of WOMBAT~v2.0 to identify patterns and changes to the diurnal cycle. Incorporating an explicit diurnal cycle into the decomposition is challenging, because the diurnal cycle itself may vary throughout the year (this is the case for CO$_2$, for example); accommodating this would appear to require many more parameters. Investigating how to parameterise the diurnal cycle and its patterns in a parsimonious and interpretable way would be a promising avenue for future work.

In our application to CO$_2$ fluxes, we considered the partition of the NEE into components of GPP and respiration. This is a strength of our work, since most CO$_2$ flux-inversion systems estimate NEE directly \citep[e.g.,][]{basuetal2013} and hence cannot determine how the net flux arose. In our inversion, the mole-fraction field on which the data are taken depends on the net flux rather than on the component fluxes directly. The decomposition of NEE into GPP and respiration must therefore rely partly on other information, such as the sign constraints on these fields and the bottom-up estimates of the spatio-temporal distribution of the fluxes. A promising avenue to improve the decomposition is to incorporate ancillary data sources such as remotely sensed observations of SIF or observations of the trace gas carbonyl sulfide \citep[COS;][]{maetal2021}, both of which can be directly linked to GPP. For example, SIF data were used by \citet{liuetal2017} to estimate changes to GPP due to the 2015--2016 \nospellcheck{El Ni\~{n}o} event, while \citet{huetal2021} used COS for the boreal and arctic regions in North America. Both SIF and COS are simulated by SiB4 \citep{kooijmansetal2021}, facilitating their use in our framework. However, one challenge to using SIF data is that its relationship to GPP appears to vary between ecosystems \citep{magneetal2020}.

Although we did not do so in Section~\ref{sec:results}, the trends in the fluxes may be projected either backwards into the past, or forwards into the future. The former would be useful for validating the bottom-up and posterior estimates, while the latter could provide some insights into future change (with limited validity since the study period only spans six years). A related possible extension is to expand the linear component of the decomposition to include some curvature, such as through a quadratic term. If combined with a longer study period, this could identify whether the increasing uptake of CO$_2$ by the ocean and the land is speeding up or slowing down. 

Changes to seasonal cycles can take forms other than the phase shifts and amplitude changes that we considered in Section~\ref{sec:seasonal_cycle_changes}. One form of change of scientific interest for CO$_2$ is the length of the growing season, a feature which corresponds roughly to the ``width'' of the trough of the GPP seasonal cycle. This type of change can be accommodated in our framework, but it is more difficult to produce metrics that identify when such a change has happened; producing such metrics would be a valuable focus for future work.

We conclude the paper by emphasising that WOMBAT~v2.0 could also be applied to infer fluxes of other greenhouse gases such as methane or nitrous oxide. The borrowing-of-strength induced by the spatially-varying time series decomposition of the fluxes is likely to be even more beneficial for these gases, for which observations tend to be sparser than for CO$_2$.

\section*{Code and data availability}

Software that implements the WOMBAT~v2.0 framework and reproduces the results in the manuscript is available online at \url{https://github.com/mbertolacci/wombat-v2-framework}. Instructions for acquiring the input data needed to reproduce the results are provided with the software. The outputs of WOMBAT~v2.0 for the application to CO$_2$ are also available. These comprise samples from the posterior distribution of the model parameters, bottom-up estimates and samples from the posterior distribution of the fluxes, and bottom-up estimates and samples from the posterior distribution of the trend/seasonality terms.

\section*{Acknowledgements}

This research was undertaken with the assistance of resources and services from the National Computational Infrastructure (\nospellcheck{NCI}), which is supported by the Australian Government, as well as with the assistance of resources funded by a 2021 University of Wollongong Major Equipment Grant. We are also grateful to the OCO-2 Flux Group and colleagues from the University of Bristol, UK, for their feedback and suggestions. The MERRA-2 data used in this study/project have been provided by the Global \nospellcheck{Modeling} and Assimilation Office (\nospellcheck{GMAO}) at NASA Goddard Space Flight \nospellcheck{Center}. The OCO-2 data were produced by the OCO-2 project at the Jet Propulsion Laboratory, California Institute of Technology. These data, as well as the \emph{in situ}/flask data used in this study, may be obtained from the OCO-2 v10 MIP website at \url{https://gml.noaa.gov/ccgg/OCO2_v10mip/download.php}.

\section*{Funding}

This project was largely supported by the Australian Research Council (ARC) Discovery Project (DP) DP190100180 and by NASA ROSES grant \nospellcheck{20-OCOST20-0004}. Zammit-Mangion was also supported by the ARC Discovery Early Career Research Award DE180100203.

\bibliography{paper}

\appendix
\numberwithin{equation}{section}
\renewcommand\thefigure{\thesection\arabic{figure}}
\renewcommand\thetable{\thesection\arabic{table}}
\renewcommand{\theHfigure}{S{\arabic{figure}}}
\renewcommand{\theHtable}{S{\arabic{table}}}
\setcounter{figure}{0}
\setcounter{table}{0}

\section{Remaining details of the hierarchical model}
\label{sec:remaining_hierarchical_model}

Section~\ref{sec:model} describes the changes from WOMBAT~v1.0 \citep{zammitmangionetal2022} to WOMBAT~v2.0. In this appendix, we give the remaining details of the hierarchical model that underlies WOMBAT~v2.0. Appendix~\ref{sec:alpha_covariance_matrix} describes the structure of the covariance matrix $\Sigmavec_\alpha$, and Appendix~\ref{sec:mole_fraction_process_model} and Appendix~\ref{sec:mole_fraction_data_model} give the details of the mole-fraction process and data models, respectively.

\subsection{\texorpdfstring{Structure of the covariance matrix of $\alpha$}{Structure of the covariance matrix of alpha}}
\label{sec:alpha_covariance_matrix}

As in WOMBAT~v1.0, the covariance matrix $\Sigmavec_\alpha$ in \eqref{eqn:alpha_prior} induces correlations between a component's residual basis-function coefficients over time. In WOMBAT~v2.0 we generalise this to also allow for correlations between the basis-function coefficients of different components. Specifically, we structure the submatrices of $\Sigmavec_\alpha$ that correspond to the trend and seasonality as follows:
\begin{equation}
  \begin{split}
    \var(\alpha_{c, j, r})
    & = 1 / \tau_c^\beta\ ;
      \quad c \in \mathcal{C};\ j = 0, 1;\ r = 1, \ldots, R, \\
    \var(\alpha_{c, j, k, r})
    & = 1 / \tau_c^\beta\ ;
      \quad c \in \mathcal{C};\ j = 2, \ldots, 5;\ k = 1, \ldots, K_c;\ r = 1, \ldots, R, \\
    \mathrm{corr}(\alpha_{c, j, r}, \alpha_{c', j, r})
    & = \rho_{c, c'}^\beta\ ;
      \quad c, c' \in \mathcal{C};\ j = 0, 1;\ r = 1, \ldots, R \\
    \mathrm{corr}(\alpha_{c, j, k, r}, \alpha_{c', j, k, r})
    & = \rho_{c, c'}^\beta\ ;
      \quad c, c' \in \mathcal{C};\ j = 2, \ldots, 5;\ k = 1, \ldots, \min(K_c, K_{c'});\ r = 1, \ldots, R,
  \end{split}
  \label{eqn:clim_dependence}
\end{equation}
where $\tau_c^\beta > 0$, $-1 < \rho_{c, c'}^\beta < 1$, and all other unspecified correlations are fixed to zero. The precision parameter $\tau_c^\beta$ governs how different the trend and seasonality can be from the bottom-up estimate, and the correlation parameter $\rho_{c, c'}^\beta$ models dependencies between any two components $c$ and $c'$. For example, with CO$_2$ we expect the adjustments to the seasonal cycle of respiration to be anti-correlated with those to the seasonal cycle of GPP. Because the fluxes from these components have opposite signs, this would correspond to a value of $\rho_{c, c'}^\beta$ close to one.

The submatrices of $\Sigmavec_\alpha$ corresponding to the residual are structured similarly, but they also account for temporal dependence:
\begin{equation}
  \begin{split}
    \var(\alpha_{c, 6, r, q})
    & = 1 / \tau_c^\epsilon\ ;
      \quad c \in \mathcal{C};\ r = 1, \ldots, R;\ q = 1, \ldots, Q, \\
    \mathrm{corr}(\alpha_{c, 6, r, q}, \alpha_{c, 6, r, q'})
    & = (\kappa_c^\epsilon)^{|q - q'|}\ ;
      \quad c \in \mathcal{C};\ r = 1, \ldots, R;\ q, q' = 1, \ldots, Q, \\
    \mathrm{corr}(\alpha_{c, 6, r, q}, \alpha_{c', 6, r, q})
    & = \rho_{c, c'}^\epsilon\ ;
      \quad c, c' \in \mathcal{C};\ r = 1, \ldots, R;\ q = 1, \ldots, Q,
  \end{split}
  \label{eqn:resid_dependence}
\end{equation}
where $\tau_c^\epsilon > 0$, $0 < \kappa_c^\epsilon < 1$, $-1 < \rho_{c, c'}^\epsilon < 1$, and all other unspecified correlations are fixed to zero. Temporal dependence within a component is captured through the parameter $\kappa_c^\epsilon$, and dependence between pairs of components is captured through the parameter $\rho_{c, c'}^\epsilon$.

Since all off-diagonal entries of $\Sigmavec_\alpha$ not associated with \eqref{eqn:clim_dependence} and \eqref{eqn:resid_dependence} are zero, the basis-function coefficients are assumed to be uncorrelated across the regions $D_1, \ldots, D_R$. This is a reasonable assumption when the regions are large. For example, almost all the spatial regions in our CO$_2$ case study in Section~\ref{sec:co2_estimation} are large relative to the spatial variability of the process. If small regions were used, one would need to explicitly model spatial correlation through $\Sigmavec_\alpha$; one option would be a classical geostatistical model \citep[e.g.,][]{michalaketal2004}.

If all the $C$ components used to construct \eqref{eqn:total_decomposition} are assumed to be unknown, and if correlations between all component processes are modelled, then there are $3C + C(C-1)$ unknown parameters that parameterise $\Sigmavec_\alpha$. We assign independent priors to these parameters as follows:
\begin{align*}
  \tau_c^\beta
  & \sim \mathrm{Gamma}(\nu_\tau^\beta, \omega_{\tau, c}^\beta),
  & \tau_c^\epsilon & \sim \mathrm{Gamma}(\nu_\tau^\epsilon, \omega_{\tau, c}^\epsilon), \\
  \quad \rho_{c, c'}^\beta
  & \sim \mathrm{Beta}(a_\rho^\beta, b_\rho^\beta),
  & \quad \rho_{c, c'}^\epsilon
  & \sim \mathrm{Beta}(a_\rho^\epsilon, b_\rho^\epsilon), \\
  &
  & \kappa_c^\epsilon
  & \sim \mathrm{Beta}(a_\kappa^\epsilon, b_\kappa^\epsilon),
\end{align*}
for $c, c' \in \mathcal{C}$, where $\mathrm{Gamma}(\nu, \omega)$ is the gamma distribution with shape parameter $\nu$ and rate parameter $\omega$, and $\mathrm{Beta}(a, b)$ is the beta distribution with shape parameters $a$ and $b$. In some cases it may be reasonable to assume known values for some of these parameters; for example, in Section~\ref{sec:co2_estimation}, we assume that both $\rho_{c,\mathrm{ocean}}^\beta$ and $\rho_{c,\mathrm{ocean}}^\epsilon$ are zero when $c$ corresponds to one of the terrestrial biosphere components of flux.

\subsection{Mole-fraction process model}
\label{sec:mole_fraction_process_model}

While the flux process is of primary interest in flux inversion, inference on the flux is made using data on the trace-gas mole fraction. In what follows, we use the notation of \citet{zammitmangionetal2022}. Denote by $Y(\svec, h, t)$ the mole-fraction field at location $\svec \in \mathbb{S}^2$, geopotential height $h \geq 0$, and time $t \in \mathcal{T}$. Let $\mathcal{T}_t = [t_0, t]$ for $t_0 \leq t \leq t_1$ be the set of all time points in the study period up to time $t$. The mole-fraction field at time $t$, $Y(\cdot, \cdot, t)$, depends on the mole-fraction field at time $t_0$, denoted by $Y(\cdot,  \cdot, t_0)$ and also known as the ``initial condition,'' and on the fluxes that occur over the period $\mathcal{T}_t$, which we denote by $X(\cdot, \mathcal{T}_t)$. The relationship between the flux and mole-fraction fields is expressed through,
\begin{equation}
  Y(\svec, h, t) = \mathcal{H}(Y(\cdot, \cdot, t_0), X(\cdot, \mathcal{T}_t); \svec, h, t), \quad \svec \in \mathbb{S}^2, t \in \mathcal{T}, h \geq 0.
  \label{eqn:mole_fraction_field}
\end{equation}
The operator $\mathcal{H}$ represents the solution to the underlying chemical transport equations that describe the movement of the trace gas. This transport operator is very close to linear for long-lived trace-gas species like CO$_2$, and therefore \eqref{eqn:mole_fraction_field} can be split up as
\begin{equation}
  Y(\svec, h, t)
  = \mathcal{H}(Y(\cdot, \cdot, t_0), 0; \svec, h, t)
    + \mathcal{H}(0, X(\cdot, \mathcal{T}_t); \svec, h, t), \quad \svec \in \mathbb{S}^2, t \in \mathcal{T}, h \geq 0;
  \label{eqn:mole_fraction_field_split}
\end{equation}
that is, the field at time $t$ is an additive combination of the impact of the initial condition and the impact of the fluxes over $\mathcal{T}_t$. Equation~\eqref{eqn:mole_fraction_field_split} cannot be used directly because the initial condition is not known exactly, and the exact transport operator $\mathcal{H}$ is not available. Instead, we use an estimate of the initial condition, $\hat{Y}(\cdot, \cdot, t_0)$, and an approximate transport operator $\hat{\mathcal{H}}$, which is also approximately linear and is computed by running a chemical transport model. In Section~\ref{sec:co2_estimation}, we use the chemical transport model GEOS-Chem. Both of these approximations introduce the potential for error, which we accommodate by adding an error term to \eqref{eqn:mole_fraction_field_split}:
\begin{equation}
  Y(\svec, h, t)
  = \hat{\mathcal{H}}(\hat{Y}(\cdot, \cdot, t_0), 0; \svec, h, t)
    + \hat{\mathcal{H}}(0, X(\cdot, \mathcal{T}_t); \svec, h, t)
    + v(\svec, h, t),  \quad \svec \in \mathbb{S}^2, t \in \mathcal{T}, h \geq 0,
  \label{eqn:mole_fraction_field_approx}
\end{equation}
where $v(\cdot, \cdot, \cdot)$ is a space-height-time process that accommodates potential errors.

Since $\hat{\mathcal{H}}$ is linear, equation \eqref{eqn:flux_basis} and equation \eqref{eqn:mole_fraction_field_approx} can be combined to yield a basis-function representation in mole-fraction space that is of the same form as the basis-function representation of the flux field:
\begin{equation}
  Y(\svec, h, t)
  = Y^0(\svec, h, t) + \hat{\psivec}(\svec, h, t)' \alphavec + v(\svec, h, t),  \quad \svec \in \mathbb{S}^2, t \in \mathcal{T}, h \geq 0,
  \label{eqn:mole_fraction_basis}
\end{equation}
where $Y^0(\svec, h, t) \equiv \hat{\mathcal{H}}(\hat{Y}(\cdot, \cdot, t_0), 0; \svec, h, t) + \hat{\mathcal{H}}(0, X^0(\cdot, \mathcal{T}_t); \svec, h, t)$ for $\svec \in \mathbb{S}^2, t \in \mathcal{T}, h \geq 0$,  and $\hat{\psivec}(\cdot, \cdot, \cdot)$ is a $(2CR + 4R \sum_{c \in \mathcal{C}} K_c + CQR)$-dimensional vector of basis functions. The entries of $\hat{\psivec}(\cdot, \cdot, \cdot)$ and $\phivec(\cdot, \cdot)$ have a one-to-one correspondence through
\[
  \hat{\psi}_l(\svec, h, t)
  = \hat{\mathcal{H}}(0, \phi_l(\cdot, \mathcal{T}_t); \svec, h, t),  \quad \svec \in \mathbb{S}^2, t \in \mathcal{T}, h \geq 0,
\]
where $\hat{\psi}_l(\cdot, \cdot, \cdot)$ and $\phi_l(\cdot, \cdot)$ are the $l$th entries of $\hat{\psivec}(\cdot, \cdot, \cdot)$ and $\phivec(\cdot, \cdot)$, respectively.

\subsection{Mole-fraction data model}
\label{sec:mole_fraction_data_model}

We denote by $Z_i$ the $i$th observation of the mole-fraction field, where $i = 1, \ldots, N$. Mole-fraction observations used in flux inversion are generally either point-referenced or column-averaged. Point-referenced observations are made at a specific location $\svec_i$ and height $h_i$, usually involve a degree of time averaging (say, over three hours), and are not subject to substantial biases or correlated measurement errors. Column-averaged observations are made at a specific location $\svec_i$ and time $t_i$, represent a weighted average of the vertical mole-fraction field at that time and location, and are typically affected by both bias and correlated measurement errors. The mole-fraction data model is therefore given by
\begin{equation}
  \begin{split}
    Z_i = \begin{cases}
      \mathcal{A}_i(Y(\svec_i, h_i, \cdot)) + \epsilon_i,
        & \text{if $Z_i$ is from a point-referencing instrument}, \\
      \mathcal{A}_i(Y(\svec_i, \cdot, t_i)) + b_i + v_{Z_i} + \epsilon_i,
        & \text{if $Z_i$ is from a column-averaging instrument},
    \end{cases}
  \end{split}
  \label{eqn:mole_fraction_data_model}
\end{equation}
where $\mathcal{A}_i$ is the observation operator for the $i$th observation, $b_i$ is a bias term, $v_{Z_i}$ is a mean-zero spatio-temporally correlated random error, and $\epsilon_i$ is a mean-zero uncorrelated random error independent of $v_{Z_i}$. For point-referenced observations, $\mathcal{A}_i$ represents averaging over time, while for column-averaged observations it represents averaging over the vertical dimension, $h$.

Equations \eqref{eqn:mole_fraction_basis} and \eqref{eqn:mole_fraction_data_model} can be combined to give a basis-function representation of the data model:
\begin{equation}
  Z_i
  = Z_i^0
    + \hat{\psivec}_i' \alphavec
    + b_i
    + \xi_i
    + \epsilon_i, \quad i=1, \ldots, N,
  \label{eqn:mole_fraction_data_basis}
\end{equation}
where $Z_i^0 \equiv \mathcal{A}_i(Y^0(\svec_i, \cdot, \cdot))$; $\hat{\psivec}_i \equiv \mathcal{A}_i(\hat{\psivec}(\svec_i, \cdot, \cdot))$, in which the operator is applied element-wise; $b_i = 0$ for point-referenced observations; and
\[
  \xi_i \equiv \begin{cases}
    \mathcal{A}_i(v(\svec_i, h_i, \cdot))
      & \text{if $Z_i$ is from a point-referencing instrument}, \\
    \mathcal{A}_i(v(\svec_i, \cdot, t_i) + v_{Z_i})
      & \text{if $Z_i$ is from a column-averaging instrument}.
  \end{cases}
\]
All sources of correlated random error are grouped into one term, $\xi_i$. The overall model can be written in matrix-vector form as
\begin{equation}
  \Zvec
  = \Zvec^0
    + \hat{\Psivec} \alphavec
    + \bvec
    + \xivec
    + \epsilonvec,
  \label{eqn:overall_model}
\end{equation}
where $\Zvec \equiv (Z_1, \ldots, Z_N)'$; $\hat{\Psivec}$ is a matrix with $N$ rows for which the $i$th row is equal to $\hat{\psivec}_i'$; $\bvec \equiv (b_1, \ldots, b_N)'$; $\xivec \equiv (\xi_1, \ldots, \xi_N)'$; and $\epsilonvec \equiv (\epsilon_1, \ldots, \epsilon_N)'$.

In both WOMBAT~v1.0 and WOMBAT~v2.0, the dataset $\Zvec$ is split into disjoint groups indexed by $g = 1, \ldots, G$, where the observations within a group have similar bias and error properties. The bias term $\bvec$, and the properties of the error terms $\epsilonvec$ and $\xivec$, are therefore modelled separately for each group. The biases are assigned linear models of the form $\bvec_g = \Avec_g \etavec_g$, where $\bvec_g$ is the vector of bias terms for group $g$, $\Avec_g$ is a group-specific design matrix in which the columns have been standardised by dividing by their standard deviation, and $\etavec_g$ is a corresponding vector of coefficients. In flux inversions, it is common to pre-determine for each observation $Z_i$ an overall ``error budget'' or variance, $V_i$, that comprises the total measurement-error variance plus model-error variance associated with the observation. To reflect this, for observations in group $g$, we let $\var(\xi_i + \epsilon_i) = (\gamma_g^Z)^{-1} V_i$, where $\gamma_g^Z > 0$ is an unknown parameter that rescales the error budgets in group $g$ as needed. To apportion the total error budget between $\xi_i$ and $\epsilon_i$, we let $\var(\xi_i) = \rho_g^Z (\gamma_g^Z)^{-1} V_i$ and $\var(\epsilon_i) = (1 - \rho_g^Z) (\gamma_g^Z)^{-1} V_i$, where $0 < \rho_g^Z < 1$. The correlations between the elements of $\xivec$ in group $g$ are specific to the application. For example, with some groups in our application to CO$_2$, these elements are divided into subgroups (in our case, each subgroup is a time series; see Appendix~\ref{sec:error_budgets}). There are correlations between elements in each sub-group, but elements in different sub-groups are assumed to be mutually independent. The correlations, where present, are governed by a common length-scale parameter, which we denote by $\ell_g^Z$.

The unknown parameters in the data model are $\etavec \equiv (\etavec_1', \ldots, \etavec_G')'$, $\gammavec^Z \equiv (\gamma_1^Z, \ldots, \gamma_G^Z)'$, $\rhovec^Z \equiv (\rho^Z_1, \ldots, \rho^Z_G)'$, and $\ellvec^Z \equiv (\ell_1^Z, \ldots, \ell_G^Z)'$. We assign independent priors to these parameters as follows. For $\etavec$, we let $\etavec \sim \Gau(\zerovec, \sigma_\eta^2 \Ivec)$, where $\sigma_\eta^2 = 100$ and $\Ivec$ is the identity matrix (recall that $\Avec_g$ is column-standardised). For the overall variance and the relative-contribution proportions, we assign the priors
\begin{align*}
  \gamma_g^Z \sim \mathrm{Gamma}(\nu_\gamma, \omega_\gamma)
  \text{ and }
  \rho_g^Z \sim \mathrm{Unif}(0, 1);
  \quad g = 1, \ldots, G,
\end{align*}
respectively, where $\mathrm{Unif}(a, b)$ is the uniform distribution on the interval $[a, b]$. For the length scales, we assign $\ell_g^Z \sim \mathrm{Gamma}(\nu_\ell, \omega_{\ell, g})$, $g = 1, \ldots, G$. All hyperparameters are fixed to reasonable values, as detailed in Appendix~\ref{sec:co2_parameter_model} for the application to CO$_2$.

\section{\texorpdfstring{CO$_2$ application details}{CO2 application details}}
\label{sec:co2_application_details}

This section gives extra details of how WOMBAT~v2.0 was configured for the application to CO$_2$ fluxes.

\subsection{Additional prior information used to improve flux estimation}
\label{sec:additional_prior_information}

As mentioned in Section~\ref{sec:co2_basis_functions}, the 11 predominantly ocean regions and the New Zealand region have small land areas. For this reason, observations are unable to reliably separate their influence from those of other regions through the spatio-temporal distribution of the mole-fraction field (i.e., the trend and seasonality in these small land areas is weakly identifiable). In preliminary investigations, when the trend and seasonality of land fluxes in these regions were left to vary, the inferred fluxes in these 12 regions were found to be unrealistic. We address this by fixing the trend and seasonality in these regions to be known, and equal to that of the bottom-up estimate; data-driven inferences on land fluxes in these regions are therefore done through the residual component.

For a similar reason, we also fix the trend and seasonality to the bottom-up estimate for the ocean component in all 23 regions. When they were allowed to vary in preliminary investigations, the estimated annual cycles of ocean fluxes tend to be implausible, with fluxes attributed to ocean regions that clearly should have been attributed to land regions. Therefore, it appears that the observations, the majority of which are made over land, are unable to reliably constrain the ocean trend and seasonality. However, the trend and seasonality of the ocean fluxes is considered to be captured fairly accurately by bottom-up methods \citep[][]{landschutzeretal2016}, so the practical impact of assuming they are known should be small.

We also found it necessary to introduce extra information to reliably constrain the ocean residual fluxes. These are also weakly identifiable because they are comparably small in magnitude when compared to land fluxes. We do this by fixing the precision of the coefficients associated with the residual component of the ocean fluxes to a very uninformative value, $\tau_\mathrm{ocean}^\epsilon = 1 / 100$, and then incorporating a multivariate pseudo observation as described below. The pseudo observation implicitly establishes the covariance matrix of the basis-function coefficients for the ocean residual fluxes in a way that constrains the scale of the residual fluxes and induces spatio-temporal correlations (i.e., smoothness), while leaving the process-model mode equal to the bottom-up estimate. Constraints on ocean fluxes are often seen as necessary in flux inversion \citep[e.g.,][]{crowelletal2019}.

The ocean pseudo observation expresses the prior belief that deviations from the ocean flux prior mean will be spatio-temporally correlated. To construct it, we aggregate the ocean flux field, $X_\mathrm{ocean}(\cdot, \cdot)$, in the same manner as for the terrestrial biosphere fields in Section~\ref{sec:constraints}. This gives $\xvec_\mathrm{ocean}$, the aggregated fluxes, and $\xvec_\mathrm{ocean}^0$, the bottom-up estimate (and prior mean) of $\xvec_\mathrm{ocean}$. As in \eqref{eqn:aggregated_flux_basis}, we can write
\begin{equation}
  \xvec_\mathrm{ocean} = \xvec_\mathrm{ocean}^0 + \Phivec_{x,\mathrm{ocean}} \alphavec_\mathrm{ocean}.
  \label{eqn:aggregated_ocean}
\end{equation}
Recall from Section~\ref{sec:co2_basis_functions} that we assume known trend and seasonality for the ocean flux field. This amounts to assuming that the elements of $\alphavec_\mathrm{ocean}$ that correspond to the trend and seasonality are equal to zero. Thus, \eqref{eqn:aggregated_ocean} simplifies to $\xvec_\mathrm{ocean} = \xvec_\mathrm{ocean}^0 + \Phivec_{x,\mathrm{ocean}}^\epsilon \alphavec_\mathrm{ocean}^\epsilon$ where $\Phivec_{x,\mathrm{ocean}}^\epsilon$ and $\alphavec_\mathrm{ocean}^\epsilon$ are equal to $\Phivec_{x,\mathrm{ocean}}$ and $\alphavec_\mathrm{ocean}$, respectively, but with the columns/elements for the trend and seasonality basis-function coefficients omitted.

We next introduce the pseudo observation as $\tilde{\xvec}_\mathrm{ocean} = \xvec_\mathrm{ocean}^0$, and take its distribution to be $(\tilde{\xvec}_\mathrm{ocean} \mid \alphavec_\mathrm{ocean}^\epsilon) \sim \Gau(\xvec_\mathrm{ocean}, \Sigmavec_{\tilde{x},\mathrm{ocean}})$. The covariance matrix $\Sigmavec_{\tilde{x},\mathrm{ocean}}$ is constructed using a stationary and separable space-time exponential covariance function with a marginal standard deviation of $1.47 \times 10^{-8}$ $\mathrm{kgCO}_2/\mathrm{m}^2/\mathrm{s}$, a spatial $e$-folding length of 1000 km, and a temporal $e$-folding length of one month \citep[these choices match those of the system ``CAMS'' in][]{crowelletal2019}. The pseudo observation is equal to its conditional mean given $\alphavec_\mathrm{ocean}^\epsilon$, so (by design) it does not inform the mean of $\alphavec_\mathrm{ocean}^\epsilon$. However, its covariance matrix, $\Sigmavec_{\tilde{x},\mathrm{ocean}}$, informs the covariance of $\alphavec_\mathrm{ocean}^\epsilon$. This can be seen through the conditional distribution of $\alphavec_\mathrm{ocean}^\epsilon$ given $\tilde{\xvec}_\mathrm{ocean}$, which is
\[
  (\alphavec_\mathrm{ocean}^\epsilon \mid \tilde{\xvec}_\mathrm{ocean})
  \sim \Gau(
    \zerovec,
    \tilde{\Sigmavec}_{\alpha,\mathrm{ocean}}^\epsilon
  ),
\]
where
\begin{equation}
  \tilde{\Sigmavec}_{\alpha,\mathrm{ocean}}^\epsilon
  \equiv \left[
      (\Phivec_{x,\mathrm{ocean}}^\epsilon)'
      \Sigmavec_{\tilde{x}, \mathrm{ocean}}^{-1}
      \Phivec_{x,\mathrm{ocean}}^\epsilon
      + (1 / 100) \Ivec
    \right]^{-1}.
  \label{eqn:ocean_conditional_prior}
\end{equation}
Note that the term $(1 / 100)\Ivec$ corresponds to the precision matrix for $\alphavec_\mathrm{ocean}^\epsilon$ under the assumptions given in Section~\ref{sec:co2_basis_functions}. Equation~\eqref{eqn:ocean_conditional_prior} shows that the covariance matrix $\Sigmavec_{\tilde{x},\mathrm{ocean}}$ encodes information about the covariance of $\alphavec_\mathrm{ocean}^\epsilon$, and therefore informs the length scales and spatio-temporal magnitudes of the inferred fluxes.

\subsection{Details of mole-fraction observations}
\label{sec:data_details}

The CO$_2$ mole-fraction observations used for our inversion include retrievals of column-averaged CO$_2$ mole-fraction by NASA's OCO-2 satellite \citep{elderingetal2017}, and point-referenced \emph{in situ} and flask measurements of CO$_2$ mole-fraction from a variety of sources described below. Both of these data sets are prescribed in the OCO-2 v10 MIP protocol.

The OCO-2 satellite measures radiances in three near-infrared spectral bands that correspond to absorbance spectra for CO$_2$ and O$_2$. These radiance measurements are used to retrieve the column-average CO$_2$ mole-fraction using an optimal estimation algorithm, and the resulting retrievals are then bias-corrected using external validation data \citep{odelletal2018}. It takes 16 days for the orbit of the satellite to give global coverage of its retrievals, and adjacent orbital tracks are around 150 kilometres apart. Retrievals are made over both land and ocean, but the ocean retrievals are considered by many inversion modellers to have undesirable biases \citep{peiroetal2022}, so in this study we only use land retrievals; due to atmospheric transport, ocean fluxes can still be inferred from these data. The OCO-2 v10 MIP protocol mandates the use of post-processed column-average CO$_2$ retrievals that are constructed by averaging the retrievals in 10-second bands. The post-processing steps are described by \citet{peiroetal2022} for the previous round of the OCO-2 MIP that used version 9 OCO-2 data (the OCO-2 v9 MIP). When using retrieved column-average CO$_2$ mole fractions in an inversion, the influence of the retrieval process has to be accounted for in the observation operator of the $i$th observation, $\mathcal{A}_i$, in \eqref{eqn:mole_fraction_data_model}; more details are given by \citet{zammitmangionetal2022}, Appendix~B. The 530,201 retrievals used in this study range were obtained from September 6, 2014 to March 31, 2021.

The \emph{in situ} and flask measurements used for our inversion are also part of the OCO-2 v10 MIP protocol and come from three collections in ObsPack format \citep{masarieetal2014}. These are the \emph{\nospellcheck{obspack\_co2\_1\_NRT\_v6.1.1\_2021-05-17}} \citep{schuldtetal2021b} collection, the \emph{\nospellcheck{obspack\_co2\_1\_GLOBALVIEWplus\_v6.1\_2021-03-01}} collection \citep{schuldtetal2021a}, and the \emph{\nospellcheck{obspack\_co2\_1\_NIES\_Shipboard\_v3.0\_2020-11-10}} collection \citep{tohjimaetal2005,naraetal2017}. The data include measurements from surface sites, from towers that have multiple sampling inlets at different altitudes, and from research and commercial ships and aircraft. There are 1,054,928 \emph{in situ} and flask measurements in total, comprising 15,577 aircraft measurements, 218,613 shipboard measurements, 408,083 surface measurements, and 412,655 tower measurements. The chosen measurements are a subset of those available that are deemed suitable for use in flux-inversion systems and, in most cases, are temporally averaged into 3-hour intervals. The details of how these choices were made for the OCO-2 v10 MIP are essentially the same as for the OCO-2 v9 MIP, which is described by \citet{peiroetal2022}.

\subsection{Error budgets, measurement groups, and bias correction}
\label{sec:error_budgets}

The prescribed OCO-2 and \emph{in situ}/flask observations for the OCO-2 v10 MIP come with suggested error variances, which we term error budgets, and which are calculated following the approach described in \citet{peiroetal2022} for the OCO-2 v9 MIP. We set our error budgets/variances $V_i$, for $i = 1, \ldots, N$, equal to the v10 MIP suggestions. Recall from Appendix~\ref{sec:mole_fraction_data_model} that observations in WOMBAT are split into groups indexed by $g = 1, \ldots, G$. Observations within a group $g$ share the parameters $\etavec_g$, $\gamma_g^Z$, $\rho_g^Z$, and $\ell_g^Z$, and they are therefore assumed to have identical biases and error properties. We divide the CO$_2$ mole-fraction observations into $G = 5$ groups: (1) OCO-2 land observations, (2) aircraft observations, (3) shipboard observations, (4) surface observations, and (5) tower observations. The OCO-2 observations are treated separately because they come from an instrument that is very different from the others; the aircraft and shipboard observations because they are from instruments that are moving and at different speeds; and the surface and tower observations because they are from instruments that are stationary and that give readings at different heights.

The OCO-2 version 10 column-averaged observations are already bias-corrected using an offline algorithm (see Appendix~\ref{sec:data_details}), and \citet{zammitmangionetal2022} found that estimating additional biases for an earlier version of these observations (version 7) made little difference when estimating fluxes. As described in Appendix~\ref{sec:mole_fraction_data_model}, the point-referenced \emph{in situ} and flask observations are already assumed to be unbiased. Consequently, although our model is capable of working with the raw non-bias corrected OCO-2 data and estimating bias, for this application we let the bias term $b_i = 0$ for all observations $i = 1, \ldots, N$ (see \eqref{eqn:mole_fraction_data_model}), and omit the bias-correction terms $\{ \etavec_g : g = 1, \ldots, G \}$.

The parameters $\{ \rho_g^Z : g = 1, \ldots, G \}$ apportion the error budgets between the correlated error term and the uncorrelated error term. For groups $g = 2, \ldots, 5$, which are the \emph{in situ} and flask observations, we assume that $\rho_g^Z = 1$. This sets the variance of the uncorrelated component of the error equal to zero, which is appropriate because these observations have negligible measurement error but are still affected by misrepresentation of transport (a correlated source of error). For group $g = 1$, which contains the OCO-2 observations, $\rho_g^Z$ is left unknown and non-negative.

To specify the covariance of the vector of correlated error term, $\xivec$, we split the data from each group into separate time series that each come from a single location or instrument. We let the OCO-2 land data be one time series, the observations from a single ship or aircraft be one time series, and the observations from a single surface site or tower inlet be one time series. This yields 192 time series. Let group $g$ contain $S_g$ time series, and let $\xivec_{g, s}$, $s = 1, \ldots, S_g$, be the vector containing the elements of $\xivec$ belonging to time series $s$ in group $g$. Following \citet{zammitmangionetal2022}, we let the correlation between elements of $\xivec_{g, s}$ be governed by an exponential function of temporal separation, with $e$-folding time scale $\ell_g^Z$. All other correlations between elements of $\xivec$ are set equal to zero.

\subsection{Transport and initial condition}
\label{sec:transport}

We implement the approximate transport operations in $\hat{\mathcal{H}}$ using the GEOS-Chem global 3-D chemical transport model, version 12.3.2 \citep{beyetal2001,yantosca2019}, driven by the MERRA-2 meteorological fields \citep{bosilovichetal2015}. We use the offline GEOS-Chem CO$_2$ simulation \citep{nassaretal2010} with a transport time step of 10 minutes and a flux time step of 20 minutes. For computational efficiency, we aggregate the spatial field for the simulations to $2\degree \times 2.5\degree$ (latitude $\times$ longitude) and 47 vertical levels from the native horizontal resolution of $0.25\degree \times 0.3125\degree$ and 72 vertical levels, respectively. All fluxes are supplied to GEOS-Chem using the \nospellcheck{HEMCO} emissions component \citep{kelleretal2014}. The approximate initial condition, $\hat{Y}(\cdot, \cdot, t_0)$, which gives the value of the mole-fraction field at September 1, 2014, is the same as that in \citet{zammitmangionetal2022}. GEOS-Chem can allow for a 3-D chemical source of CO$_2$ due to oxidation of other trace gases, but this source was omitted for compatibility with the OCO-2 MIP.

\subsection{Hyperparameters}
\label{sec:co2_parameter_model}

Priors on the process-model parameters for the basis-function coefficients (see Section~\ref{sec:alpha_prior}) follow \citet{zammitmangionetal2022}, where the choices $\nu_\tau^\beta = \nu_\tau^\epsilon = 0.354$, $\omega_{\tau, c}^\beta = \omega_{\tau, c}^\epsilon = 0.0153$, $\nu_\gamma = 1.62702$, and $\omega_\gamma = 2.171239$ are explained. We assume that $\kappa_\mathrm{gpp}^\epsilon = \kappa_\mathrm{resp}^\epsilon \equiv \kappa_\mathrm{bio}^\epsilon$, so that temporal dependencies in the residual of the GPP and respiration components are captured by the single parameter $\kappa_\mathrm{bio}^\epsilon$. We let $a_\rho^\beta = b_\rho^\beta = a_\rho^\epsilon = b_\rho^\epsilon = a_\kappa^\epsilon = b_\kappa^\epsilon = 1$. For the prior distribution on the length scales in the data model (see Appendix~\ref{sec:mole_fraction_data_model}), we let $\nu_\ell = 1$, $\omega_{\ell, 1} = 1\ \text{min}^{-1}$ (for the OCO-2 data), and $\omega_{\ell, g} = 1\ \text{day}^{-1}$ for $g = 2, 3, 4$, and 5.

\subsection{Computation}
\label{sec:computation}

Computation of the basis functions took several weeks on the \nospellcheck{Gadi} supercomputer at the Australian National Computational Infrastructure. Once these were finished, the MCMC sampler that performs the inversion was run for 5,000 iterations, 1,000 of which were discarded as warm up. This took two days on a high-performance computing machine with 56 computing cores. Figure~\ref{fig:traceplots} in Appendix~\ref{sec:additional_figures} shows traceplots for the parameters of the process models, as well as those for four basis-function coefficients. The effective sample size \citep{gelmanetal2013} is at least 1,000 for all parameters, and is greater than 3,500 for 95\% of the parameters.

\section{Markov chain Monte Carlo algorithm}
\label{sec:mcmc}

As discussed in Section~\ref{sec:inference}, the first stage of inference finds estimates $\hat{\rhovec}^Z$ and $\hat{\ellvec}^Z$ of the parameters $\rhovec^Z$ and $\ellvec^Z$, respectively, and the second stage uses an MCMC algorithm to sample from the full posterior distribution of the remaining unknown parameters. This appendix gives the details of the MCMC algorithm.

Define $\rhovec^\beta \equiv (\rho_{c_1, c_2}^\beta, \rho_{c_1, c_3}^\beta, \ldots, \rho_{c_{C - 1}, c_C}^\beta)'$, $\tauvec^\beta \equiv (\tau_{c_1}^\beta, \ldots, \tau_{c_C}^\beta)'$, $\kappavec^\epsilon \equiv (\kappa_{c_1}^\epsilon, \ldots, \kappa_{c_C}^\epsilon)'$, $\rhovec^\epsilon \equiv (\rho_{c_1, c_2}^\epsilon, \rho_{c_1, c_3}^\epsilon, \ldots, \rho_{c_{C - 1}, c_C}^\epsilon)'$, and $\tauvec^\epsilon \equiv (\tau_{c_1}^\epsilon, \ldots, \tau_{c_C}^\epsilon)'$. Let $\Zvec_g$, $g = 1, \ldots, G$, be a vector containing the elements of $\Zvec$ that correspond to group $g$. We split $\alphavec$ into $\alphavec^\beta$ and $\alphavec^\epsilon$, vectors which contain the elements of $\alphavec$ corresponding to the trend/seasonality basis functions and to the residual basis functions, respectively. With this notation, the posterior distribution of the unknown parameters is $p(\alphavec, \etavec, \rhovec^\beta, \tauvec^\beta, \kappavec^\epsilon, \rhovec^\epsilon, \tauvec^\epsilon, \gammavec^Z \mid \hat{\rhovec}^Z, \hat{\ellvec}^Z, \Zvec)$. We use a Gibbs sampling scheme to sample from this distribution, which iteratively samples from the full conditional distributions of the unknown parameters in the following order:
\begin{multicols}{2}
  \begin{enumerate}
    \item
      $p(\rho_{c, c'}^\beta \mid \alphavec^\beta, \tauvec^\beta)$ for $c, c' \in \mathcal{C}$;

    \item
      $p(\tau_c^\beta \mid \alphavec^\beta, \rhovec^\beta)$ for $c \in \mathcal{C}$;

    \item
      $p(\kappa_c^\epsilon \mid \alphavec^\epsilon, \tauvec^\epsilon, \rhovec^\epsilon)$ for $c \in \mathcal{C}$;

    \item
      $p(\rho_{c, c'}^\epsilon \mid \alphavec^\epsilon, \tauvec^\epsilon, \kappavec^\epsilon)$ for $c, c' \in \mathcal{C}$;

    \item
      $p(\tau_c^\epsilon \mid \alphavec^\epsilon, \kappavec^\epsilon, \rhovec^\epsilon)$ for $c \in \mathcal{C}$;

    \item
      $p(\gamma_g^Z \mid \alphavec, \hat{\rho}_g^Z, \hat{\ell}_g^Z, \Zvec_g)$ for $g = 1, \ldots, G$;

    \item
      $p(\alphavec, \etavec \mid \Sigmavec_\alpha, \gammavec^Z, \hat{\rhovec}^Z, \hat{\ellvec}^Z, \Zvec)$.
  \end{enumerate}
\end{multicols}
\noindent In the above, variables that do not alter the full conditional distributions are omitted from the conditioning, and $\Sigmavec_\alpha$ in step 7 depends on $\rhovec^\beta, \tauvec^\beta, \kappavec^\epsilon, \rhovec^\epsilon$ and $\tauvec^\epsilon$. The details of each step follow.

\textbf{Steps 1 and 2}: The conditional distributions in steps 1 and 2 are proportional to
\begin{equation}
  \begin{split}
    p(\alphavec^\beta, \rhovec^\beta, \tauvec^\beta)
    & \propto
      p(\alphavec^\beta \mid \rhovec^\beta, \tauvec^\beta) p(\rhovec^\beta) p(\tauvec^\beta) \\
    & =
      |\Sigmavec_\alpha^\beta|^{-1 / 2} (\alphavec^\beta)' (\Sigmavec_\alpha^\beta)^{-1} \alphavec^\beta
      \prod_{c, c' \in \mathcal{C}, c \neq c'} p(\rho_{c, c'}^\beta)
      \prod_{c \in \mathcal{C}} p(\tau_c^\beta),
  \end{split}
  \label{eqn:clim_joint_distribution}
\end{equation}
where $\Sigmavec_\alpha^\beta$ is the submatrix of $\Sigmavec_\alpha$ corresponding to the entries of $\alphavec^\beta$ ($\Sigmavec_\alpha^\beta$ depends on $\rhovec^\beta$ and $\tauvec^\beta$), and the prior densities for $\rho_{c, c'}^\beta$ and $\tau_c^\beta$ are described in Appendix~\ref{sec:alpha_covariance_matrix}. We perform steps 1 and 2 by using a slice sampler \citep{neal2003} for each unknown parameter in turn, with the target density calculated using \eqref{eqn:clim_joint_distribution}.

\textbf{Steps 3, 4, and 5}: The conditional distributions in steps 3, 4, and 5 are proportional to
\begin{equation}
  \begin{split}
    & p(\alphavec^\epsilon, \kappavec^\epsilon, \rhovec^\epsilon, \tauvec^\epsilon) \\
    & \quad \propto
      p(\alphavec^\epsilon \mid \kappavec^\epsilon, \rhovec^\epsilon, \tauvec^\epsilon) p(\kappavec^\epsilon) p(\rhovec^\epsilon) p(\tauvec^\epsilon) \\
    & \quad =
      |\Sigmavec_\alpha^\epsilon|^{-1 / 2} (\alphavec^\epsilon)' (\Sigmavec_\alpha^\epsilon)^{-1} \alphavec^\epsilon
      \prod_{c \in \mathcal{C}} p(\kappa_c^\epsilon),
      \prod_{c, c' \in \mathcal{C}, c \neq c'} p(\rho_{c, c'}^\epsilon)
      \prod_{c \in \mathcal{C}} p(\tau_c^\epsilon),
  \end{split}
  \label{eqn:resid_joint_distribution}
\end{equation}
where $\Sigmavec_\alpha^\epsilon$ is the submatrix of $\Sigmavec_\alpha$ corresponding to the entries of $\alphavec^\epsilon$. As in steps 1 and 2, we use slice sampling to perform steps 3, 4, and 5.

\textbf{Step 6}: The full conditional distribution for step 6 is available in closed form, so we sample from it directly. The distribution is
\begin{equation}
  (\gamma_g^Z \mid \alphavec, \hat{\rho}_g^Z, \hat{\ell}_g^Z, \Zvec_g)
  \sim \mathrm{Gamma}(\nu_{\gamma, g}^*, \omega_{\gamma, g}^*),
  \quad g = 1, \ldots, G,
\end{equation}
where $\nu_{\gamma, g}^* \equiv \nu_\gamma + \frac{1}{2} N_g$ and $\omega_{\gamma, g}^* \equiv \omega_g + \frac{1}{2} (\Zvec_g - \Psivec_g \alphavec)' \Dvec_g^{-1} (\Zvec_g - \Psivec_g \alphavec)$. Here, $N_g$ is the number of observations in group $g$; $\Psivec_g$ is a matrix that contains the rows of $\Psivec$ corresponding to observations in group $g$; and
\begin{equation}
  \Dvec_g \equiv \Vvec_g^{1 / 2} \left[
    \hat{\rho}_g^Z \Rvec_{\xi_g} + (1 - \hat{\rho}_g^Z) \Ivec
  \right] \Vvec_g^{1 / 2},
\end{equation}
where $\Vvec_g$ is a diagonal matrix containing the error budgets $V_i$ for group $g$ on its diagonal, $\Rvec_{\xi_g} \equiv \mathrm{corr}(\xivec_g \mid \hat{\rho}_g^Z, \hat{\ell}_g^Z)$, and $\xivec_g$ is a vector containing the elements of $\xivec$ that belong to group $g$.

\textbf{Step 7}: Define $\thetavec \equiv (\alphavec', \etavec')'$; $\Sigmavec_\theta \equiv \mathrm{bdiag}(\Sigmavec_\alpha, \sigma_\eta^2 \Ivec)$, where $\mathrm{bdiag}(\cdot)$ constructs a block-diagonal matrix from its arguments; $\Hvec \equiv \mathrm{bdiag}(\Psivec, \Avec_1, \ldots, \Avec_g)$; $\Sigmavec_Z \equiv \mathrm{var}(\Zvec \mid \gammavec^Z, \hat{\rhovec}^Z, \hat{\ellvec}^Z)$; and $F_\theta = \{ \thetavec : \alphavec \in F_\alpha \}$, the extension of the constraint region $F_\alpha$ to the domain of $\thetavec$. The conditional distribution in step 7 is
\begin{equation}
  (\thetavec \mid \Sigmavec_\alpha, \gammavec^Z, \hat{\rhovec}^Z, \hat{\ellvec}^Z, \Zvec)
    \sim
    \mathrm{ConstrGau}(\muvec_\theta^*, \Sigmavec_\theta^*, F_\theta),
  \label{eqn:theta_conditional}
\end{equation}
where $\Sigmavec_\theta^* = (\Hvec' \Sigmavec_Z^{-1} \Hvec + \Sigmavec_\theta)^{-1}$ and $\muvec_\theta^* = \Sigmavec_\theta^* \Hvec' \Sigmavec_Z^{-1} \Zvec$. In Section~\ref{sec:constraints}, we describe a system of linear constraints $F_\alpha$ that approximate the physical constraints relevant to the CO$_2$ application. Following \citet{stelletal2022}, we use the method described by \citet{pakmanpaninski2014} to sample from \eqref{eqn:theta_conditional}. This method can sample from a multivariate Gaussian distribution under both linear and quadratic constraints. It uses a HMC sampling scheme with exact Hamiltonian dynamics, which accommodates constraints along a trajectory by reflecting the trajectory off the constraint boundary; \citeauthor{pakmanpaninski2014} show how reflections can be performed while preserving the target distribution. The algorithm is computationally efficient for sparse constraint matrices such as the matrix $\Phivec$ in Section~\ref{sec:constraints}, and we find it mixes well for our application.

\clearpage
\section{Additional tables and figures}
\label{sec:additional_figures}

\begin{figure}[ht]
  \begin{center}
    \includegraphics{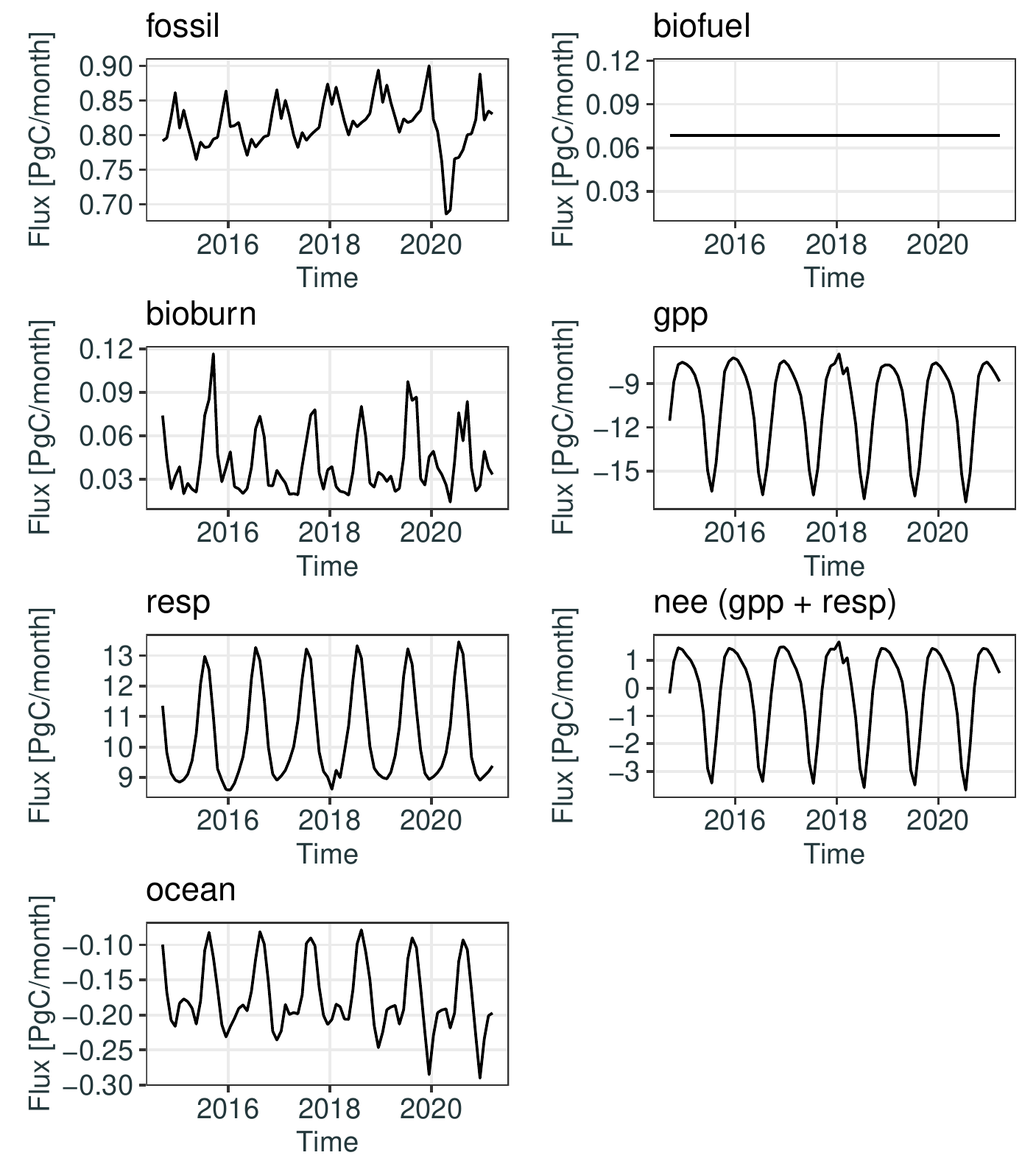}
  \end{center}
  \caption{
    Monthly-average global total fluxes for the bottom-up estimates for all the flux components. The bottom-up estimate of the biofuel fluxes has spatial but not temporal variability, so its global total is a constant.
  }
  \label{fig:global_bottom_up}
\end{figure}

\begin{figure}[t]
  \begin{center}
    \includegraphics{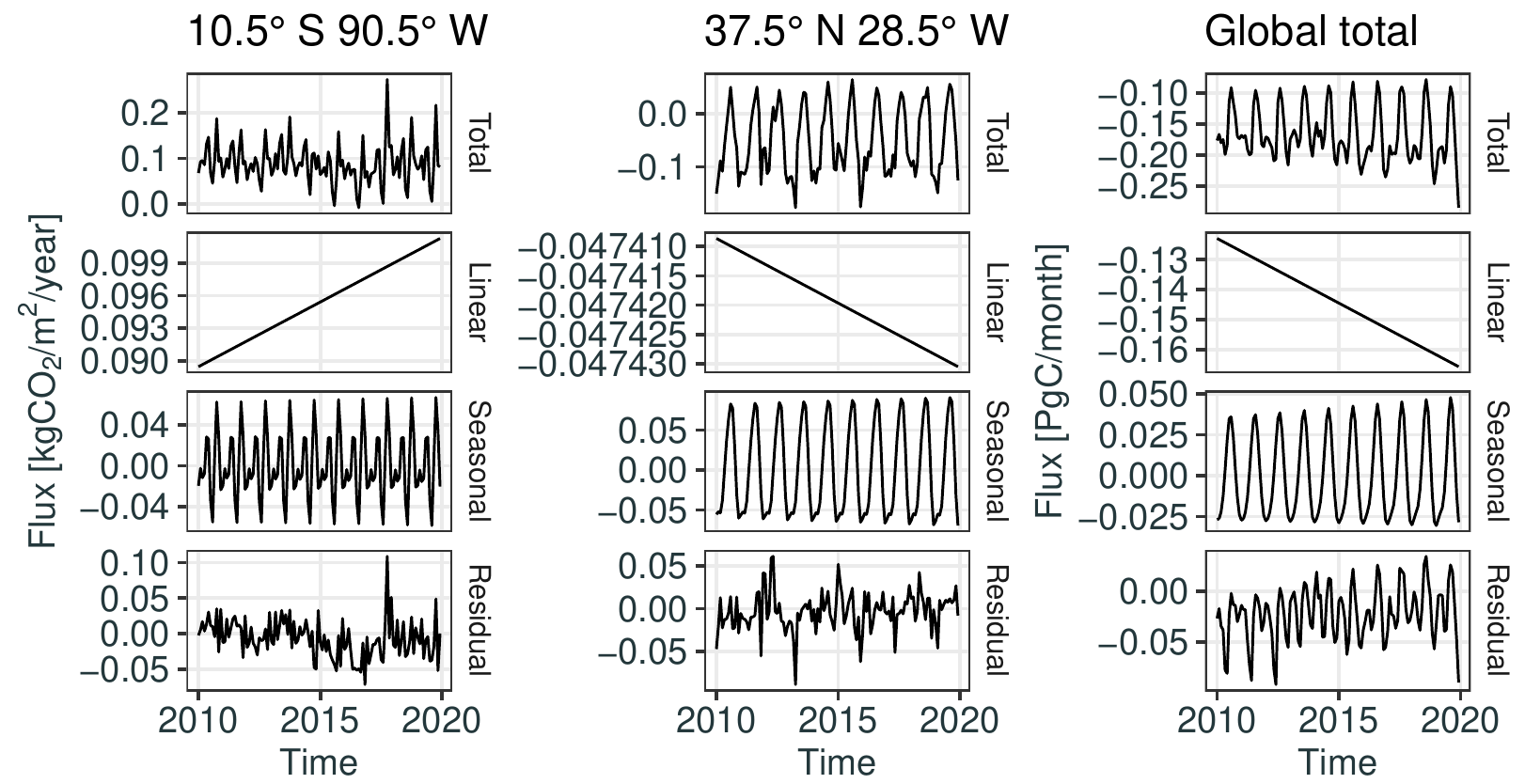}
  \end{center}

  \caption{
    Estimated ocean fluxes from \citet{landschutzeretal2016}, $X_\mathrm{ocean}^0(\cdot, \cdot)$. From left to right, fluxes are shown, respectively, for a grid cell in the eastern tropical Pacific Ocean (coordinates $10.5\degree$ S, $90.5 \degree$ W), a grid cell in the northern Atlantic Ocean ($37.5\degree$ S, $28.5\degree$ W), and the global total ocean fluxes. The first row gives the total fluxes, the second row the linear component of the fluxes, the third row the seasonal component, and the last row gives the residual fluxes, $\epsilon_c^0(\cdot, \cdot)$. Only the last 10 years of fluxes, which span 38 years, are shown. For the grid-cell fluxes, the scale is in $\mathrm{kgCO}_2/\mathrm{m}^2/\mathrm{year}$, while the global total ocean flux is given in $\mathrm{PgC}/\mathrm{month}$.
  }
  \label{fig:landschutzer_fluxes}
\end{figure}

\begin{figure}
  \begin{center}
    \includegraphics{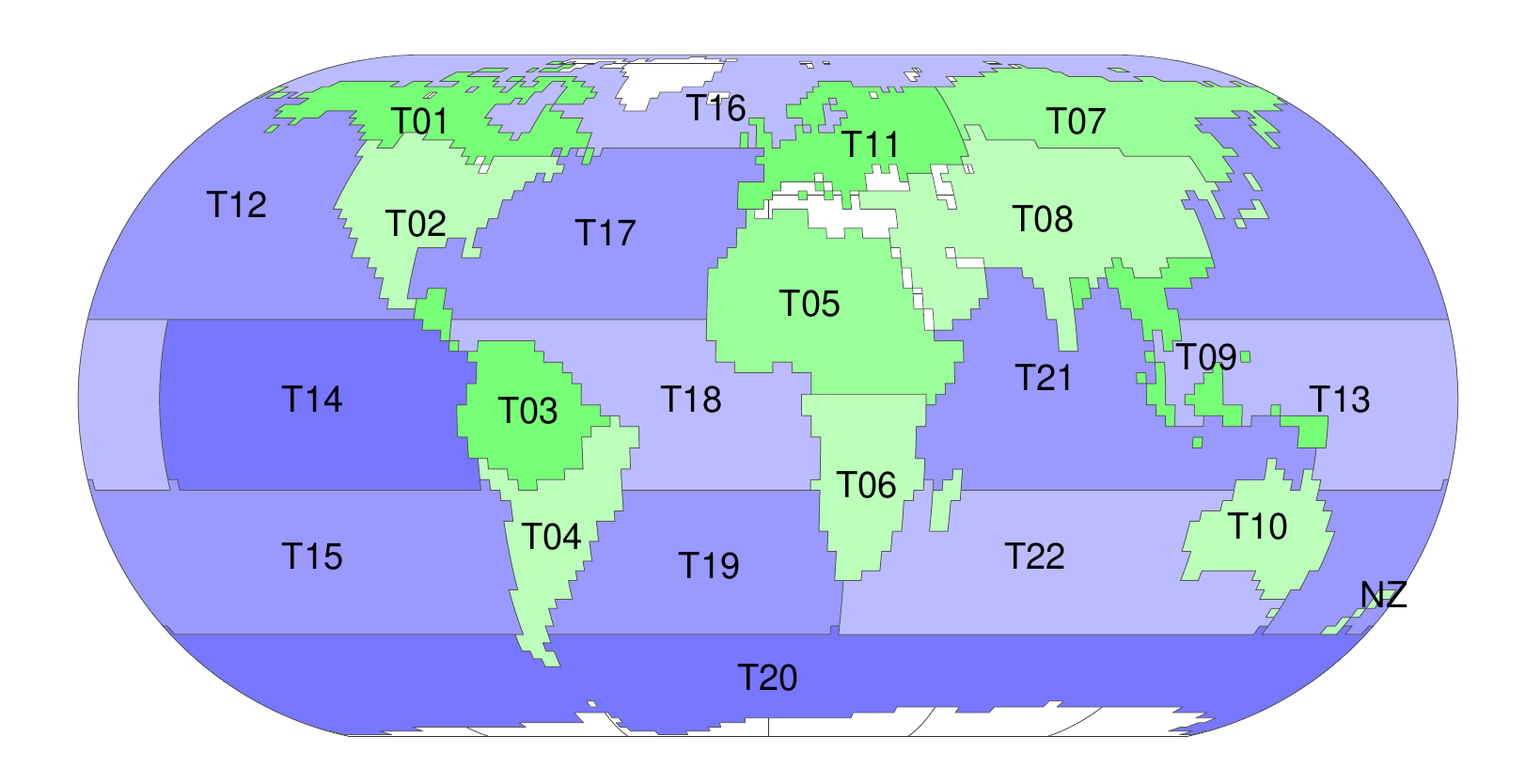}
  \end{center}
  \caption{
    Map of the 22 TransCom3 regions (T01--T22) and the New Zealand region (NZ) used to create basis functions. The full names of these regions are given in Table~\ref{tab:region_table}. The white parts of the map correspond to areas assumed to have zero CO$_2$ surface flux.
  }
  \label{fig:region_map}
\end{figure}

\begin{figure}
  \begin{center}
    \includegraphics{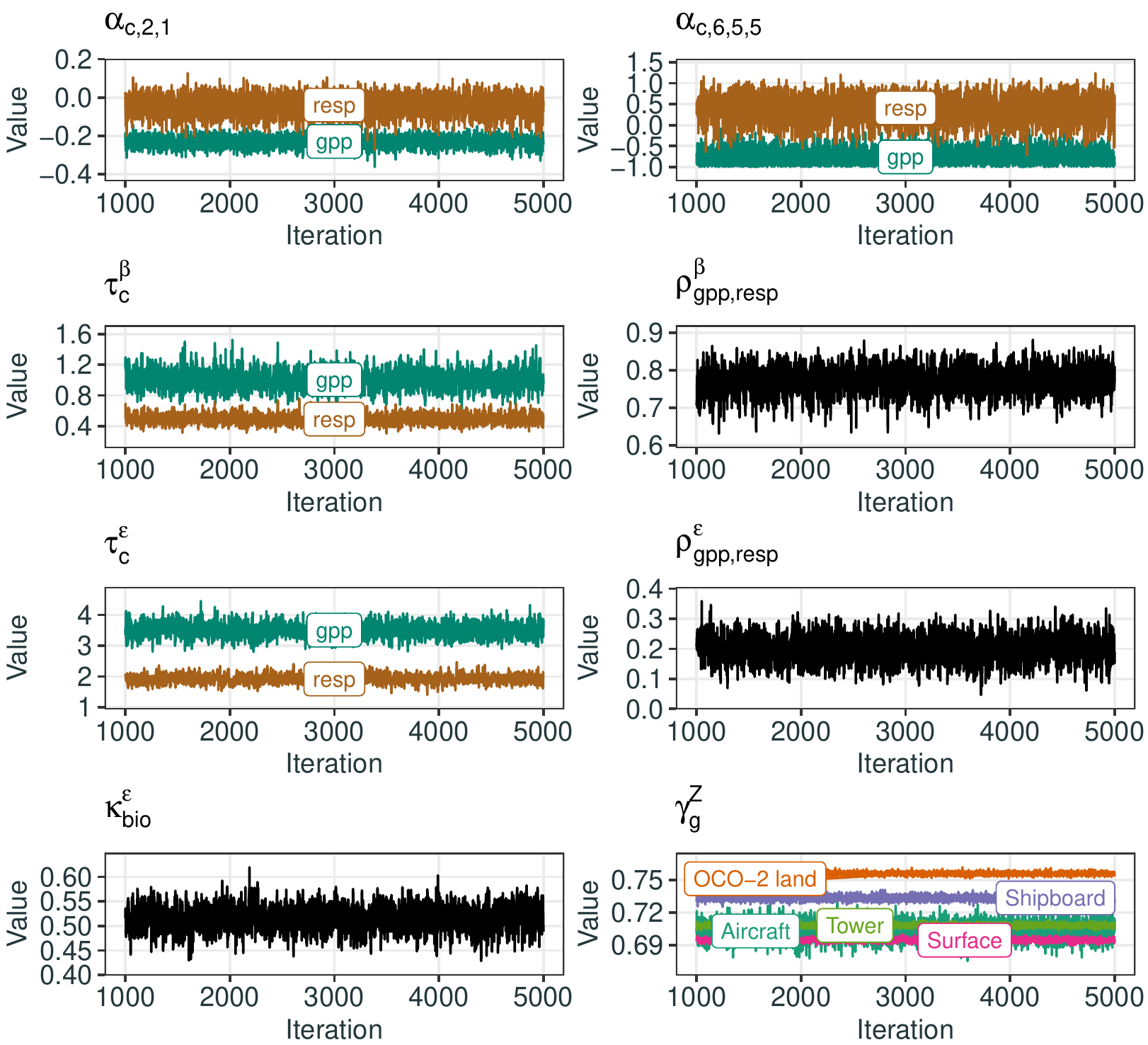}
  \end{center}
  \caption{
    Traceplots for the parameters of the process models, as well as for four basis-function coefficients, following the MCMC warm-up phase of 1000 iterations.
  }
  \label{fig:traceplots}
\end{figure}

\begin{table}
  \begin{center}
  \begin{tabular}{lll||lll}
    \hline
    Code & Name & Type
      & Code & Name & Type \\ \hline
    T01 & North American Boreal & Land
      & T12 & North Pacific Temperate & Ocean \\
    T02 & North American Temperate & Land
      & T13 & West Pacific Tropical & Ocean \\
    T03 & Tropical South America & Land
      & T14 & East Pacific Tropical & Ocean \\
    T04 & South American Temperate & Land
      & T15 & South Pacific Temperate & Ocean \\
    T05 & Northern Africa & Land
      & T16 & Northern Ocean & Ocean \\
    T06 & Southern Africa & Land
      & T17 & North Atlantic Temperate & Ocean \\
    T07 & Eurasia Boreal & Land
      & T18 & Atlantic Tropical & Ocean \\
    T08 & Eurasia Temperate & Land
      & T19 & South Atlantic Temperate & Ocean \\
    T09 & Tropical Asia & Land
      & T20 & Southern Ocean & Ocean \\
    T10 & Australia & Land
      & T21 & Indian Tropical & Ocean \\
    T11 & Europe & Land
      & T22 & South Indian Temperate & Ocean \\
    NZ & New Zealand & Land \\
    \hline
    \end{tabular}
  \end{center}
  \caption{
    The code, name, and type of the 22 TransCom3 regions and the New Zealand region used in our study. A map showing these regions is given in Figure~\ref{fig:region_map}.
  }
  \label{tab:region_table}
\end{table}

\begin{figure}
  \begin{center}
    \includegraphics{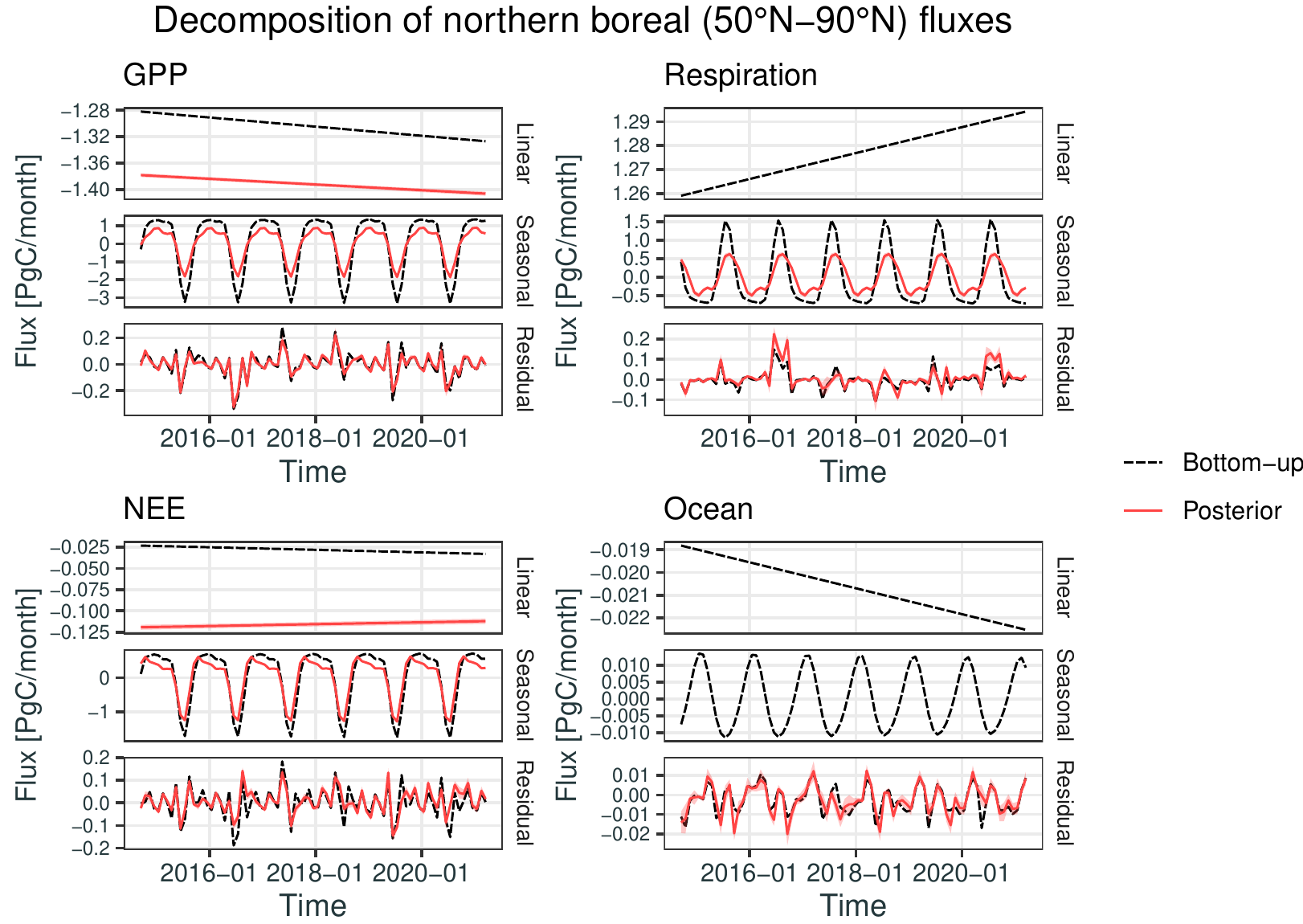} \\[0.4cm]
    \includegraphics{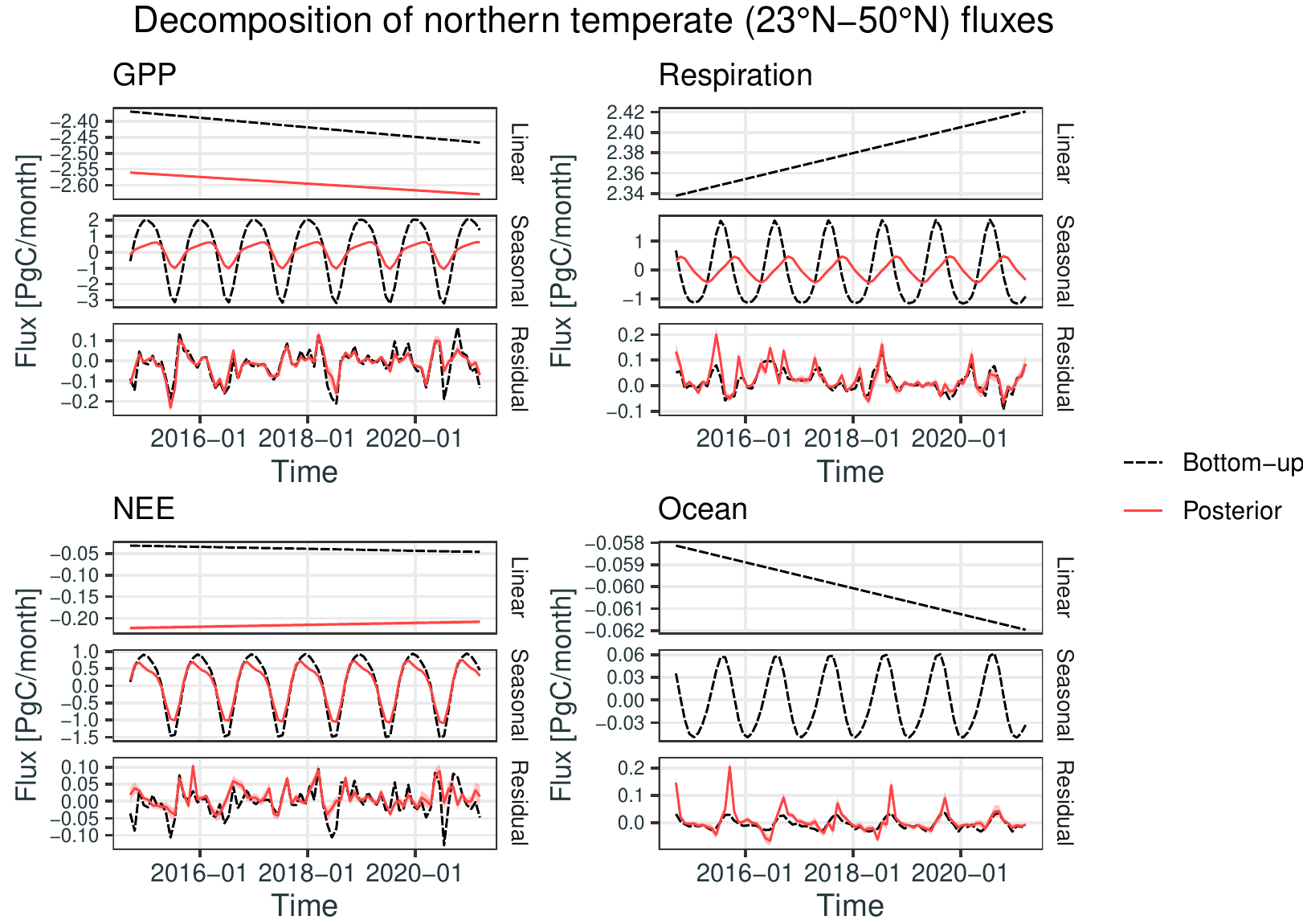}
  \end{center}
  \caption{
    As in Figure~\ref{fig:global_components}, but for the northern boreal latitudes (50$\degree$N--90$\degree$N, top four plots), and for the northern temperate latitudes (23$\degree$N--50$\degree$N, bottom four plots).
  }
  \label{fig:flux_components_n_extratropics}
\end{figure}

\begin{figure}
  \begin{center}
    \includegraphics{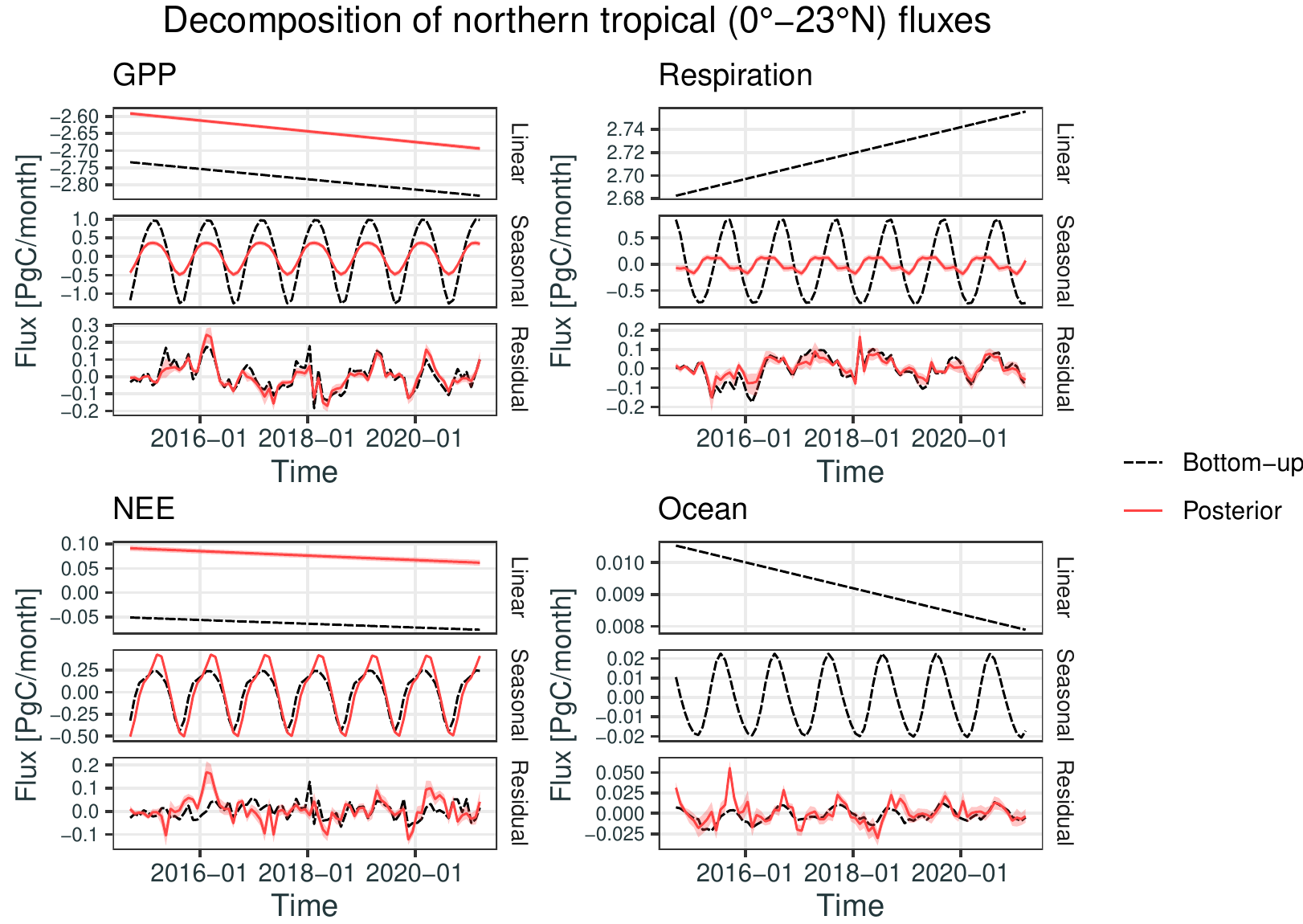} \\[0.4cm]
    \includegraphics{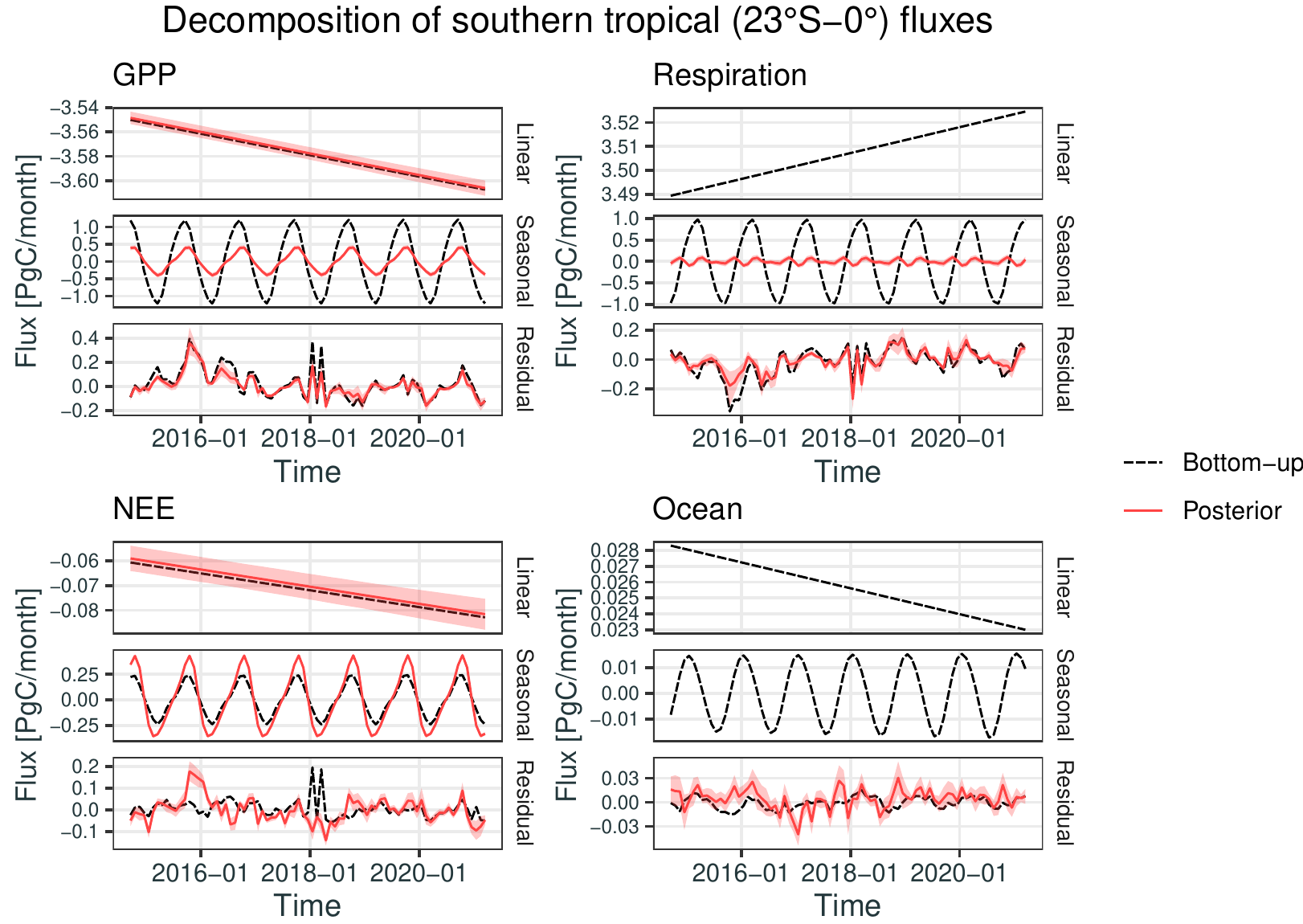}
  \end{center}
  \caption{
    As in Figure~\ref{fig:global_components}, but for the northern tropical latitudes (0$\degree$--23$\degree$N, top four plots), and for the southern tropical latitudes (23$\degree$S--0$\degree$, bottom four plots).
  }
  \label{fig:flux_components_tropical}
\end{figure}

\begin{figure}
  \begin{center}
    \includegraphics{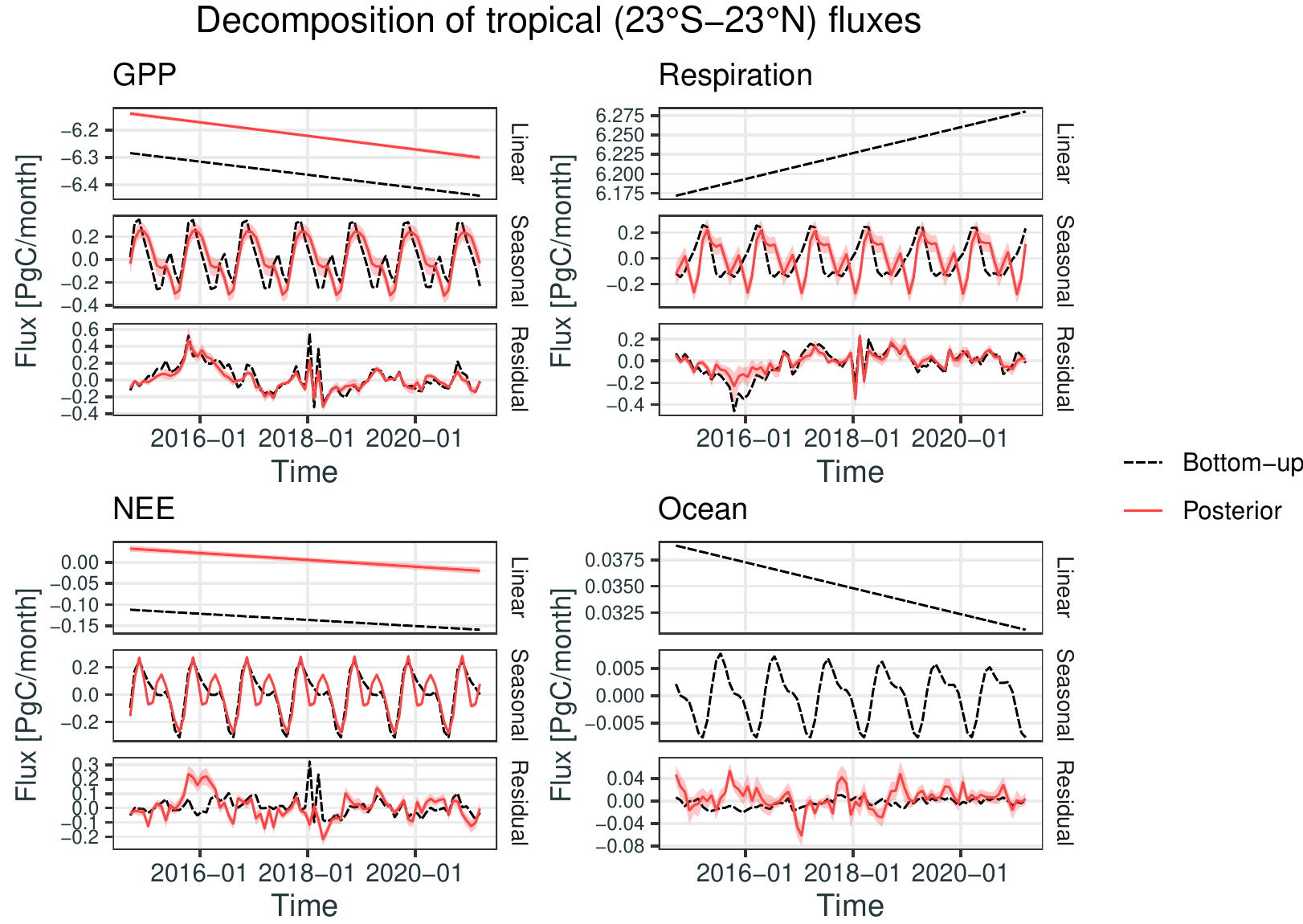} \\[0.4cm]
    \includegraphics{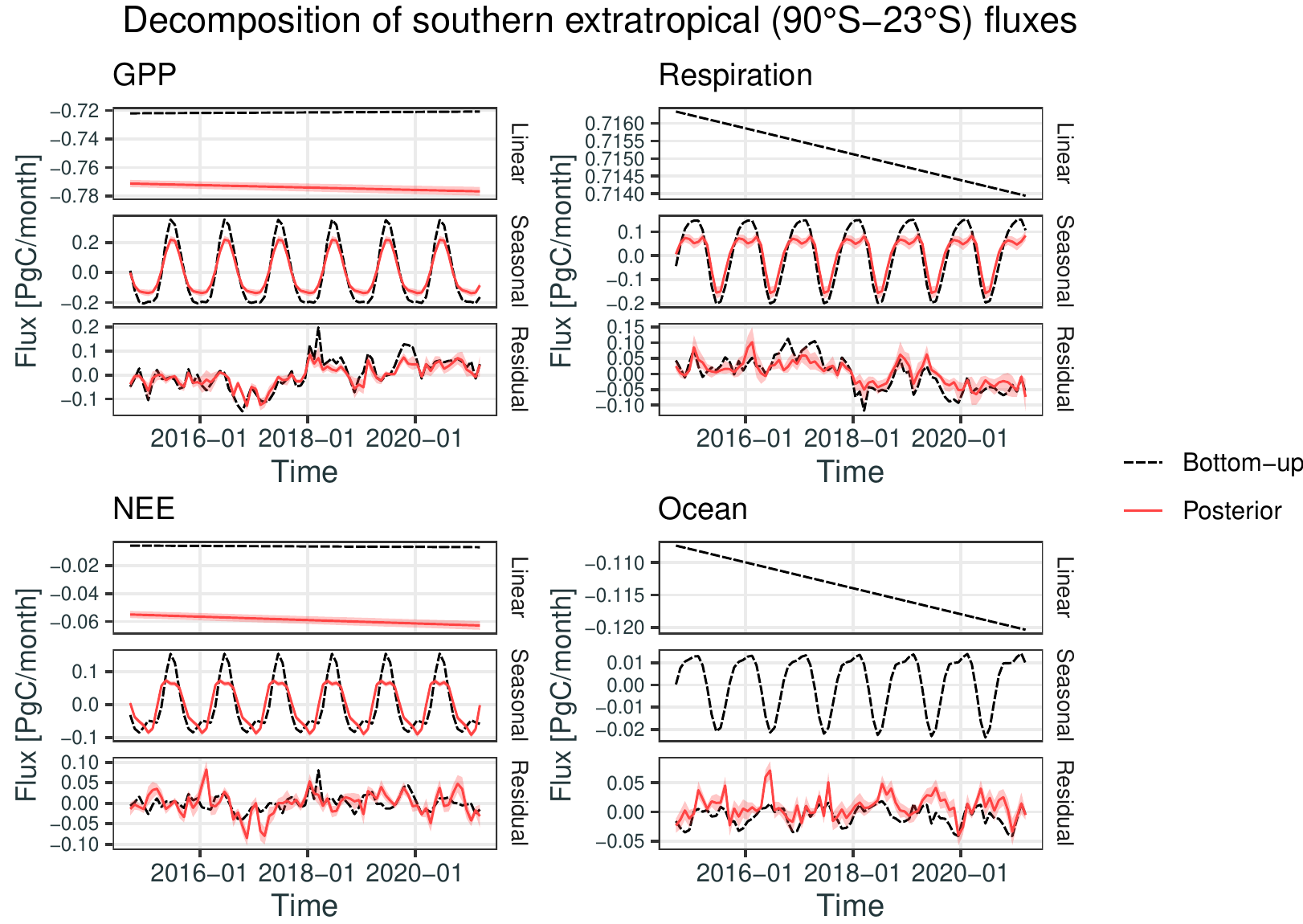}
  \end{center}
  \caption{
    As in Figure~\ref{fig:global_components}, but for the tropics (23$\degree$S--23$\degree$N, top four plots), and for the southern extratropical latitudes (90$\degree$S--23$\degree$S, bottom four plots).
  }
  \label{fig:flux_components_tropical_s_extratropics}
\end{figure}

\begin{figure}
  \begin{center}
    \includegraphics{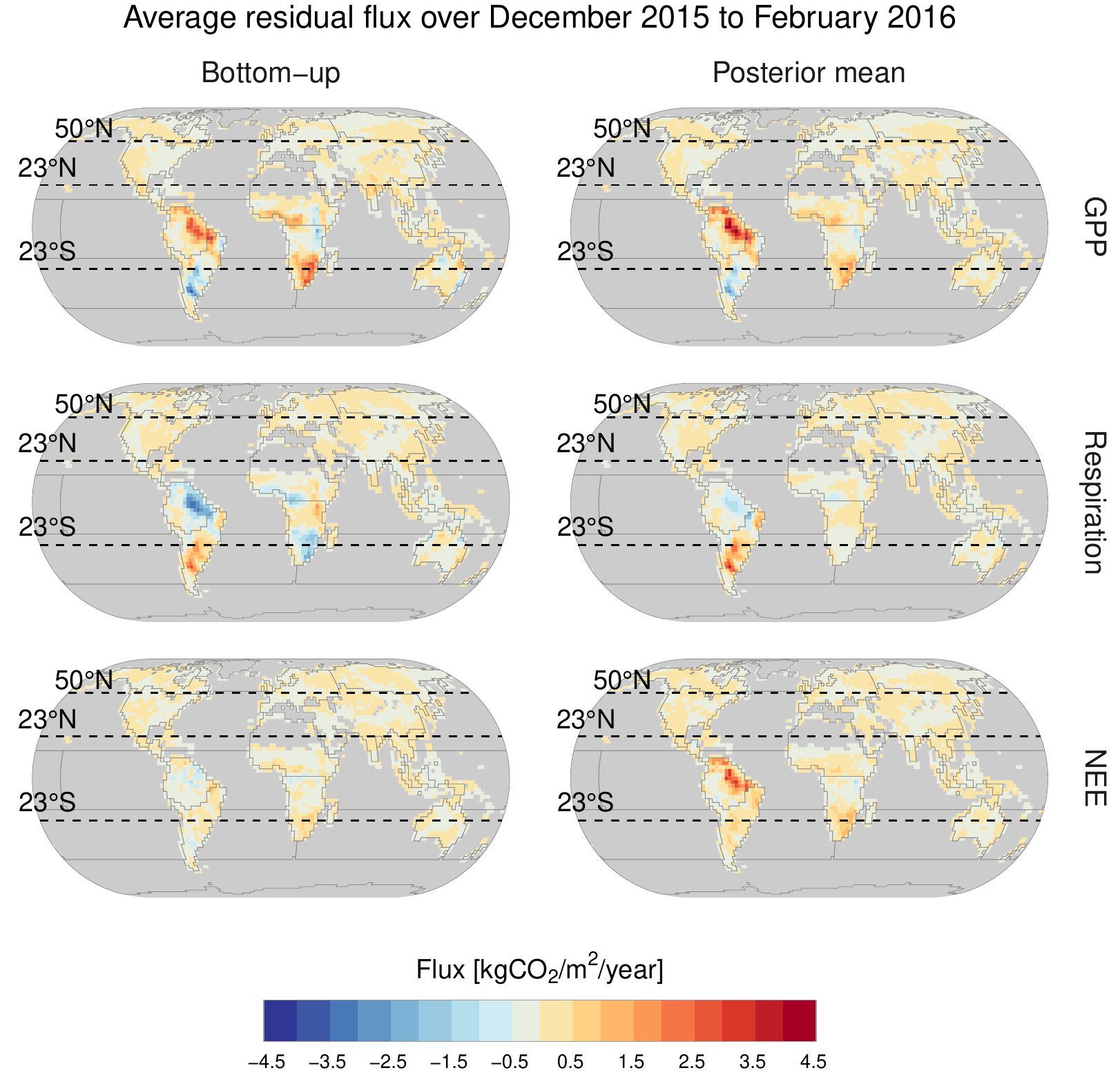}
  \end{center}
  \caption{
    Estimates of the average residual flux in each grid cell over the \nospellcheck{El Ni\~{n}o} period from December 2015 to February 2016. The top row shows the estimates for GPP, the middle row those for respiration, and the bottom row those for NEE. The left column shows the bottom-up estimates of the flux, and the right column shows the posterior mean estimates.
  }
  \label{fig:el_nino_2015_residual}
\end{figure}

\begin{figure}
  \begin{center}
    \includegraphics{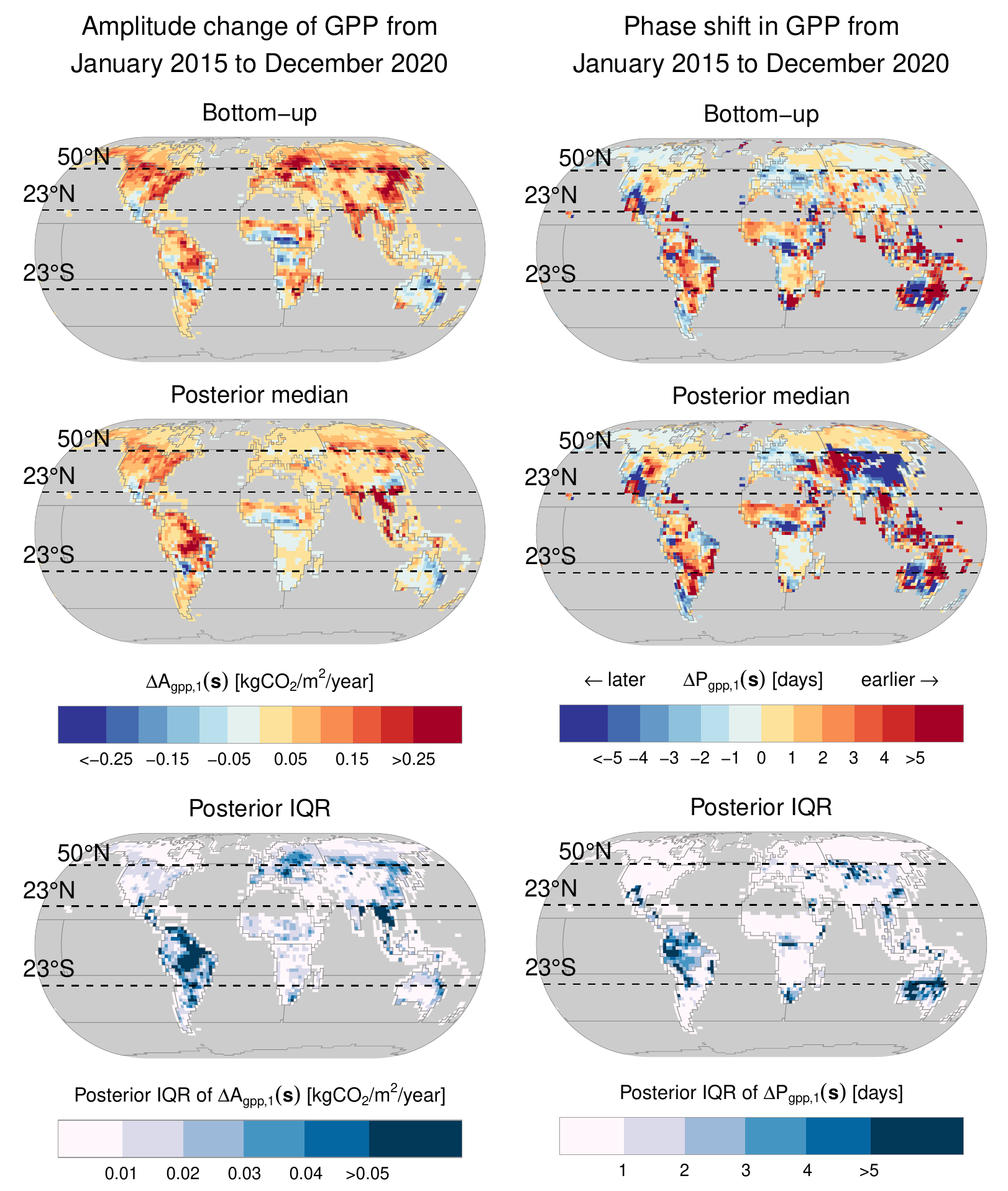}
  \end{center}
  \caption{
    As in Figure~\ref{fig:harmonic_shift_map_nee}, but for the GPP component.
  }
  \label{fig:harmonic_shift_map_gpp}
\end{figure}

\begin{figure}
  \begin{center}
    \includegraphics{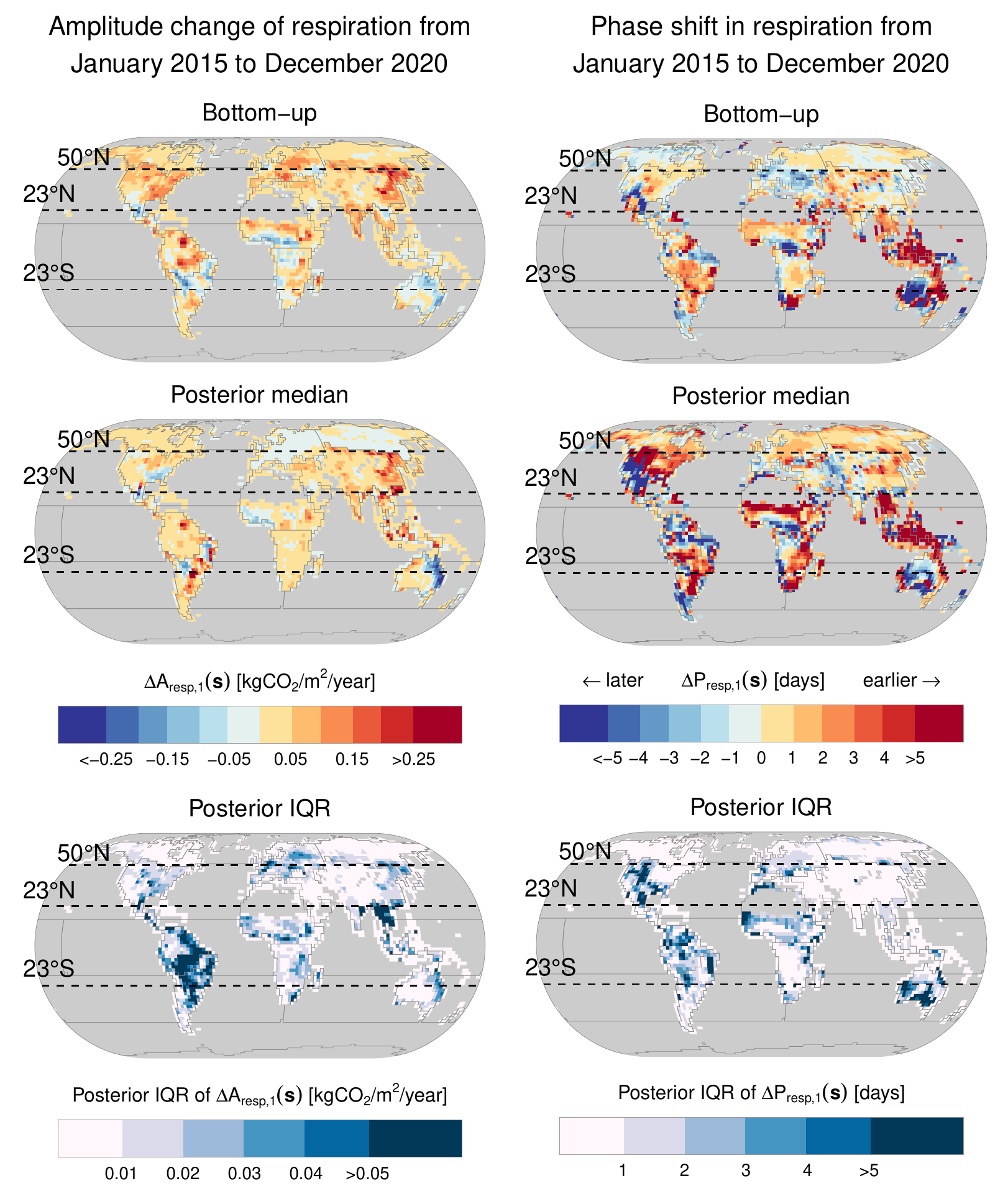}
  \end{center}
  \caption{
    As in Figure~\ref{fig:harmonic_shift_map_nee}, but for the respiration component.
  }
  \label{fig:harmonic_shift_map_resp}
\end{figure}

\end{document}